\documentclass[journal]{IEEEtran}

\usepackage[utf8]{inputenc} 
\usepackage[T1]{fontenc}
\usepackage{url}              
\usepackage{cite}             

\usepackage[cmex10]{amsmath}  

\usepackage{graphicx}         
\usepackage{booktabs}         

\usepackage[caption=false,font=footnotesize]{subfig}

\usepackage{xcolor}
\usepackage{amssymb}
\usepackage{epsfig,verbatim}
\usepackage{mathtools}
\usepackage{bbm}
\usepackage{enumitem}

\usepackage{caption}
\usepackage{subfig}

\usepackage{hyperref}       

\usepackage{pgfplots}
  \pgfplotsset{compat=newest}
\usetikzlibrary{plotmarks}
\usetikzlibrary{arrows.meta}
\usepackage{grffile}
\usepackage{amsmath}

\usepackage{mathtools, cuted}
\usepackage{array}

\long\def\comment#1{}

\newfont{\bbb}{msbm10 scaled 700}

\newfont{\bb}{msbm10 scaled 1100}

\newcommand{\EE}{\mathbb{E}}

\newcommand{\NN}{\mathbb{N}}

\newcommand{\RR}{\mathbb{R}}

\newcommand{\ZZ}{\mathbb{Z}}
\newcommand{\av}{{\bf a}}

\newcommand{\ev}{{\bf e}}

\newcommand{\uv}{{\bf u}}

\newcommand{\vv}{{\bf v}}
\newcommand{\xv}{{\bf x}}
\newcommand{\yv}{{\bf y}}
\newcommand{\zv}{{\bf z}}
\newcommand{\zerov}{{\bf 0}}

\newcommand{\Nm}{{\bf N}}

\newcommand{\Xm}{{\bf X}}
\newcommand{\Ym}{{\bf Y}}
\newcommand{\Zm}{{\bf Z}}

\newcommand{\Ac}{{\cal A}}
\newcommand{\Bc}{{\cal B}}
\newcommand{\Cc}{{\cal C}}

\newcommand{\Ec}{{\cal E}}

\newcommand{\Hc}{{\cal H}}
\newcommand{\Ic}{{\cal I}}

\newcommand{\Nc}{{\cal N}}

\newcommand{\Pc}{{\cal P}}

\newcommand{\Rc}{{\cal R}}
\newcommand{\Sc}{{\cal S}}

\newcommand{\Uc}{{\cal U}}

\newcommand{\Vc}{{\cal V}}
\newcommand{\Xc}{{\cal X}}
\newcommand{\Yc}{{\cal Y}}

\newcommand{\hsf}{{\mathsf h}}

\newcommand{\Hsf}{{\mathsf H}}

\newcommand{\Msf}{{\mathsf M}}

\newcommand{\Vsf}{{\mathsf V}}

\newcommand{\Zsf}{{\mathsf Z}}

\newcommand{\rs}{{\boldsymbol r}}

\newcommand{\epsilonv}{\hbox{\boldmath$\epsilon$}}

\newcommand{\var}{{\hbox{Var}}}

\newtheorem{theorem}{Theorem}

\newtheorem{definition}{Definition}

\newtheorem{lemma}{Lemma}

\newtheorem{corollary}{Corollary}

\newtheorem{remark}{Remark}

\newtheorem{proposition}{Proposition}

\newcommand{\argmax}{\operatornamewithlimits{argmax}}
\newcommand{\argmin}{\operatornamewithlimits{argmin}}

\newtheorem{example}{Example}

\begin{document}

\title{A Comprehensive Study on Ziv-Zakai \\ Lower Bounds on the MMSE} 

\author{Minoh~Jeong,
        Alex~Dytso,~\IEEEmembership{Senior~Member,~IEEE,}
        and~Martina~Cardone,~\IEEEmembership{Senior~Member,~IEEE}
\thanks{At the time of the manuscript submission, M. Jeong was with the Department of Electrical and Computer Engineering at the University of Minnesota. He is now with the Department of Electrical and Computer Engineering, University of Michigan, Ann Arbor, MI 48109, USA e-mail: minohj@umich.edu}
\thanks{A. Dytso is with Qualcomm Flarion Technology, Inc., Bridgewater, NJ 08807, USA e-mail: odytso2@gmail.com}
\thanks{M. Cardone is with the Department of Electrical and Computer Engineering at the University of Minnesota, Minneapolis, MN 55455, USA e-mail: mcardone@umn.edu}
\thanks{
Part of this work was presented at the 2023 IEEE International Symposium on Information Theory~\cite{MinohISIT2023}.
}
\thanks{Manuscript received April 19, 2005; revised August 26, 2015.}
}

\markboth{Journal of \LaTeX\ Class Files,~Vol.~14, No.~8, August~2015}%
{Shell \MakeLowercase{\textit{et al.}}: Bare Demo of IEEEtran.cls for IEEE Journals}

\maketitle

\begin{abstract}
This paper explores Bayesian lower bounds on the minimum mean squared error (MMSE) that belong to the well-known Ziv-Zakai family. The Ziv-Zakai technique relies on connecting the bound to an $\Msf$-ary hypothesis testing problem. There are three versions of the Ziv-Zakai bound (ZZB): the first version relies on the so-called \emph{valley-filling function}, the second one is a relaxation of the first bound which omits the valley-filling function, and the third one, namely the single-point ZZB (SZZB), replaces the integration present in the first two bounds with a single point maximization.

The first part of this paper focuses on providing the most general version of the bounds. 
It is shown that these bounds hold without any assumption on the distribution of the estimand. This makes the bounds applicable to discrete and mixed distributions. 
Then, the SZZB is extended to an $\Msf$-ary setting and a version of it that holds for the multivariate setting is provided. 

In the second part,  general properties of these bounds are provided. First, unlike the Bayesian \emph{Cram\'er-Rao bound}, it is shown that all the versions of the ZZB \emph{tensorize}.  Second, a characterization of the \emph{high-noise} asymptotic is provided, which is used to argue about the  tightness of the bounds. Third, a complete \emph{low-noise} asymptotic is provided under the assumptions of mixed-input distributions and Gaussian additive noise channels.   
In the low-noise, it is shown that the ZZB is generally tight, but there are examples for which 
the SZZB is not tight. 

In the third part, the tightness of the bounds is evaluated.  First, it is shown that in the low-noise regime the ZZB without the valley-filling function, and, therefore, also the ZZB with the valley-filling
function, are tight for mixed-input distributions and Gaussian additive noise channels.  Second, for discrete inputs it is  shown that the ZZB with the valley-filling function is always sub-optimal, and equal to zero without the valley-filling function. Third, unlike for the ZZB, an example is shown for which the SZZB is tight to the MMSE for discrete inputs. Fourth, sufficient and necessary conditions for the tightness of the bounds are provided.  Finally, some examples are provided in which the bounds in the Ziv-Zakai family outperform other well-known Bayesian  lower bounds, namely the Cram\'er-Rao bound and the maximum entropy bound.

\end{abstract}



\section{Introduction}
\label{sec:intro}
\IEEEPARstart{T}{he} mean squared error (MSE) is the standard fidelity metric in  estimation problems. In a Bayesian framework, its minimum value, i.e., the minimum mean squared error (MMSE), is attained by the conditional expectation of a parameter of interest. 
Regrettably, a closed-form characterization of the MMSE is often difficult due to the fact that the conditional expectation for an arbitrary estimation problem is difficult to 
characterize. 
Because of this, one often needs to rely on lower bounds on the MMSE.

In the literature, a plethora of different MMSE lower bounds have been  developed and analyzed, which can be loosely grouped into several families of which we mention the most popular. 
A first family, namely the so-called Weiss-Weinstein family~\cite{weinstein1988general}, consists of lower bounds derived from the Cauchy-Schwarz inequality. This family includes the ubiquitous Cram\'er-Rao bound (also known as the van Trees bound~\cite{van2004detection}) and several others, such as  the Bobrovsky–Zakai bound~\cite{bobrovsky1976lower,koike2021improvement} and the Bhattacharyya bound~\cite{bhattacharyya1946some}.
A second family of Bayesian MMSE lower bounds is the one that relies on the maximum entropy principle~\cite{Cover:InfoTheory}, which connects the MMSE and the conditional entropy. 
A third family of bounds leverages the solution to the  rate-distortion function~\cite{goblick1965theoretical}, provided that such a solution exists in closed form; see also~\cite{merhav2012data} for a more generic approach.
A fourth family of bounds utilizes functional inequalities, such as the Poincar\'e inequality~\cite{ISIT2022_Poincare} and the log-Sobolev inequality~\cite{aras2019family}. 
A fifth family of bounds is based on the variational representation of information divergences, such as the $f$-divergence; see~\cite{saito2022meta,esposito2021lower,chen2016bayes,xu2016information}.
A sixth family minimizes the MMSE over all priors that satisfy certain constraints, such as priors with a bounded relative entropy to the Gaussian distribution~\cite{KL_bounds,fauss2021variational}; this approach often leads to closed-form expressions.
The final family is under consideration in this work: the Ziv-Zakai family, which provides a lower bound on the MMSE using a connection to an $\Msf$-ary hypothesis testing problem~\cite{ZZbound_one_of_the_originals,Seidman1970,Chazan1975,Bellini1974,bell1995performance,Bell1997}.

There are several different versions of the Ziv-Zakai bound (ZZB)~\cite{ZZbound_one_of_the_originals,Seidman1970,Chazan1975,Bellini1974,bell1995performance,Bell1997}, all of which rely on the same core technique, but have different secondary  steps. Specifically, there are three versions of the ZZB: the first version relies on the so-called \emph{valley-filling function}, the second one is a relaxation of the first bound which omits the valley-filling function, and the third one, namely the single-point ZZB (SZZB), replaces the integration present in the first two bounds with a single point maximization. 
For a \emph{scalar} parameter $X\in\RR$ estimation problem, the first two versions of the ZZB~\cite{Bell1997} on the MMSE are given by the following chain of inequalities, 
\begin{align}\label{eq:basicZZB}
    & \EE[|X - \hat{X}|^2] \nonumber \\
    & \geq \frac{1}{2} \int_0^\infty \!\!\! t \Vc_t\left\{\int_{-\infty}^\infty (f_X(x) \! +\! f_X(x+t))P_e(x,x+t) {\rm d}x \right\} {\rm d}t \nonumber \\
    & \geq \frac{1}{2} \int_0^\infty \!\!\! t \! \int_{-\infty}^\infty (f_X(x) \!+\! f_X(x+t))P_e(x,x+t) \ {\rm d}x \ {\rm d}t,
\end{align}
where $\hat{X}$ is the estimate of $X$, $\Vc_t\{g(t)\} = \sup_{u:u\geq t} g(u)$ is the valley-filling function, and $P_e(x,x+t)$ is the minimum (Bayes) error probability of the binary hypothesis testing problem between $\Hc_0: X = x$ and $\Hc_1: X = x+t$ with prior probabilities $\Pr(\Hc_0) = \frac{f_X(x)}{f_X(x)+f_X(x+t)}$ and $\Pr(\Hc_1) = 1 - \Pr(\Hc_0)$. The first inequality in~\eqref{eq:basicZZB} is the ZZB with the valley-filling function, and the second is the ZZB without the valley-filling function.
To the best of our knowledge, the most general and tightest version of the two ZZBs in~\eqref{eq:basicZZB} was presented by Bell et al. in~\cite{bell1995performance}. In particular, the main contributions of~\cite{bell1995performance} were: 1) to extend the ZZB to the vector case; prior to this, the bound worked only for the scalar parameter case, and 2) to increase the number of hypotheses for $P_e$ in the ZZB beyond the binary hypothesis setting, which tightens the bound. 

For a scalar parameter estimation, the third version of the ZZB, namely the SZZB, is defined as
\begin{align}\label{eq:basicSZZB}
    & \EE[|X - \hat{X}|^2] \nonumber \\
    & \geq \max_{t>0} \frac{t^2}{2} \int_{-\infty}^\infty (f_X(x) + f_X(x+t))P_e(x,x+t) {\rm d}x.
\end{align}
In contrast to the ZZB, the SZZB~\cite{bell1995performance} has not been explored much. To the best of our knowledge,~\eqref{eq:basicSZZB} is the most general version of the SZZB~\cite{bell1995performance}, which is only for a scalar parameter case and a binary hypothesis setting. 

The ZZB has several appealing advantages. 
Unlike the other bounds, in fact, the ZZB only requires one regularity condition, that is, the parameter under estimation needs to have a probability density function (PDF); this is one of the key advantages of the ZZB. Hence, the ZZB has a broader applicability than, for instance, the Cram\'er-Rao bound, which requires several smoothness assumptions on the PDF of the estimand. 
Moreover, the ZZB is one of the tightest bounds in the literature, leading to numerous applications of it; see~\cite{ZZ_high_noise} for an analysis on the tightness in the high-noise regime. For instance, the ZZB has been leveraged in estimating the quantum parameter~\cite{tsang2012ziv,PhysRevLett.Giovannetti2012,Gao_2012,PhysRevA.Zhang2014,berry2015quantum,Rubio_2018,PhysRevResearch.Zhang2022,PhysRevLett.Zhuang2022}, time delay~\cite{Mishra2017TDE,9794611}, time of arrival~\cite{Driusso2015TOA,Laas2021TOA,Wang2022TOA,Gifford2022TOA,graff2024zivzakaioptimal}, position~\cite{Keskin2016VL,Closas2017,gusi2018ziv}, direction of arrival~\cite{Khan2010DOA,Gupta2019DOA,Alexander2019DOA,zhang2022ziv,ZZdoaGPE,2D_DoA}, and in MIMO radar systems~\cite{Chiriac2010Radar,chiriac2015ziv}.

The ZZB is derived by bounding the \emph{outage probability} (i.e., the probability that the estimation error is above a given threshold) using a hypothesis testing approach. There are, however, other ways of bounding the outage probability. In particular, the authors of~\cite{GenOutageBoundsApproach} provided one of the most general settings by lower bounding the outage probability using the \emph{reverse H\"older's inequality} with an auxiliary function that can be optimized. This approach is akin in spirit to the Weiss-Weinstein method, however the auxiliary functions are introduced at the MMSE level and not at the outage probability level. The authors of~\cite{GenOutageBoundsApproach} showed that with a proper choice of the auxiliary function, one can recover the scalar version of the ZZB in~\cite{bell1995performance}. 
Interestingly, the results of~\cite{GenOutageBoundsApproach} also allow for a version of the ZZB with an infinite number of hypotheses.

In spite of the wide applicability of the ZZB, several important questions still remain unanswered. 
First of all, {\em is the ZZB applicable to discrete or mixed prior distributions?} The ZZB, in fact, holds under the regularity condition that the estimand needs to have a PDF, which makes the ZZB not applicable to the cases of discrete or mixed distributions. Discrete and mixed distributions play an important role both practically and theoretically in a variety of applications; see, for example,~\cite{wu2012optimal} where mixed distributions are used in compressed sensing, or~\cite{mohajer2025poincarelowerboundapproach} where mixed distributions are used for integrated sensing and communication. Thus, eliminating this condition would broaden the applicability of the ZZB and make it universal in the sense that it will not require any regularity conditions.
Second, {\em how does the bound behave in important asymptotic regimes}? 
Understanding whether a lower bound is tight to the MMSE is significant since it determines when and whether the bound should be used.
Some preliminary results on the tightness have been derived in~\cite{bell1995performance} and~\cite{ZZ_high_noise}.
Due to the cumbersome expression of the ZZB, however, such an analysis on the tightness of the bound is often intractable, and hence, an asymptotic analysis of the bound could help simplify its form and understand how the bound behaves. In particular, the high-noise regime is practically relevant, and thus an analysis in this regime would provide meaningful insights into the performance of the bound.
Third, {\em among the three ZZB-type of bounds, which bound should we use?} 
For example, the introduction of the valley-filling function tightens the bound, but it also introduces one additional computational step in the evaluation of it.
It is also not clear how the valley-filling function relates to the parameter $\mathsf{M}$ of the bound and which of these contributes the most to the refinement in the tightness of the ZZB. It is even unknown whether the ZZB with the valley-filling function is tighter than the SZZB. 
Lastly, {\em do the ZZBs tensorize?} It is well-known that bounds such as the Bayesian Cram\'er-Rao do not tensorize, which limits their effectiveness in high-dimensional settings. Given the rather involved structure of the ZZBs, it is not immediately clear 
if this property holds.

\subsection{Contributions and Organization}

\begin{itemize}
	 \item In Section~\ref{sec:pre}, we start with a few definitions that we will leverage to make the expressions of the lower bounds in the Ziv-Zakai family consistent and concise. 
     We also  formulate the problem of interest. 
	 
  \item In Section~\ref{sec:ZZB}, we seek to present the most general version of the bounds in the Ziv-Zakai family. In particular,
 \begin{itemize}
 \item  In Section~\ref{sec:gen_Form_bound}, Theorem~\ref{thm:BZZ_Mary_vec_mmse} generalizes ZZB in a measure-theoretic way to avoid the regularity condition of the original ZZB that required that the prior distribution needs to be continuous. This generalization leads to lower bounds on the MMSE applicable to discrete and mixed distributions. Similarly,  Theorem~\ref{thm:SZZB} presents a general measure theoretic form for the SZZB.
 In addition,   previously the SZZB relied only on binary hypothesis testing;  we here extend the SZZB to the $\mathsf{M}$-ary case. Importantly, we also extend the SZZB to the vector parameter case. 
 \item In Section~\ref{sec:tensorization}, Proposition~\ref{prop:tensorization} shows that all the three versions of the bound are tensorizable.  
 \end{itemize}

  \item  In Section~\ref{sec:asymp}, we characterize the asymptotics of the ZZB family. In particular,
  \begin{itemize}
  \item In Section~\ref{sec:High_Noise_Assymptotics}, Theorem~\ref{thm:zzb_sig_inf_M}  presents the {\em high-noise} behavior of the ZZB and of the SZZB for a rather general set of noise distributions. 
  \item In Section~\ref{sec:Low_Noise_Assymptotics}, Theorem~\ref{thm:low_noise} derives the {\em low-noise} asymptotics of the ZZB for the practically relevant additive Gaussian noise model and any prior that can be written as a mixture of a discrete and a continuous distribution.  Importantly, using this result, we prove that the ZZB has the same rate of convergence as the MMSE implying the tightness of the ZZB. To the best of our knowledge, this is the first Bayesian MMSE lower bound holding for discrete and mixed prior distributions, which is tight to the MMSE for the additive Gaussian channel in the low-noise regime.
  Moreover, Proposition~\ref{prop:assum_low_SZZB} derives a lower bound and an upper bound for the rate of convergence of the SZZB in the low-noise regime for the aforementioned setting.  Furthermore, Example~\ref{ex:szzb_low_noise_gaussian} shows that the SZZB is not tight in general in the low-noise regime.
  \end{itemize}

\item In Section~\ref{sec:prop_comp}, we analyze the tightness of the ZZB and of the SZZB under several aspects. In particular,
\begin{itemize}
    \item  In Section~\ref{sec:discted_tight},  effects  of the valley-filling function are considered.  Specifically, Proposition~\ref{prop:discrete_zero_zzb} shows that the ZZB without the valley-filling function is always equal to zero for discrete distributions; this shows that the valley-filling function is necessary for this family of distributions.  Moreover, Theorem~\ref{thm:Discrete_not_tight} shows that for discrete distributions even with the valley-filling function, the ZZB is always sub-optimal.  Finally, as opposed to the ZZB, Example~\ref{ex:bern} shows an example of a discrete input for which the SZZB is tight.
    \item In Section~\ref{sec:tighntess_cont}, Proposition~\ref{prop:mmse=ZZB} characterizes necessary and sufficient conditions for the tightness of the ZZB without the valley-filling function.  For the case of a univariate estimand, this improves the result in~\cite{bell1995performance}: it shows that a sufficient condition in~\cite{bell1995performance} is also necessary for the bound to be tight (see Corollary~\ref{cor:unimodal_symmetric}).  
    For the SZZB, Proposition~\ref{prop:SZZB_condtion} presents a necessary condition for the tightness. This condition implies that for continuous prior and channel distributions, the SZZB is always sub-optimal. 
    \item In Section~\ref{ZZB vs. SZZB}, it is demonstrated (see Example~\ref{ex:unif_mixed_unif}) that the ZZB with the valley-filling function is {\em not} always tighter than the SZZB. 
    This is a somewhat surprising result since a more computationally involved ZZB has been typically thought to be tighter than the SZZB.
    \item In Section~\ref{sec:comp_MEB},  numerical evaluations are presented.   Specifically,  numerical examples are provided that show that the bounds in the Ziv-Zakai family outperform the Bayesian Cram\'er-Rao bound~\cite{van2004detection} and the maximal entropy bound~\cite{Cover:InfoTheory}. 
\end{itemize}
	    
    \item In Section~\ref{sec:Conclusion}, we conclude the paper.
\end{itemize}

\section{Preliminaries and problem formulation}\label{sec:pre}

\subsection{Notation}
\begin{figure}[t]
    \centering
    \includegraphics[width=\linewidth]{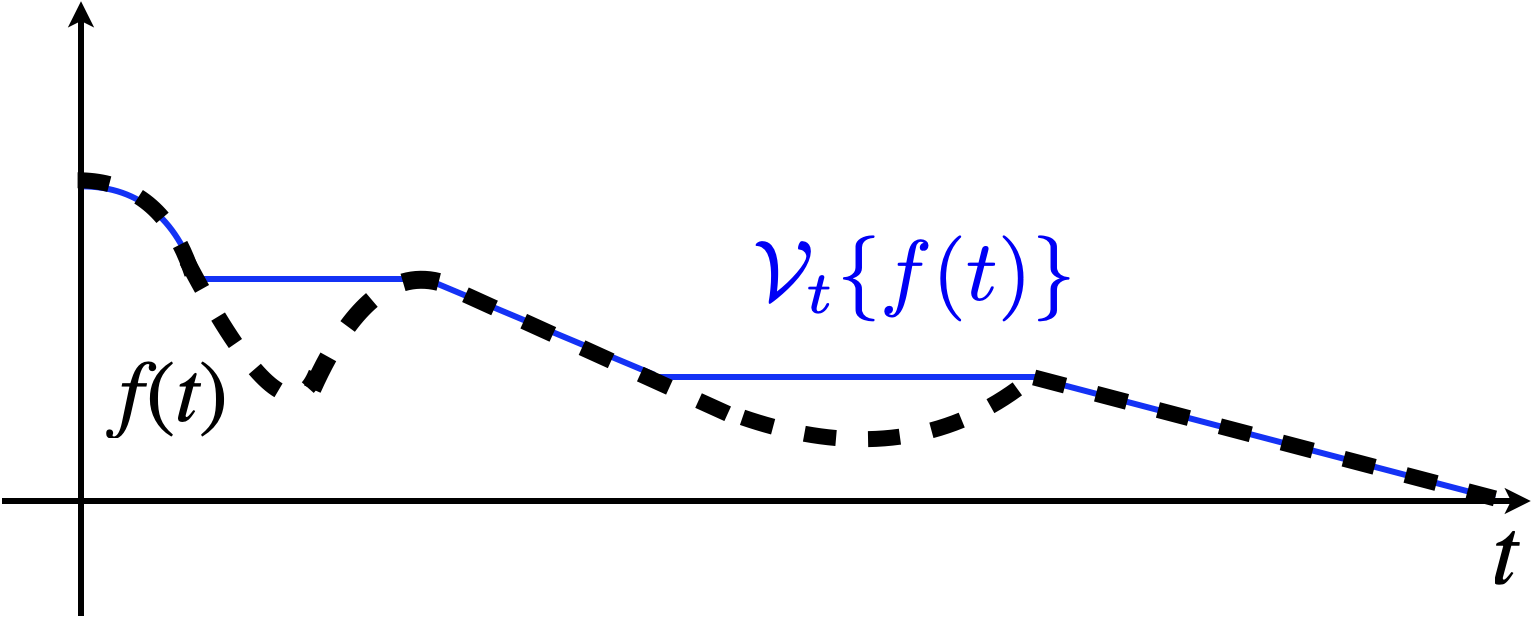}
    \caption{An example of $f(t)$ and $\Vc_t\{f(t)\}$.}
    \label{fig:valley_filling}
\end{figure}

Boldface upper case letters $\mathbf{X}$ denote random vectors; 
the boldface lower case letter $\mathbf{x}$ indicates a specific realization of $\mathbf{X}$; 
$X_i$ and $x_i$ (or $(\xv)_i$) denote the $i$th element of $\Xm$ and $\xv$, respectively; 
$\ev_i\in\RR^d$ is the $i$th standard basis vector that contains a one in the $i$th entry and a zero in all of the other entries; 
$[n_1: n_2]$ is the set of integers from $n_1$ to $n_2 \geq n_1$;
calligraphic letters $\Xc$ indicate sets/events; 
$\varnothing$ is the empty set;
$\left< \xv, \yv \right>$ is the inner product between $\xv$ and $\yv$; 
$I_n$ is the identity matrix of dimension $n$ and $\zerov_n$ is the $n$-dimensional vector of all zeros;
$\mathbbm{1}\{\mathcal{S}\}$  is the indicator function that yields 1 if $\mathcal{S}$ is true and 0 otherwise.
For a pair of random vectors $(\Xm,\Ym)$, we let $P_{\Xm}(\cdot)$ denote the distribution of $\Xm$, and $P_{\Ym|\Xm}(\cdot|\xv)$ be the distribution that governs the noisy observation model $\Ym|\Xm=\xv$. 
We denote by $d$ the dimension of $\Xm$.
For a function $f:\RR \to \RR$, the valley-filling function is defined as
\begin{equation}
\label{eq:VF}
	\Vc_t \{ f(t) \}
	 = \sup_{u: u\geq t} f(u).
\end{equation}
An example of a function and of its valley-filling function is provided in Fig.~\ref{fig:valley_filling}.

\subsection{Problem Formulation}
Consider the problem of estimating a random vector $\Xm \in \Xc$ from a noisy observation $\Ym \in \mathcal{Y}$, where $P_{\Xm,\Ym}$ denotes the joint distribution of $\Xm$ and $\Ym$. The alphabet $\Xc$ is assumed to be $\RR^d$.\footnote{$\Xc$ is assumed to be a Hilbert space for more general versions of our results in Appendix~\ref{app:gen_zzb_szzb}.} The alphabet $\mathcal{Y}$ of the observation $\Ym$ is assumed to be some abstract space, and it will be specified only when needed. As it is standard, we refer to the marginal distribution $P_{\Xm}$ as the prior on $\Xm$ and to the conditional distribution $P _{\Ym| \Xm}$ as the channel. The quality of estimating $\Xm $ from $\Ym$ is often captured by the MMSE, defined as 
\begin{align}\label{eq:def_mmse}
	{\rm mmse}(\Xm|\Ym)
	& = \EE \left[ \| \Xm - \EE[\Xm | \Ym ] \|^2\right] \nonumber \\
	& =  \sum_{i=1}^{d} \EE\left[ ( X_i - \EE[X_i | \Ym] )^2 \right].
\end{align}
As already mentioned, our goal is to derive the most general version of the lower bounds on the MMSE~\eqref{eq:def_mmse} that belong to the Ziv-Zakai family. In addition, we  seek to understand the settings under which this family of bounds is tight. 
The presentation of such bounds requires defining elements of  an $\mathsf{M}$-ary hypothesis testing setting.
\begin{definition}
\label{def:MHT}
Suppose that we are given a channel $P_{\Ym| \Xm}$.  For  an integer $\mathsf{M} \ge 2$ and  $\xv \in  \Xc$, consider the following { \em $\mathsf{M}$-ary hypothesis testing problem}, 
\begin{subequations}
\label{eq:hypotest}
\begin{equation}
	\Hc_i: \Ym \sim P_{\Ym|\Xm=\xv+\uv_i},  \ i \in [0:\mathsf{M}-1], 
\end{equation}
where $\{\uv_i\}_{i=0}^{\mathsf{M}-1}$ is some collection of points in $\Xc$, 
and
\begin{equation}
\Pr(\Hc_i)  = p_i .
\end{equation}
\end{subequations}
For a collection of probabilities $\mathcal{P}=\{ p_i \}_{i=0}^{\mathsf{M}-1}$ and a collection $\mathcal{U}=\{\uv_i\}_{i=0}^{\mathsf{M}-1}$,
{\em the minimum probability of error} for this $\mathsf{M}$-ary hypothesis testing problem is denoted by $P_e \left ({\xv}; \mathcal{P}, \mathcal{U}  \right )$.
\hfill $\square$
\end{definition}

An essential characteristic of the bounds belonging to the Ziv-Zakai family consists of connecting estimation and detection problems.
The function $\hsf_{\Msf}$ defined below captures such a connection and makes the expressions of the ZZB and of the SZZB consistent and concise. 

\begin{definition}\label{def:h_func}
For a set $\Uc= \{ \uv_k \}_{k=0}^{\mathsf{M}-1}\subset \Xc$ 
and $\xv\in \Xc$, we define a measure (denoted by $\mu_\Uc$) and a set of probability densities evaluated at $\xv$ (denoted by $\Pc_{\Uc}(\xv)$) as\footnote{By construction, the Radon-Nikodym derivative is always well-defined.}
\begin{subequations}
\begin{align}
	\mu_\Uc 
	& = \sum_{{k=0}}^{\mathsf{M}-1}  P_{\Xm-\uv_{k} },\label{eq:def_mu_U} \\
	\Pc_\Uc(\xv) 
	&= \left\{ p_i : p_i = \frac{ P_{\Xm - \uv_i}({\rm d} \xv)}{ \mu_\Uc ({\rm d} \xv)}, \uv_i\in\Uc \right\}.\label{eq:def_P_U}
\end{align}
\end{subequations}
For $t>0$, an integer $\Msf\geq2$, and $\av\in \Xc$, we define
 \begin{equation}\label{eq:BZZ_Mary_vec_mmse_h}
	\hsf_{\Msf}(t,\av,P_{\Xm,\Ym}) 
	 \! = \!\! \sup\limits_{\substack{  \mathcal{U}  \subset {\Xc}: \\ \mathcal{U}= \{ \uv_k \}_{k=0}^{\mathsf{M}-1} , \\  \left<\uv_k,\av\right>= kt,~\forall k}}  \, \int    P_e \left ({\xv}; \mathcal{P}_\mathcal{U}(\xv), \mathcal{U}  \right )  \,    \mu_{\mathcal{U}} ( {\rm{d}} \xv),
\end{equation}
where $P_e \left ({\xv}; \Pc_\Uc(\xv), \Uc \right )$ is the minimum error probability of the $\Msf$-ary hypothesis testing defined in Definition~\ref{def:MHT}. \hfill $\square$
\end{definition}

Dividing the measure $\mu_\Uc$ in~\eqref{eq:def_mu_U} by $\Msf$ makes it a probability measure, which  becomes a mixture distribution of $P_{\Xm-\uv_k},~k\in[0:\Msf-1]$ with uniform weights. Hence,  $\hsf_{\Msf}(t,\av,P_{\Xm,\Ym}) $ can be re-written as follows,
\begin{equation}\label{eq:alt_h}
    \hsf_{\Msf}(t,\av,P_{\Xm,\Ym}) 
     = \Msf \sup\limits_{\substack{  \mathcal{U}  \subset {\Xc}: \\ \mathcal{U}= \{ \uv_k \}_{k=0}^{\mathsf{M}-1} , \\  \left<\uv_k,\av\right>= kt,~\forall k}} \!\!\!\EE_{\Xm\sim \frac{\mu_\Uc }{\Msf}} [P_e \left ({\Xm}; \mathcal{P}_\mathcal{U}(\Xm), \mathcal{U}  \right )].
\end{equation}
As shown in~\eqref{eq:alt_h}, $\hsf_{\Msf}(t,\av,P_{\Xm,\Ym})$ is proportional to the maximum value of the expected error probability  over all possible  choices of $\Uc$ along $\Msf$ $(d-1)$-dimensional, parallel, equi-spaced hyperplanes, perpendicular to $\av$, intersecting the ray $\rs(s) = s \cdot \av,~ s\in \RR_+$ at $\rs(kt),~k\in[0:\Msf-1]$.
The function $ \hsf_{\Msf}$ appears, albeit in different ways, in all members of the ZZ family of bounds, and it will be instrumental in our analysis.

\section{Ziv-Zakai bounds}\label{sec:ZZB}

In this section, we present the general expressions of the ZZB and the SZZB, and we explore their tensorization properties.

\subsection{General Forms of the Bounds}
\label{sec:gen_Form_bound}
We start by generalizing the ZZB to hold without any restriction on the joint distribution $P_{\Xm,\Ym}$ (e.g., continuity). The proof largely depends on the ideas developed in~\cite{bell1995performance}, which require $\Xm$ to have a PDF.  By doing a more careful accounting of the terms in the Lebesgue integral, we generalize these results to any probability measure in $\mathbb{R}^d$.

\begin{theorem}\label{thm:BZZ_Mary_vec_mmse} 
For any integer $\mathsf{M}\geq2$, the following holds
\begin{equation}
	{\rm mmse}(\Xm|\Ym)
	\geq   \overline{\mathsf{ZZ}}(P_{\Xm,\Ym},\mathsf{M})
	\geq {\mathsf{ZZ}}(P_{\Xm,\Ym},\mathsf{M}),
\end{equation}
with 
\begin{align}
	\overline{\mathsf{ZZ}}(P_{\Xm,\Ym},\mathsf{M})
	& = \sum_{i=1}^{ d }  \int_0^\infty \frac{t}{2} \, \Vc_t \left\{ \frac{ \hsf_\Msf(t,\ev_i,P_{\Xm,\Ym}) }{\mathsf{M}-1} \right\}  \ {\rm d} t,  \label{eq:BZZ_Mary_vec_mmse_sub1} \\
	{\mathsf{ZZ}}(P_{\Xm,\Ym},\mathsf{M})
	& = \sum_{i=1}^{d}  \int_0^\infty \frac{t}{2} \,  \frac{{\mathsf{h}_\Msf }(t,\ev_i,P_{\Xm,\Ym})}{\mathsf{M}-1}  \ {\rm d} t,\label{eq:BZZ_Mary_vec_mmse_sub2}
\end{align}
where $\hsf_\Msf(t,\ev_i,P_{\Xm,\Ym})$ is defined in Definition~\ref{def:h_func} and $\Vc_t\{\cdot\}$ is the valley-filling function in~\eqref{eq:VF}.
\end{theorem}
\begin{IEEEproof}
The proof is provided in Appendix~\ref{app:gen_zzb}.
\end{IEEEproof}

Next, we generalize the SZZB in~\cite{bell1995performance} to hold for any distribution in $\mathbb{R}^d$ and arbitrary $\Msf\geq2$. 
\begin{theorem}\label{thm:SZZB}
For any integer $\Msf\geq 2$, the SZZB is given by
\begin{equation}
	{\rm mmse}(\Xm|\Ym)
	 \geq {\mathsf{ZZ}_{\rm sp}}(P_{\Xm,\Ym},\mathsf{M}),
\end{equation}
with
\begin{equation}\label{eq:SZZB}
	{\mathsf{ZZ}_{\rm sp}}(P_{\Xm,\Ym},\mathsf{M})
	 = \sum_{i=1}^d \sup_{\Delta>0} \frac{\Delta^2 \, \hsf_{\Msf}(\Delta, \ev_i,P_{\Xm,\Ym})}{2(\Msf-1)} .
\end{equation}
\end{theorem}
\begin{IEEEproof}
The proof is provided in Appendix~\ref{app:GSZZB}.
\end{IEEEproof}
\begin{remark}
In Appendix~\ref{app:gen_zzb} and Appendix~\ref{app:GSZZB}, we present more general versions of the bounds in Theorem~\ref{thm:BZZ_Mary_vec_mmse} and Theorem~\ref{thm:SZZB} that work when $\Xm \in \Xc$, where $\Xc$ is assumed to be a Hilbert space.  For ease of exposition this generality is, however, not pursued in the main body of the paper.  Finally, we note that all the three versions of the ZZB do not require any regularity condition on the priors, e.g., bounded  moments.~\hfill~$\square$
\end{remark}

\subsection{On Tensorization}
\label{sec:tensorization}
The expressions of the ZZB in Theorem~\ref{thm:BZZ_Mary_vec_mmse} and of the SZZB in Theorem~\ref{thm:SZZB} are quite cumbersome, particularly because of the presence of an inner layer of optimization in~\eqref{eq:BZZ_Mary_vec_mmse_h}. It is, therefore, important to understand whether these bounds admit any simplifications. 
For instance, it is well-known that under the assumption that the pairs $(X_i,\Ym_i)_{i=1}^d$ are generated independently and identically distributed, we have that the MMSE tensorizes, i.e., ${\rm mmse}(\Xm|\Ym) = \sum_{i=1}^d {\rm mmse}(X_i|\Ym_i)$.
The next proposition, which also concludes this section, indeed shows that these bounds tensorize.

\begin{proposition}\label{prop:tensorization}
If $P_\Xm = \prod_{i=1}^d P_{X_i}$ and $P_{\Ym|\Xm} = \prod_{i=1}^d P_{\Ym_i|X_i}$, it holds that
\begin{subequations}
\begin{align}
	& \overline{\mathsf{ZZ}}(P_{\Xm,\Ym},\mathsf{M})
	=  \sum_{i=1}^{d} \overline{\mathsf{ZZ}}(P_{X_i,{\Ym_i}},\mathsf{M}), \\
	& {\mathsf{ZZ}}(P_{\Xm,\Ym},\mathsf{M})
	=  \sum_{i=1}^{d} {\mathsf{ZZ}}(P_{X_i,{\Ym_i}},\mathsf{M}) , \\
	& \mathsf{ZZ}_{\rm sp} (P_{\Xm,\Ym}, \Msf)
	= \sum_{i=1}^d  \mathsf{ZZ}_{\rm sp} (P_{X_i,{\Ym_i}}, \Msf). 
\end{align}
\end{subequations}
\end{proposition}
\begin{IEEEproof}
The proof is provided in Appendix~\ref{app:sec:tensorization}.
\end{IEEEproof}
The fact that the ZZB in Theorem~\ref{thm:BZZ_Mary_vec_mmse} and the SZZB in Theorem~\ref{thm:SZZB} do tensorize showcases another advantage of these bounds  over other bounds such as the Bayesian Cram\'er-Rao bound, which does not tensorize in general (see Appendix~\ref{app:CRBNoTens} for an example).

\section{Asymptotics}\label{sec:asymp}
In this section, we analyze the high-noise and low-noise asymptotics of the ZZB in Theorem~\ref{thm:BZZ_Mary_vec_mmse} and of the SZZB in Theorem~\ref{thm:SZZB}.
\subsection{High-Noise Asymptotics}
\label{sec:High_Noise_Assymptotics}
Here, we study the ZZB in Theorem~\ref{thm:BZZ_Mary_vec_mmse} and the SZZB in Theorem~\ref{thm:SZZB} in the high-noise regime. 
In particular, we characterize the asymptotics under the following natural assumptions~\cite{ZZ_high_noise}:
\begin{itemize}
	\item {\bf A1:} $P_{\Ym|\Xm}$ can be parameterized by $\eta\geq0$, which is referred to as the {\em noise level}, i.e., $P_{\Ym|\Xm}(\yv|\xv;\eta)$ for all $(\xv,\yv)\in(\Xc,\Yc)$. 
	In order to highlight the dependence on~$\eta$, we let $P_e \left ({\xv}; \mathcal{P}_\mathcal{U}(\xv), \mathcal{U}  \right ) = P_e \left (\eta,{\xv}; \mathcal{P}_\mathcal{U}(\xv), \mathcal{U}  \right ) $ denote the optimal error probability in Definition~\ref{def:MHT}. \label{item:A1}
	\item {\bf A2:} $P_e \left (\eta,{\xv}; \mathcal{P}_\mathcal{U}(\xv), \mathcal{U}  \right ) $ is non-decreasing in $\eta$. \label{item:A2}
	\item {\bf A3:} $\lim\limits_{\eta\to\infty} P_e \left (\eta, {\xv}; \mathcal{P}_\mathcal{U}(\xv), \mathcal{U}  \right )  = 1 -  \max\limits_{p\in\Pc_\Uc(\xv)} p $. \label{item:A3}
\end{itemize}
We highlight that most, if not all, practical noise models satisfy these conditions.

We also note that under the assumptions above, in the high-noise regime the MMSE converges to the variance of $\Xm$, i.e.,
\begin{equation}
\label{eq:HighNoiseMMSEVar}
	\lim_{\eta \to \infty} {\rm mmse}(\Xm | \Ym) 
	= \sum_{i=1}^{d}\var(X_i).
\end{equation}
We will use the  expression in~\eqref{eq:HighNoiseMMSEVar} in later sections for comparisons and analysis of the tightness of the bounds. Moreover,~\cite{ZZ_high_noise}  leveraged \eqref{eq:HighNoiseMMSEVar} and  
assumptions \textbf{A1-A3} to provide a few examples for the univariate case for which the ZZB is tight and not tight in the high-noise regime.
We now define the high-noise asymptotics of the ZZB and SZZB.
\begin{definition}\label{def:high_noise}
The high-noise asymptotics of the ZZB and SZZB are defined as follows,
\begin{subequations}
\begin{align}
	& \overline{\mathsf{V}}(P_\Xm,\mathsf{M}) 
	= \lim_{\eta\to\infty} \overline{\mathsf{ZZ}}(P_{\Xm,\Ym},\mathsf{M}) , \\
	& {\mathsf V}(P_\Xm,\mathsf{M}) 
	= \lim_{\eta\to\infty} {\mathsf{ZZ}}(P_{\Xm,\Ym},\mathsf{M})  ,\\
	& \Vsf_{\rm sp}(P_\Xm,\Msf) 
	= \lim_{\eta\to\infty} {\mathsf{ZZ}_{\rm sp}}(P_{\Xm,\Ym},\mathsf{M}).
\end{align}
\end{subequations}
\end{definition}
The next theorem characterizes the asymptotics in Definition~\ref{def:high_noise}.

\begin{theorem}\label{thm:zzb_sig_inf_M}
For any $P_\Xm$ and any $\mathsf{M}\ge 2$, we have that
\begin{subequations}
\begin{align}
	& \overline{\mathsf V}(P_\Xm,\mathsf{M})
	=   \sum_{i=1}^{d}  \int_0^\infty \frac{t}{2} \Vc_t\left\{ \frac{ \mathsf{M} - \mathsf{H}_\Msf(t,\ev_i,P_\Xm)  }{\mathsf{M}-1} \right\} {\rm d} t, \label{eq:high_noise_valley_ZZB}\\
	& {\mathsf V}(P_\Xm,\mathsf{M})
	=   \sum_{i=1}^{d} \int_0^\infty \frac{t}{2}  \frac{ \mathsf{M} - \Hsf_\Msf(t,\ev_i,P_\Xm) }{\mathsf{M}-1} \ {\rm d} t, \label{eq:high_noise_ZZB}\\
	& \Vsf_{\rm sp}(P_\Xm, \Msf)
	= \sum_{i=1}^{d} \sup_{\Delta>0} \frac{\Delta^2}{2} \frac{\Msf - \Hsf_\Msf(\Delta,\ev_i,P_\Xm)}{\Msf-1},\label{eq:high_noise_SZZB}
\end{align}
\end{subequations}
where
\begin{equation}
\label{eq:h_high_noise}
	\mathsf{H}_\Msf(t,\ev_i,P_\Xm)
	 =  \inf\limits_{\substack{  \mathcal{U}  \subset \mathcal{X}: \\ \mathcal{U}= \{ \uv_k \}_{k=0}^{\mathsf{M}-1}  \\   (\uv_k)_i =kt,~\forall k }}  \int  \max_{j\in[0:\mathsf{M}-1]}  P_{\Xm-\uv_{j} }( {\rm{d}}\xv)  .
\end{equation}
\end{theorem}

\begin{IEEEproof}
The proof is provided in Appendix~\ref{app:high_noise}.
\end{IEEEproof}
In subsequent sections, we will rely on the quantities $\overline{\mathsf V}(P_\Xm,\mathsf{M}),  {\mathsf V}(P_\Xm,\mathsf{M})$ and $\Vsf_{\rm sp}(P_\Xm, \Msf)$ in Theorem~\ref{thm:zzb_sig_inf_M} to asses the tightness of the bounds by comparing them to the limit in~\eqref{eq:HighNoiseMMSEVar}.  

\subsection{Low-Noise Asymptotics}
\label{sec:Low_Noise_Assymptotics}
Here, we study the ZZB in Theorem~\ref{thm:BZZ_Mary_vec_mmse} and the SZZB in Theorem~\ref{thm:SZZB} in the low-noise regime, i.e., when $\eta \to 0$. 
Unlike the high-noise regime, in the low-noise regime the convergence rate highly depends on the noise distribution.  Therefore, one needs to fix a noise model to provide any quantitative statements in the low-noise regime.
We here focus on the practically relevant additive Gaussian noise channel. In particular, we consider the following channel model
\begin{equation}\label{eq:AWGN}
	\Ym = \Xm +\Nm, \text{ where } \Nm\sim\Nc(\zerov_d, {\eta I_d}),
\end{equation}
where $d$ is the dimension of $\Xm$ and $\eta$ is the noise level defined in Section~\ref{sec:High_Noise_Assymptotics}. We note that, for the channel model in~\eqref{eq:AWGN}, it holds that $\lim_{\eta\to 0} {\rm mmse}(\Xm | \Ym) = 0$, which motivates our aim to characterize the rate of convergence. 

For the channel model in~\eqref{eq:AWGN}, several MMSE results in the low-noise regime exist. It is known that whenever $\Xm$ is a discrete random variable, then the MMSE decays much faster than linearly in $\eta$~\cite[Section IV-B]{MMSEdim}.
Differently, when $\Xm$ is a continuous random variable, then the MMSE decays linearly in $\eta$~\cite[Section IV-C]{MMSEdim}.
A fairly general result on the behavior of the MMSE is given  in~\cite{David2016,MMSEdim,barletta2025estimation}:   for any $P_\Xm$ of the following form,
\begin{equation} 
	P_{\Xm} = \alpha P_{\Xm_C} + (1-\alpha)P_{\Xm_D},~0\leq \alpha \leq 1, \label{eq:mixture_def}
\end{equation}
where $P_{\Xm_C}$ is an absolutely continuous distribution with respect to a $d$-dimensional Lebesgue measure, and $P_{\Xm_D}$ is discrete (i.e., atomic), it holds that
\begin{equation}
\label{eq:MMSELowNoiseAsymptotic}
	\lim_{\eta \to0} \frac{{\rm mmse}(\Xm|\Ym)}{\eta} 
	= \alpha d .
\end{equation}
The next theorem proves that the low-noise asymptotic of the ZZB in Theorem~\ref{thm:BZZ_Mary_vec_mmse} for the channel model in~\eqref{eq:AWGN} is indeed equal to~\eqref{eq:MMSELowNoiseAsymptotic}.

\begin{theorem}\label{thm:low_noise}
Consider any $P_\Xm$ of the form in~\eqref{eq:mixture_def}.
Then, for the channel model in~\eqref{eq:AWGN}, it holds that
\begin{equation}\label{eq:low_noise_zzb}
	\lim_{\eta \to0} \frac{ {\mathsf{ZZ}}(P_{\Xm,\Ym},2 )}{\eta} 
	 = \alpha d.
\end{equation}

\end{theorem}
\begin{IEEEproof}
The proof is provided in Appendix~\ref{app:low_noise_proof}.
\end{IEEEproof}

\begin{remark}
Theorem~\ref{thm:low_noise} demonstrates that the ZZB is tight in the low-noise regime under the channel model in~\eqref{eq:AWGN} and with a prior that can be decomposed as in \eqref{eq:mixture_def}. Moreover, in the low-noise regime, the ZZB can be used in its simplest form in~\eqref{eq:BZZ_Mary_vec_mmse_sub2} with $\mathsf{M}=2$ and without the valley-filling function.  
Furthermore, we note that while for a continuous $\Xm$, there are many bounds that are tight in the low-noise regime (e.g., Bayesian Cram\'er-Rao), we are not aware of any lower bounds that are tight in the low-noise regime for mixed distributions other than the ZZB. Note that mixed distributions are important in compressed sensing applications (see, for example,~\cite{wu2012optimal}). 
\hfill $\square$
\end{remark}

We now turn our attention to the SZZB in Theorem~\ref{thm:SZZB}, and we seek to understand if it is tight in the low-noise regime.  We have the following partial answer.
\begin{proposition} \label{prop:assum_low_SZZB} For $P_{\Xm}$ in~\eqref{eq:mixture_def} and channel model in~\eqref{eq:AWGN}, we have that 
   \begin{equation}\label{eq:low_noise_szzb}
    \gamma   \alpha d
     \leq \lim_{\eta\to0} \frac{\Zsf\Zsf_{\rm sp}(P_{\Xm,\Ym},2)}{\eta} 
    \leq \alpha d,
\end{equation} 
where $\gamma = 4 \sup_{t>0}t^2 Q(t) \approx 0.662$ with $Q(t) = \frac{1}{\sqrt{2\pi}} \int_{t}^{\infty}{\rm{e}}^{-\frac{x^2}{2}} {\rm d}x $.
\end{proposition}
\begin{IEEEproof}
    The proof is provided in Appendix~\ref{app:low_noise_szzb}.
\end{IEEEproof}
We conjecture that the upper bound in~\eqref{eq:low_noise_szzb} is strict.  The next example, which demonstrates that the lower bound in~\eqref{eq:low_noise_szzb}  can be tight (see Appendix~\ref{app:szzb_low_noise_gaussian} for the computation), shows that, unlike the ZZB, the SZZB is not tight in general in the low-noise regime. 
\begin{example}\label{ex:szzb_low_noise_gaussian}
Let $P_X= \Nc(0,1)$ and $P_{Y|X}(\cdot|x)=\Nc(x,\eta )$. Then, for every $\Msf \ge 2$,
\begin{equation}
    \lim_{\eta\to0}\frac{\Zsf\Zsf_{\rm sp}(P_{X,Y},\Msf)}{\eta}
     = \gamma,
\end{equation}
which agrees with the lower bound in~\eqref{eq:low_noise_szzb}.
\end{example}

\section{Properties and comparison}\label{sec:prop_comp}
In this section, we analyze properties of the ZZB in Theorem~\ref{thm:BZZ_Mary_vec_mmse} and of the SZZB in Theorem~\ref{thm:SZZB}, and we investigate how well the ZZB family performs with respect to other standard Bayesian MMSE lower bounds.

\subsection{ZZB for Discrete Inputs and the Need for the Valley-Filling Function}
\label{sec:discted_tight}
The two expressions in Theorem~\ref{thm:BZZ_Mary_vec_mmse} differ only in the valley-filling function. Thus, a natural question arises: Under which conditions, does the valley-filling function provide an improvement?

In the literature, the ZZB with valley-filling function has been generally studied under the assumption that the distribution of $\Xm$ has a PDF~\cite{bell1995performance}. The case of discrete random variables is indeed typically ignored and  often erroneously assumed to be as trivial as replacing the PDF with the corresponding probability mass function (PMF). We here demonstrate and emphasize that care needs to be taken when dealing with discrete distributions. 
To verify this claim, we present the following  series of results. First, we prove that for a discrete $\Xm$, the ZZB without the valley-filling function is always equal to zero. Second, we present a simple example that shows that the bound with the valley-filling function is not equal to zero. These two results imply that for discrete inputs, the valley-filling function is an indispensable component of the ZZB. Third, the final result in this series shows that even with the valley-filling function, the ZZB is strictly sub-optimal for discrete inputs. 
The next proposition shows that for a discrete $\Xm$, the ZZB without the valley-filling function is always equal to zero.
\begin{proposition}\label{prop:discrete_zero_zzb}
Suppose that  $\Xm$ is discrete and $P_{\Ym|\Xm}$ is arbitrary. Then, for every $\Msf \ge 2$, 
\begin{equation}
	{\mathsf{ZZ}}(P_{\Xm,\Ym},\mathsf{M})
	 = 0.
\end{equation}
\end{proposition}

\begin{IEEEproof}
    The proof is provided in Appendix~\ref{app:discrete_zero_zzb}.
\end{IEEEproof}
We highlight that the result in Proposition~\ref{prop:discrete_zero_zzb} holds for all noise distributions.

To show the remaining two results in the series, we focus on the case of a scalar input in the  high-noise regime,
i.e., we consider $\overline{\mathsf V}(P_X, \mathsf{M}) $ 
in Theorem~\ref{thm:zzb_sig_inf_M}.
The next example (proof in Appendix~\ref{app:ex_discrete_ber})
shows that, in general, $\overline{\mathsf V}(P_X,\mathsf{M}) \neq 0$.
For completeness, the next example also evaluates $\Vsf_{\rm sp}(P_X,\Msf)$.

\begin{example}\label{ex:bern}
Let $X\sim{\rm Ber}(p)$ be a Bernoulli random variable with parameter $0<p<1$. 
Then, from~\eqref{eq:HighNoiseMMSEVar},  the high-noise behavior of ${\rm mmse}(X|\Ym)$ with arbitrary $P_{\Ym|X}$ satisfying the assumptions \textbf{A1-A3} in Section~\ref{sec:High_Noise_Assymptotics} is given by
\begin{align}
    \lim_{\eta\to\infty} {\rm mmse}(X|\Ym)
    = \var(X) = p(1-p),
\end{align} 
and for any integer $\Msf\geq2$, we have that $\overline{\mathsf V}(P_X,\Msf)$ and $\Vsf_{\rm sp}(P_X,\Msf)$ are given by
\begin{subequations}\label{eq:Bern}
\begin{align}
	\overline{\mathsf V}(P_X,\Msf) & =  \frac{1}{4}\min\{p,1-p\} \text{ and } \\
    \Vsf_{\rm sp}(P_X,\Msf) & = \frac{1}{2}\min\{p,1-p\}.
\end{align}
\end{subequations}
\end{example}
The above example demonstrates that unlike $\Zsf\Zsf(P_{\Xm,\Ym},\Msf)$, both $\overline{\Zsf\Zsf}(P_{\Xm,\Ym},\Msf)$ and $\Zsf\Zsf_{\rm sp}(P_{\Xm,\Ym},\Msf)$ are not trivial bounds for discrete inputs. Moreover, somewhat surprisingly, the SZZB is tight for $p=\frac{1}{2}$. We next show that, even with the valley-filling function, the ZZB for a discrete $\Xm$ is strictly sub-optimal in high-noise.
We accomplish this by comparing the ZZB with the valley-filling function with the variance of the input, i.e., the high-noise behavior of the MMSE in~\eqref{eq:HighNoiseMMSEVar}.
\begin{theorem}\label{thm:Discrete_not_tight}
Let $X$ be a discrete random variable  such that $\EE[X] = 0$ and $\inf\limits_{x\in \Sc_X}|x| >0$, where $\Sc_X$ is the support of $X$.
Then, for any integer $\mathsf{M}\geq2$, it holds that
\begin{equation}
	\overline{\mathsf V}(P_X, \mathsf{M}) < \var(X).
\end{equation}
\end{theorem}
\begin{IEEEproof}
    The proof is provided in Appendix~\ref{app:Discrete_not_tight}.
\end{IEEEproof}

\begin{remark}
In Theorem~\ref{thm:Discrete_not_tight}, the assumption that the support of $X$ has no accumulation point at zero was made to make the proof easier and most likely can be removed. 
Moreover, the result in Theorem~\ref{thm:Discrete_not_tight} shows that, when working with discrete inputs, the ZZB might not be the best bound to use, especially in the practically relevant high-noise regime. \hfill $\square$
\end{remark}

Thus, through the series of the above results, we have demonstrated that the valley-filling function is necessary for 
discrete priors, yet it is not sufficient in the sense that 
it does not guarantee the tightness to the MMSE.

\begin{table*}[t]
\centering
\caption{High-noise asymptotics for the examples in~\eqref{eq:Examples}, where $i \in \{1,2\}$.}
\begin{tabular}{|c||c|c|c|c|c|c|}
\hline
 & $\var({X_i})$ & $\overline{\mathsf V}(P_{{X_i}},\Msf)$ & ${\mathsf V}(P_{{X_i}},\Msf)$ & ${\mathsf V}_{\rm sp}(P_{{X_i}},\Msf)$  & \cite{GenOutageBoundsApproach} with valley-filling  & \cite{GenOutageBoundsApproach}  \\ \hline\hline
$X_1$ & $\frac{1}{12}$ & $\frac{1}{12}$ & $\frac{1}{12}$ & $\frac{2}{27}$ & $\frac{1}{12}$ & $\frac{1}{12}$ \\ \hline
$X_2$ & $\frac{13}{48}$ & $\begin{cases}
\frac{77}{384}  & \text{ if } \Msf = 2\\
\frac{20\Msf^2 -43 \Msf +26}{96 (\Msf-1)^2} & \text{ if } \Msf \geq 3 \end{cases}$ & $\frac{71\Msf^3-232\Msf^2 + 251\Msf - 86}{384(\Msf-1)^3}$ & $\frac{1}{4}$ &  $\frac{5}{24}$ &  $\frac{71}{384}$ \\ \hline
\end{tabular}
\label{table:two_examples}
\end{table*}
\subsection{ZZ Bounds Tightness}
\label{sec:tighntess_cont}

We now focus on
conditions under which the ZZB or the SZZB are tight to the MMSE.
We next present necessary and sufficient conditions on the PDF of $P_{\Xm|\Ym}$ for which not only the valley-filling function is not needed, but the ZZB without the valley filling function is indeed tight.
\begin{proposition}\label{prop:mmse=ZZB}
The following two conditions are equivalent:
\begin{enumerate}
    \item 
    \begin{equation}
        {\rm mmse}(\Xm|\Ym) = {\mathsf{ZZ}}(P_{\Xm,\Ym} ,\mathsf{M}). \label{eq:mmse=ZZB}
    \end{equation}
    \item  For every $t>0$ there exist $\Uc_i = \{\uv_{i,k}\}_{k=0}^{\mathsf{M}-1} \subset \Xc,~\forall i\in[1: d]$, each of which satisfies $( \uv_{i,k})_i= kt,~k\in[0:\mathsf{M}-1]$ and for all $(\xv,\yv)\in\Xc\times\Yc$,\footnote{We remind the reader that, in general,  the $\argmax$ function outputs a set.}
\begin{align}\label{eq:suff_necess_cond}
	 &  \argmax_{k\in[0:\mathsf{M}-1]} f_{\Xm|\Ym}(\xv+\uv_{i,k} |  \yv) \nonumber \\
	 & \bigcap \argmin_{k\in[0:\mathsf{M}-1]} \left|\EE[X_i|\Ym = \yv] - x_i - kt \right|  \neq \varnothing.
\end{align}
\end{enumerate}
\end{proposition}
\begin{IEEEproof}
    The proof is provided in Appendix~\ref{app:mmse=ZZB}.
\end{IEEEproof}

\begin{remark}
In words, the condition in~\eqref{eq:suff_necess_cond} implies that the ZZB becomes tight if for all $\yv\in\Yc$, the $i$th element of the MAP decision of $\Hc_k$, $k\in [0:\Msf-1]$, is the closest to the MMSE estimate of $X_i$ (i.e., $\EE[X_i|\Ym]$) compared to the $i$th element of the other $\Msf-1$ candidates. This is because the $i$th element of the MAP decision of $\Hc_k$, $k\in [0:\Msf-1]$ is given by $x_i + ( \uv_{i,k})_i= x_i + kt,~k\in[0:\mathsf{M}-1]$, which is precisely the term that we would like to be close to $\EE[X_i|\Ym]$.
Note that for the condition 2), $(\uv_{i,k})_j,~j\neq i$ can be arbitrary, if $(\uv_{i,k})_i=kt$ and~\eqref{eq:suff_necess_cond} hold.
As an example, consider an estimation problem with $\Msf=2$, $P_\Xm\sim\Nc(\zerov_2, I_2)$ and $P_{\Ym|\Xm}\sim \Nc(\Xm, I_2)$. This yields $f_{\Xm|\Ym}(\xv|\yv) = \frac{1}{\pi} {\rm e}^{- \|\xv-\frac{1}{2}\yv\|^2 }$ and $\EE[\Xm|\Ym=\yv] = \frac{1}{2}\yv$, from which it is not difficult to see that the condition 2) in Proposition~\ref{prop:mmse=ZZB} holds with $\Uc_i = \{\uv_{i,k}: \uv_{i,k} = kt\ev_i , k\in \{0,1\}\},~i\in\{1,2\}$.
\hfill $\square$
\end{remark}

By leveraging Proposition~\ref{prop:mmse=ZZB}, we now strengthen a result in~\cite{bell1995performance}, namely we prove that the sufficient condition in~\cite{bell1995performance} for the tightness of the ZZB is indeed also necessary.
\begin{corollary}\label{cor:unimodal_symmetric}
Let $X\in\RR$ be a continuous random variable.  Then, for any integer $\Msf\geq2$, it holds that
\begin{equation}\label{eq:unimodal_symmetric}
	{\rm mmse}(X|\Ym) = {\mathsf{ZZ}}(P_{X, \Ym},\mathsf{M}),
\end{equation}
if and only if, for all $\yv \in \Yc$,  the PDF $f_{X|\Ym}(x|\yv)$  is unimodal\footnote{
A PDF is said to be unimodal if it has at most one local maximum.} and symmetric with respect to its mode.
\end{corollary}
\begin{IEEEproof}
The proof is provided in Appendix~\ref{app:CorIff}.
\end{IEEEproof}

The series of results in Theorem~\ref{thm:Discrete_not_tight}, Proposition~\ref{prop:mmse=ZZB}, and Corollary~\ref{cor:unimodal_symmetric} demonstrate that the ZZB does not offer a tight bound to the MMSE for discrete inputs. However, in contrast to the ZZB, we observe from Example~\ref{ex:bern} that the SZZB can be tight to the MMSE for discrete inputs.
This triggers the following question: Under what conditions is the SZZB tight? The next proposition answers this question.
\begin{proposition}\label{prop:SZZB_condtion}
Suppose that 
\begin{equation}
    {\rm mmse}(\Xm|\Ym) 
     = \Zsf\Zsf_{\rm sp}(P_{\Xm,\Ym},\Msf).
\end{equation}
Then, for every $i\in[1:d]$, the marginal distribution of $|\EE[X_i|\Ym] - X_i| $ is discrete with at most two mass points. 
\end{proposition}

\begin{IEEEproof}
    The proof is provided in Appendix~\ref{app:SZZB_condtion}.
\end{IEEEproof}

\begin{remark}\label{rem:SZZB_never_tight}
The condition in Proposition~\ref{prop:SZZB_condtion} is extremely restrictive and is rarely satisfied in practice. 
It is not difficult to see that if 
$P_\Xm$ is a continuous distribution, then the absolute error $|\EE[X_i|\Ym] - X_i|$ is a continuous random variable, unless the channel distribution is trivial. In addition, if the channel is Gaussian then the error $|\EE[X_i|\Ym] - X_i|$  is a continuous random variable for all distributions of $\bf{X}$ \cite{dytso2021distribution}. Thus, for a continuous $P_\Xm$ and a continuous $P_{\Ym|\Xm}$, the SZZB is, in general, sup-optimal.
\hfill $\square$
\end{remark}

\subsection{ZZB vs. SZZB}
\label{ZZB vs. SZZB} 
Theorem~\ref{thm:BZZ_Mary_vec_mmse} and Theorem~\ref{thm:SZZB} provide two different Ziv-Zakai-type of lower bounds on the MMSE. In particular, $\overline{\Zsf\Zsf}$ in Theorem~\ref{thm:BZZ_Mary_vec_mmse} is obtained by integrating $t \Vc_t\{\hsf_\Msf(t,\ev_i,P_{\Xm,\Ym})\}$ over $t>0$, whereas $\mathsf{ZZ}_{\rm sp}$ in Theorem~\ref{thm:SZZB} requires to find a supremum of $t^2 \hsf_\Msf(t,\ev_i,P_{\Xm,\Ym})$ over $t>0$. Both bounds can be used in practice, however, there is no guidance as to when one bound should be preferred over the other. 
Moreover,  since $\overline{\Zsf\Zsf}$ in Theorem~\ref{thm:BZZ_Mary_vec_mmse} is more involved than $\mathsf{ZZ}_{\rm sp}$ in Theorem~\ref{thm:SZZB}, one may erroneously conclude (and this is indeed a common belief) that $\overline{\Zsf\Zsf}$ in Theorem~\ref{thm:BZZ_Mary_vec_mmse} is tighter than $\mathsf{ZZ}_{\rm sp}$ in Theorem~\ref{thm:SZZB}.
To clarify these points, in addition to Example~\ref{ex:bern}, it is also instructive to consider the following example.

\begin{example}\label{ex:unif_mixed_unif}
Consider the random variables $X_1$ and $X_2$ with the following PDFs,
\begin{subequations}\label{eq:Examples}
    \begin{align}
    f_{X_1}(x) &=   \mathbbm{1} \{x\in [0,1]\},\\
    f_{X_2}(x) &=  \mathbbm{1}\{x \in [0,1/2]\} + \mathbbm{1}\{x\in[1,3/2]\}.
    \end{align}
\end{subequations}
Table~\ref{table:two_examples} shows the high-noise asymptotics for the above two inputs for every $\Msf \ge 2$; the derivation can be found in Appendix~\ref{app:no_ineq}. 
\end{example}

Combining Example~\ref{ex:bern} and Example~\ref{ex:unif_mixed_unif}, we observe that the following scenarios are possible:
\begin{itemize}
\item From $X_1$:  \emph{For all $\Msf$, the ZZB without the valley-filling function is tight but the SZZB is not tight, that is,}
\begin{equation}
   \var({X_1})=\overline{\mathsf V}(P_{{X_1}},\Msf)={\mathsf V}(P_{{X_1}},\Msf) > {\mathsf V}_{\rm sp}(P_{{X_1}},\Msf).
\end{equation}
\item From $X_2$: \emph{For all $\Msf$, none of the bounds is tight or equal, and the SZZB outperforms the ZZB with the valley-filling function, that is,}
\begin{equation}
   \var({X_2})> {\mathsf V}_{\rm sp}(P_{{X_2}},\Msf) >\overline{\mathsf V}(P_{{X_2}},\Msf)> {\mathsf V}(P_{{X_2}},\Msf) .
\end{equation}
\item  From Example~\ref{ex:bern} and Theorem~\ref{thm:Discrete_not_tight} : \emph{For $X\sim{\rm Ber}(p)$ with $p=1/2$,} the SZZB 
is tight and  it outperforms the ZZB with the valley-filling function for every $\Msf$, that is,
\begin{align}
	\! \var(X) & = {\mathsf V}_{\rm sp}(P_{X},2) \nonumber \\
	& = {\mathsf V}_{\rm sp}(P_{X},\Msf) > \overline{\mathsf V}(P_{X},\Msf) \geq{\mathsf V}(P_{X},\Msf) .
\end{align}
\end{itemize}  
The above examples demonstrate that neither of the bounds is always tighter than the other. From our experience, the SZZB tends to be tighter than the ZZB for distributions that tend to concentrate  over a finite number of regions.
In the remainder of this section, we provide a few more numerical examples and compare the ZZB and the SZZB to other well-known bounds. 

\begin{remark}
From the analysis above, we observe that, while increasing $\Msf$ helps to tighten the bounds (i.e., the bounds are monotonically non-decreasing in $\Msf$), it does not necessarily lead to a tight bound to the MMSE, even when $\Msf \to \infty$. In all the examples that we considered, we indeed observed that if a bound in the Ziv-Zakai family is not tight for $\Msf=2$, then it is not tight even for larger values of $\Msf$. 
In~\cite{GenOutageBoundsApproach}, for the scalar case, the authors derived a version of the ZZB with a countable number of hypotheses, which is potentially tighter than ours. Examples and some comparisons of the  bound in~\cite{GenOutageBoundsApproach} are given in Table~\ref{table:two_examples}. We suspect that by increasing $\Msf$, we can approach the bound in~\cite{GenOutageBoundsApproach} arbitrarily close.     We were, however, not able to find an example where the bound with an infinite number of hypotheses is equal to the MMSE but the bound with finite $\Msf$ is not. 
Finding an example for which a bound in the Ziv-Zakai family becomes tight to the MMSE when $\Msf$ grows (e.g., when $\Msf \to \infty$) is, therefore, an interesting open question, worth of further investigation.
\hfill $\square$
\end{remark}

 \begin{figure*}[t]
   \centering
   \subfloat[$\mu=1$.]{%
     \label{subfig:mu1}
%
%
\definecolor{mycolor1}{rgb}{1.00000,0.00000,1.00000}%
\begin{tikzpicture}

\begin{axis}[%
width=6cm,
height=6cm,
at={(1.444in,0.917in)},
scale only axis,
xmin=0,
xmax=1,
xtick distance=0.1,
xlabel={$\omega$},
ymin=0.6,
ymax=2.2,
ytick distance=0.2,
axis background/.style={fill=white},
xmajorgrids,
ymajorgrids,
legend style={legend cell align=left, align=left, draw=white!15!black}
]
\addplot [color=black, thick]
  table[row sep=crcr]{%
0	1\\
0.05	1.19\\
0.1	1.36\\
0.15	1.51\\
0.2	1.64000000000001\\
0.25	1.75000000000001\\
0.3	1.84000000000001\\
0.35	1.91000000000001\\
0.4	1.96000000000001\\
0.45	1.99000000000001\\
0.5	2.00000000000001\\
0.55	1.99000000000001\\
0.6	1.96000000000001\\
0.65	1.91000000000001\\
0.7	1.84000000000001\\
0.75	1.75000000000001\\
0.8	1.64000000000001\\
0.85	1.51\\
0.9	1.36\\
0.95	1.19\\
1	1\\
};
\addlegendentry{MMSE}

\addplot [color=red, dashdotted, thick, mark=o, mark options={solid, red}]
  table[row sep=crcr]{%
0	0.999998907363772\\
0.05	1.1785442463258\\
0.1	1.33366035718785\\
0.15	1.47211661674042\\
0.2	1.59580445420997\\
0.25	1.70523154525114\\
0.3	1.80011236593654\\
0.35	1.87948389238169\\
0.4	1.94160787975348\\
0.45	1.98357848255942\\
0.5	1.99999863275543\\
0.55	1.98357848255942\\
0.6	1.94160787975348\\
0.65	1.87948389238169\\
0.7	1.80011236593654\\
0.75	1.70523154525114\\
0.8	1.59580445420997\\
0.85	1.47211661674042\\
0.9	1.33366035718785\\
0.95	1.1785442463258\\
1	0.999998907363772\\
};
\addlegendentry{ZZB}

\addplot [color=teal, dashdotted,  thick, mark=+, mark options={solid, teal}]
  table[row sep=crcr]{%
0	0.662866464104418\\
0.05	0.74886019334617\\
0.1	0.84848552898358\\
0.15	0.95763831890029\\
0.2	1.06758627334421\\
0.25	1.16977786229572\\
0.3	1.2586304103157\\
0.35	1.33086776716855\\
0.4	1.38421931967349\\
0.45	1.4169975966601\\
0.5	1.42806185181789\\
0.55	1.4169975966601\\
0.6	1.38421931967349\\
0.65	1.33086776716855\\
0.7	1.2586304103157\\
0.75	1.16977786229572\\
0.8	1.06758627334421\\
0.85	0.95763831890029\\
0.9	0.84848552898358\\
0.95	0.74886019334617\\
1	0.662866464104418\\
};
\addlegendentry{SZZB}

\addplot [color=blue, dashed, thick, mark=x, mark options={solid, blue}]
  table[row sep=crcr]{%
0	1\\
0.05	1.14675960969599\\
0.1	1.26892987961137\\
0.15	1.38108073847689\\
0.2	1.48444863717726\\
0.25	1.57785376349454\\
0.3	1.6591709142819\\
0.35	1.72592352916341\\
0.4	1.77569047615588\\
0.45	1.80644934184617\\
0.5	1.81685884501783\\
0.55	1.80644934184617\\
0.6	1.77569047615588\\
0.65	1.72592352916341\\
0.7	1.6591709142819\\
0.75	1.57785376349454\\
0.8	1.48444863717726\\
0.85	1.38108073847689\\
0.9	1.26892987961137\\
0.95	1.14675960969599\\
1	1\\
};
\addlegendentry{CRB}

\addplot [color=mycolor1, dashed, thick, mark=diamond, mark options={solid, mycolor1}]
  table[row sep=crcr]{%
0	1\\
0.05	1.17472278142137\\
0.1	1.32848714883785\\
0.15	1.46771747824304\\
0.2	1.59234228625222\\
0.25	1.70115360911351\\
0.3	1.79268596777982\\
0.35	1.86552902462843\\
0.4	1.91847927288237\\
0.45	1.95062660478891\\
0.5	1.96140615236085\\
0.55	1.95062660478891\\
0.6	1.91847927288237\\
0.65	1.86552902462843\\
0.7	1.79268596777982\\
0.75	1.70115360911351\\
0.8	1.59234228625222\\
0.85	1.46771747824304\\
0.9	1.32848714883785\\
0.95	1.17472278142137\\
1	1\\
};
\addlegendentry{MEB}

\end{axis}
\end{tikzpicture}%
     }
   \hfil
   \subfloat[$\mu=1.5$.]{%
     \label{subfig:mu1.5}
%
%
\definecolor{mycolor1}{rgb}{1.00000,0.00000,1.00000}%
\begin{tikzpicture}

\begin{axis}[%
width=6cm,
height=6cm,
at={(1.444in,0.917in)},
scale only axis,
xmin=0,
xmax=1,
xtick distance=0.1,
xlabel={$\omega$},
ymin=0.5,
ymax=3.5,
ytick distance=0.5,
axis background/.style={fill=white},
xmajorgrids,
ymajorgrids,
legend style={legend cell align=left, align=left, draw=white!15!black}
]
\addplot [color=black, thick]
  table[row sep=crcr]{%
0	1\\
0.05	1.4275\\
0.1	1.81000000000001\\
0.15	2.14750000000001\\
0.2	2.44000000000001\\
0.25	2.68750000000001\\
0.3	2.89000000000001\\
0.35	3.04750000000001\\
0.4	3.16000000000001\\
0.45	3.22750000000001\\
0.5	3.25000000000001\\
0.55	3.22750000000001\\
0.6	3.16000000000001\\
0.65	3.04750000000001\\
0.7	2.89000000000001\\
0.75	2.68750000000001\\
0.8	2.44000000000001\\
0.85	2.14750000000001\\
0.9	1.81000000000001\\
0.95	1.4275\\
1	1\\
};
\addlegendentry{MMSE}

\addplot [color=red, dashdotted, thick, mark=o, mark options={solid, red}]
  table[row sep=crcr]{%
0	0.999999726332701\\
0.05	1.35627002427364\\
0.1	1.65989766724202\\
0.15	1.933698780749\\
0.2	2.18307638087196\\
0.25	2.40963145871846\\
0.3	2.61226618930942\\
0.35	2.78813193029179\\
0.4	2.93232742157954\\
0.45	3.03649954308557\\
0.5	3.08263341831206\\
0.55	3.03649954308557\\
0.6	2.93232742157954\\
0.65	2.78813193029179\\
0.7	2.61226618930942\\
0.75	2.40963145871846\\
0.8	2.18307638087196\\
0.85	1.933698780749\\
0.9	1.65989766724202\\
0.95	1.35627002427364\\
1	0.999999726332701\\
};
\addlegendentry{ZZB}

\addplot [color=teal, dashdotted, thick, mark=+, mark options={solid, teal}]
  table[row sep=crcr]{%
0	0.662866574673545\\
0.05	0.786683853877976\\
0.1	0.988705851468329\\
0.15	1.29096960006355\\
0.2	1.57518019535335\\
0.25	1.8275062918679\\
0.3	2.04574303801125\\
0.35	2.22718878876401\\
0.4	2.36721464663188\\
0.45	2.45834728744473\\
0.5	2.49062973270304\\
0.55	2.45834728744473\\
0.6	2.36721464663188\\
0.65	2.22718878876401\\
0.7	2.04574303801125\\
0.75	1.8275062918679\\
0.8	1.57518019535335\\
0.85	1.29096960006355\\
0.9	0.988705851468329\\
0.95	0.786683853877976\\
1	0.662866574673545\\
};
\addlegendentry{SZZB}

\addplot [color=blue, dashed, thick, mark=x, mark options={solid, blue}]
  table[row sep=crcr]{%
0	1\\
0.05	1.17268925127227\\
0.1	1.29627708556488\\
0.15	1.40466241617024\\
0.2	1.50152413033894\\
0.25	1.58696716330909\\
0.3	1.65991368058965\\
0.35	1.7188650203531\\
0.4	1.76229658493941\\
0.45	1.78892508050592\\
0.5	1.79790029321029\\
0.55	1.78892508050592\\
0.6	1.76229658493941\\
0.65	1.7188650203531\\
0.7	1.65991368058965\\
0.75	1.58696716330909\\
0.8	1.50152413033894\\
0.85	1.40466241617024\\
0.9	1.29627708556488\\
0.95	1.17268925127227\\
1	1\\
};
\addlegendentry{CRB}

\addplot [color=mycolor1, dashed, thick, mark=diamond, mark options={solid, mycolor1}]
  table[row sep=crcr]{%
0	1\\
0.05	1.32063123416933\\
0.1	1.60244436084734\\
0.15	1.86590169749956\\
0.2	2.10894201179497\\
0.25	2.32669936783937\\
0.3	2.51380766023803\\
0.35	2.665221407173\\
0.4	2.77665472759171\\
0.45	2.84486278978005\\
0.5	2.86782703701157\\
0.55	2.84486278978005\\
0.6	2.77665472759171\\
0.65	2.665221407173\\
0.7	2.51380766023803\\
0.75	2.32669936783937\\
0.8	2.10894201179497\\
0.85	1.86590169749956\\
0.9	1.60244436084734\\
0.95	1.32063123416933\\
1	1\\
};
\addlegendentry{MEB}

\end{axis}
\end{tikzpicture}%
     } 
     \hfil
   \subfloat[$\mu=2.5$.]{%
     \label{subfig:mu2.5}
%
%
\definecolor{mycolor1}{rgb}{1.00000,0.00000,1.00000}%
\begin{tikzpicture}

\begin{axis}[%
width=6cm,
height=6cm,
at={(1.444in,0.917in)},
scale only axis,
xmin=0,
xmax=1,
xtick distance=0.1,
xlabel={$\omega$},
ymin=0,
ymax=8,
ytick distance=1,
axis background/.style={fill=white},
xmajorgrids,
ymajorgrids,
legend style={legend cell align=left, align=left, draw=white!15!black, }
]
\addplot [color=black, thick]
  table[row sep=crcr]{%
0	1.00000000000036\\
0.05	2.18749999999982\\
0.1	3.24999999999979\\
0.15	4.18749999999975\\
0.2	4.99999999999973\\
0.25	5.6874999999997\\
0.3	6.24999999999968\\
0.35	6.68749999999967\\
0.4	6.99999999999966\\
0.45	7.18749999999965\\
0.5	7.24999999999965\\
0.55	7.18749999999965\\
0.6	6.99999999999966\\
0.65	6.68749999999967\\
0.7	6.24999999999968\\
0.75	5.6874999999997\\
0.8	4.99999999999973\\
0.85	4.18749999999975\\
0.9	3.24999999999979\\
0.95	2.18749999999982\\
1	1.00000000000036\\
};
\addlegendentry{MMSE}

\addplot [color=red, dashdotted, thick, mark=o, mark options={solid, red}]
  table[row sep=crcr]{%
0	0.999999725528263\\
0.05	1.7255207953909\\
0.1	2.3240848758505\\
0.15	2.85485047973326\\
0.2	3.33008163896102\\
0.25	3.75367012012089\\
0.3	4.12547766234939\\
0.35	4.4422965019109\\
0.4	4.69731595236678\\
0.45	4.87795763242497\\
0.5	4.95612700230625\\
0.55	4.87795763242497\\
0.6	4.69731595236678\\
0.65	4.4422965019109\\
0.7	4.12547766234939\\
0.75	3.75367012012089\\
0.8	3.33008163896102\\
0.85	2.85485047973326\\
0.9	2.3240848758505\\
0.95	1.7255207953909\\
1	0.999999725528263\\
};
\addlegendentry{ZZB}

\addplot [color=teal, dashdotted, thick, mark=+, mark options={solid, teal}]
  table[row sep=crcr]{%
0	0.662866073175694\\
0.05	1.01540899829693\\
0.1	1.86891908191241\\
0.15	2.64381130922729\\
0.2	3.35626429624522\\
0.25	4.01178115174918\\
0.3	4.61084064401527\\
0.35	5.15005617137701\\
0.4	5.62125626496413\\
0.45	6.00711551876401\\
0.5	6.25001885887187\\
0.55	6.00711551876401\\
0.6	5.62125626496413\\
0.65	5.15005617137701\\
0.7	4.61084064401527\\
0.75	4.01178115174918\\
0.8	3.35626429624522\\
0.85	2.64381130922729\\
0.9	1.86891908191241\\
0.95	1.01540899829693\\
1	0.662866073175694\\
};
\addlegendentry{SZZB}

\addplot [color=blue, dashed, thick, mark=x, mark options={solid, blue}]
  table[row sep=crcr]{%
0	0.999999999999996\\
0.05	1.04741374381784\\
0.1	1.07071648478067\\
0.15	1.08791475141654\\
0.2	1.10136068729509\\
0.25	1.11199521530561\\
0.3	1.12030040114611\\
0.35	1.12655360990178\\
0.4	1.13092265408134\\
0.45	1.13350821513745\\
0.5	1.13436437452902\\
0.55	1.13350821513745\\
0.6	1.13092265408134\\
0.65	1.12655360990178\\
0.7	1.12030040114611\\
0.75	1.11199521530561\\
0.8	1.10136068729509\\
0.85	1.08791475141654\\
0.9	1.07071648478067\\
0.95	1.04741374381784\\
1	0.999999999999996\\
};
\addlegendentry{CRB}

\addplot [color=mycolor1, dashed, thick, mark=diamond, mark options={solid, mycolor1}]
  table[row sep=crcr]{%
0	1.00000000000003\\
0.05	1.46754532029967\\
0.1	1.87901182002397\\
0.15	2.27451119229707\\
0.2	2.64835993030023\\
0.25	2.99015126395747\\
0.3	3.28865310705963\\
0.35	3.5332858060737\\
0.4	3.71500349757248\\
0.45	3.8269127067534\\
0.5	3.86470423338664\\
0.55	3.8269127067534\\
0.6	3.71500349757248\\
0.65	3.5332858060737\\
0.7	3.28865310705963\\
0.75	2.99015126395747\\
0.8	2.64835993030023\\
0.85	2.27451119229707\\
0.9	1.87901182002397\\
0.95	1.46754532029967\\
1	1.00000000000003\\
};
\addlegendentry{MEB}

\end{axis}
\end{tikzpicture}%
     } 
     \hfil
   \subfloat[$\mu=3$.]{%
     \label{subfig:mu3}
%
%
\definecolor{mycolor1}{rgb}{1.00000,0.00000,1.00000}%
\begin{tikzpicture}

\begin{axis}[%
width=6cm,
height=6cm,
at={(1.444in,0.917in)},
scale only axis,
xmin=0,
xmax=1,
xtick distance=0.1,
xlabel={$\omega$},
ymin=0,
ymax=12,
ytick distance=1,
axis background/.style={fill=white},
xmajorgrids,
ymajorgrids,
legend style={legend cell align=left, align=left, draw=white!15!black}
]
\addplot [color=black, thick]
  table[row sep=crcr]{%
0	1.00000000000002\\
0.05	2.71000000000092\\
0.1	4.24000000000095\\
0.15	5.59000000000098\\
0.2	6.760000000001\\
0.25	7.75000000000121\\
0.3	8.56000000000127\\
0.35	9.19000000000105\\
0.4	9.64000000000106\\
0.45	9.91000000000106\\
0.5	10.0000000000011\\
0.55	9.91000000000106\\
0.6	9.64000000000106\\
0.65	9.19000000000105\\
0.7	8.56000000000127\\
0.75	7.75000000000121\\
0.8	6.760000000001\\
0.85	5.59000000000098\\
0.9	4.24000000000095\\
0.95	2.71000000000092\\
1	1.00000000000002\\
};
\addlegendentry{MMSE}

\addplot [color=red, dashdotted, thick, mark=o, mark options={solid, red}]
  table[row sep=crcr]{%
0	0.99999988102728\\
0.05	1.88392249437476\\
0.1	2.60789309456779\\
0.15	3.24873213827955\\
0.2	3.82191648925847\\
0.25	4.33243502296382\\
0.3	4.78028163927719\\
0.35	5.16169514610763\\
0.4	5.46856520481005\\
0.45	5.68583754282179\\
0.5	5.7798036121259\\
0.55	5.68583754282179\\
0.6	5.46856520481005\\
0.65	5.16169514610763\\
0.7	4.78028163927719\\
0.75	4.33243502296382\\
0.8	3.82191648925847\\
0.85	3.24873213827955\\
0.9	2.60789309456779\\
0.95	1.88392249437476\\
1	0.99999988102728\\
};
\addlegendentry{ZZB}

\addplot [color=teal, dashdotted, thick, mark=+, mark options={solid, teal}]
  table[row sep=crcr]{%
0	0.66286646615837\\
0.05	1.34256336742405\\
0.1	2.49979496907078\\
0.15	3.56633249544296\\
0.2	4.56087897018397\\
0.25	5.48992234032591\\
0.3	6.3543246697785\\
0.35	7.15079029165861\\
0.4	7.8713081951042\\
0.45	8.50004075229097\\
0.5	8.99999908532546\\
0.55	8.50004075229097\\
0.6	7.8713081951042\\
0.65	7.15079029165861\\
0.7	6.3543246697785\\
0.75	5.48992234032591\\
0.8	4.56087897018397\\
0.85	3.56633249544296\\
0.9	2.49979496907078\\
0.95	1.34256336742405\\
1	0.66286646615837\\
};
\addlegendentry{SZZB}

\addplot [color=blue, dashed, thick, mark=x, mark options={solid, blue}]
  table[row sep=crcr]{%
0	1.00000000000002\\
0.05	1.01498863110454\\
0.1	1.02168743860911\\
0.15	1.02647185233771\\
0.2	1.03012738239255\\
0.25	1.03296879052832\\
0.3	1.03515843272356\\
0.35	1.03679052899792\\
0.4	1.0379225832298\\
0.45	1.03858935906413\\
0.5	1.03880963465217\\
0.55	1.03858935906413\\
0.6	1.0379225832298\\
0.65	1.03679052899792\\
0.7	1.03515843272356\\
0.75	1.03296879052832\\
0.8	1.03012738239255\\
0.85	1.02647185233771\\
0.9	1.02168743860911\\
0.95	1.01498863110454\\
1	1.00000000000002\\
};
\addlegendentry{CRB}

\addplot [color=mycolor1, dashed, thick, mark=diamond, mark options={solid, mycolor1}]
  table[row sep=crcr]{%
0	0.999999999999981\\
0.05	1.48282318638227\\
0.1	1.90743648386192\\
0.15	2.31655241360566\\
0.2	2.70407784168897\\
0.25	3.05898019425257\\
0.3	3.36935821997348\\
0.35	3.62399465343605\\
0.4	3.8132907294155\\
0.45	3.92992658210947\\
0.5	3.96932424981358\\
0.55	3.92992658210947\\
0.6	3.8132907294155\\
0.65	3.62399465343605\\
0.7	3.36935821997348\\
0.75	3.05898019425257\\
0.8	2.70407784168897\\
0.85	2.31655241360566\\
0.9	1.90743648386192\\
0.95	1.48282318638227\\
1	0.999999999999981\\
};
\addlegendentry{MEB}

\end{axis}

\end{tikzpicture}%
}  
     \hfil
   \caption{Comparisons of Bayesian MMSE lower bounds for $P_X$ in~\eqref{eq:WeightedGaussian}. In every evaluation, we set $\Msf=2$ for the bounds in the Ziv-Zakai family.}
   \label{fig:comp}
 \end{figure*}
 
\subsection{Comparison with Other  MMSE Lower Bounds}\label{sec:comp_MEB}
We here provide a few notable examples for which either the ZZB or the SZZB are tighter than other standard Bayesian MMSE lower bounds, such as the Bayesian Cram\'er-Rao bound (CRB)~\cite{van2004detection,Polyanskiy_Wu_2024} and the maximal entropy bound (MEB)~\cite{Cover:InfoTheory}. 
We consider the high-noise regime in which, for a prior density function $f_X$, the Bayesian CRB and the MEB are defined~as
\begin{align}
    \mathsf{CRB}
    & = \left( \int \frac{(f^\prime_X(x))^2}{f_X(x)} {\rm d}x \right)^{-1},
\end{align}
and
\begin{align}
    \mathsf{MEB}
    & = \frac{1}{2\pi {\rm{e}}} {\rm{e}}^{2 h(X)},
\end{align}
where $h(X) = \EE[ -\log f_X(X)]$ is the differential entropy of $X$. Moreover, from~\eqref{eq:HighNoiseMMSEVar}, we have that the $\rm mmse(X|\Ym)$ in the high-noise regime is given by $\var(X)$. For the high-noise regime, the ZZBs are given 
in Theorem~\ref{thm:zzb_sig_inf_M}.
We set the prior distribution $P_X$ to be a weighted sum of two Gaussian distributions, i.e., 
\begin{equation}
\label{eq:WeightedGaussian}
P_X = \omega\Nc(-\mu,1) + (1-\omega)\Nc(\mu,1), 
\end{equation}
where $\omega\in[0,1]$ and $\mu \in \mathbb{R}$.
In all the settings that we considered, the ZZB is always tighter than the CRB and the MEB, while the performance of the SZZB depends on the parameters $\mu$ and~$\omega$. 
When $\omega\in\{0,1\}$, then $P_X$ in~\eqref{eq:WeightedGaussian} reduces to a Gaussian distribution; for these two cases, from Fig.~\ref{fig:comp} we observe that the ZZB, CRB, and MEB are all tight to the MMSE, as we expect. However, the SZZB is not tight for these two cases; this shows a different behavior of the SZZB.
Interestingly, the SZZB outperforms the other lower bounds, including the ZZB, for some values of the parameters $\omega$ and $\mu$ (see Fig.~\ref{subfig:mu2.5} and Fig.~\ref{subfig:mu3}).
These two examples suggest that the SZZB performs well when the probability measure $P_{X}$ strongly concentrates on each mode. Under such a condition instead the other lower bounds do not perform well. This consideration strengthens the applicability of the SZZB, since multimodal-like distributions are practically relevant prior distributions~\cite{heath2018foundations,everitt2013finite}.

\section{Conclusion}
\label{sec:Conclusion}
In this paper, we have provided concise expressions for general Ziv-Zakai bounds, which require no regularity conditions. In particular, we have removed the continuity assumption, and the bounds now hold for any input distribution. 
We have presented general properties of the bounds, such as tensorization, and high-noise and low-noise asymptotics.

The first key observation here is that in the low-noise regime, the ZZB is tight in its simplest form (i.e., without the valley-filling function, and with the number of hypotheses being set to two) for mixed-input distributions and under additive Gaussian noise.  Thus, the ZZB bound should be preferred over the SZZB bound in low-noise scenarios.  The fact that the SZZB is not tight in low-noise is a bit surprising since even simple bounds, such as the Bayesian Cram\'er-Rao, are tight in the low-noise regime albeit with more regularity conditions. 

The second key observation is that in the high-noise regime, there are distributions for which neither of the bounds are tight.  For instance, for discrete inputs, we have shown that the ZZB, in general, is sub-optimal in the high-noise regime. 
For continuous distributions, we have provided necessary and sufficient conditions for the tightness of the ZZB without the valley-filling function.
In contrast to the ZZB, the SZZB can be tight for discrete priors, but it is always sub-optimal for continuous distributions.  This suggests the following `rule-of-thumb': the SZZB should be preferred over the ZZB  bound when the   priors are `peaky'.  

Finally, we have provided evidence on the effectiveness of the bounds in the Ziv-Zakai family by showing examples in which the ZZB and the SZZB outperform other well-known Bayesian MMSE lower bounds, namely the Cram\'er-Rao bound and the maximum entropy bound. 

There are several interesting future directions. First, one could continue studying the properties of ZZ bounds.
For example, one would be to also explore a high-dimensional asymptotic and see which form of the bound is superior.  Additionally, since discrete inputs pose challenges to most known bounds in the literature,  it would be interesting to either improve the bounds in the Ziv-Zakai family or design a new family of bounds that would be well-suited for discrete inputs; the interested reader is referred to~\cite{ExpoAndPoincare} for a bound that is tight for discrete inputs in the high-noise regime. Second, it would be interesting to extend our analysis beyond the MMSE and consider other loss functions such $L_p$ losses~\cite{mmpe} or cross-entropy base losses~\cite{courtade2013multiterminal, Focal2024lossy}.  Third, bounds like CRB  via the I-MMSE relationship~\cite{guo2005mutual} are used to provide lower bounds on the achievable rates~\cite{e19080409}, and it would be interesting to see what achievability bounds are possible with ZZ bounds given that these are tighter than CRB.

 \appendices

\section{Ziv-Zakai Bounds for Generalized MMSE}\label{app:gen_zzb_szzb}
We here assume a general setting where the alphabet $\mathcal{X}$, such that   $\Xm \in \mathcal{X}$, is a Hilbert space endowed with the inner product $\left<\cdot,\cdot\right>$, and show the bounds in the Ziv-Zakai family at this level of generality.  In the main body of the paper, for the ease of exposition, we, however,  assume that $\mathcal{X} = \mathbb{R}^d$.   
To begin, we define the following directional generalization of the MMSE: given some fixed $\av\in\Xc$, let 
\begin{equation}\label{eq:def_err}
	\Ec_\av(\Xm|\Ym)
	 = \inf_{\phi\in\Phi} \EE\left[  \left<\av, \Xm - \phi(\Ym) \right>^2 \right],
\end{equation}
where $\Phi$ is a set of measurable functions $\phi:\Yc\to\Xc$.

The results in Theorem~\ref{thm:BZZ_Mary_vec_mmse} and Theorem~\ref{thm:SZZB} are direct consequences of the results in Theorem~\ref{thm:BZZ_Mary_vec} and Theorem~\ref{thm:SZZB_general} in the subsequent subsections. 
In particular, since Theorem~\ref{thm:BZZ_Mary_vec} and Theorem~\ref{thm:SZZB_general} provide Ziv-Zakai lower bounds on $\Ec_\av(\Xm|\Ym)$ in~\eqref{eq:def_err} for any $\av\in\Xc$ and any inner product $\left<\cdot,\cdot\right>$, we can obtain a lower bound on ${\rm mmse}(\Xm|\Ym)$ in~\eqref{eq:def_mmse} by setting $\phi(\Ym) = \EE[\Xm | \Ym ]$, taking the inner product $\left<\xv,\yv\right>=\sum_{j=1}^d x_j y_j$, setting $\av = \ev_i$, and summing it over $i\in[1:d]$. This would lead us to the MMSE lower bounds in Theorem~\ref{thm:BZZ_Mary_vec_mmse} and Theorem~\ref{thm:SZZB}.

\subsection{Generalized ZZB}\label{app:gen_zzb}

\begin{theorem}
\label{thm:BZZ_Mary_vec}
Let $\epsilonv = \phi(\Ym) - \Xm$ be the estimation error when an estimator $\phi(\Ym)\in \mathcal{X}$ is used to estimate $\Xm\in\Xc$.
Then, for every $ \av \in \mathcal{X}$, any estimator $\phi:\Yc\to\Xc$, and any integer $\mathsf{M}\geq2$, we have that
\begin{equation}\label{eq:thmBBZ_Mary_vec}
	\EE\left[  \left< \av, \epsilonv \right>^2 \right]
	\geq \int_0^\infty \frac{t}{2} \Vc_t \left\{ \frac{\hsf_{\Msf}(t,\av,P_{\Xm,\Ym}) }{\mathsf{M}-1} \right\} \ {\rm d} t,
\end{equation}
where $\hsf_{\Msf}(t,\av,P_{\Xm,\Ym}) $ is defined in Definition~\ref{def:h_func}.
\end{theorem}

\begin{IEEEproof}
We start by defining a few terms. First, we note that
\begin{equation}\label{eq:BZZ_Mary_vec_proof1}
	\EE[  \left< \av, \epsilonv \right>^2 ]
	 = \int_0^\infty \frac{t}{2} \Pr\left( |\left< \av, \epsilonv \right>| \geq \frac{t}{2} \right) \ {\rm d} t.
\end{equation}
Second, we fix some $\mathsf{M} \ge 2$,  we choose a  set of vectors $\Uc = \{ \uv_i \}_{i=0}^{\mathsf{M}-1} \subset \mathcal{X}$ such that $\left<\av, \uv_i\right> = it$, and we define $\mathcal{P}_\mathcal{U}(\xv)$ as in Definition~\ref{def:h_func}.
We now note that 
\begin{align}\label{eq:gen_ZZB_proof1}
	& \Pr\left( |\left< \av, \epsilonv \right>| \geq \frac{t}{2} \right) \nonumber \\
	& =  \int \Pr \left( \left< \av, \phi(\Ym) - \Xm \right>  \!\geq \!  \frac{t}{2} \; \middle | \; \Xm=\xv  \right)   P_{\Xm}({\rm{d}}\xv)  \nonumber \\
	& \quad +   \int \Pr \left( \left< \av, \phi(\Ym) - \Xm \right> \! \leq \! - \frac{t}{2} \; \middle | \; \Xm=\xv \right) P_{\Xm}({\rm{d}}\xv) \nonumber \\
	& =  \frac{1}{\mathsf{M}-1} \!\sum_{i=1}^{\mathsf{M}-1} \!\left \{ \! \int \! \Pr\! \left( \left< \av, \phi(\Ym) -\Xm \right>  \!\geq \! \frac{t}{2} \!\; \middle | \;\! \Xm=\xv  \right) \! P_{\Xm}({\rm{d}}\xv) \right. \nonumber \\
	& \quad \left. +\!   \int \!    \Pr \left( \left< \av, \phi(\Ym) \!-\! \Xm \right> \! \leq \! - \frac{t}{2} \; \middle | \; \Xm\!=\!\xv \right)   P_{\Xm}({\rm{d}}\xv)\right\} ,
\end{align}
where the first equality follows by applying the law of total probability and by substituting $\epsilonv = \phi(\Ym) - \Xm$.
The summation in~\eqref{eq:gen_ZZB_proof1} is then given by~\eqref{eq:Getting_to_the_error_prob_step}, at the top of the next page,
\begin{figure*}[h]
\begin{align}\label{eq:Getting_to_the_error_prob_step}
& \sum_{i=1}^{\mathsf{M}-1} \left \{ \! \int  \!\Pr \left( \!\left< \av, \phi(\Ym) \!-\! \Xm \right>  \!\geq \! \frac{t}{2} \; \middle | \; \Xm=\xv  \right)  P_{\Xm}({\rm{d}}\xv) +  \!\! \int  \! \Pr \left( \!\left< \av, \phi(\Ym) \!-\! \Xm\right> \! \leq \! -\frac{t}{2} \; \middle | \; \Xm=\xv \right)  P_{\Xm}({\rm{d}}\xv)  \right\} \notag \\
& \stackrel{{\rm{(a)}}}{=}  \sum_{i=1}^{\mathsf{M}-1} \left \{  \int \! \Pr \left( \left< \av, \phi(\Ym) \!-\! \Xm \right>  \!\geq \! \frac{t}{2} \; \middle | \; \Xm=\xv+\uv_{i-1}  \right)  P_{\Xm-\uv_{i-1} }({\rm{d}}\xv)  \right. \nonumber \\
& \qquad \qquad \left . +   \int \! \Pr \left( \left< \av, \phi(\Ym) \!-\! \Xm \right> \! \leq \! - \frac{t}{2} \; \middle | \; \Xm=\xv+\uv_{i} \right)  P_{\Xm-\uv_{i}}({\rm{d}}\xv) \right\} \notag \\
&\stackrel{{\rm{(b)}}}{=}   \sum_{i=1}^{\mathsf{M}-1} \left \{  \int  \Pr \left( \left< \av, \phi(\Ym) \!-\! \xv \right>  \!\geq \! {\left( i-\frac{1}{2} \right)}t\; \middle | \; \Xm=\xv+\uv_{i-1}  \right)   P_{\Xm-\uv_{i-1} }( {\rm{d}}\xv)  \right. \notag\\
& \qquad \qquad +\left .    \int    \Pr \left( \left< \av, \phi(\Ym) \!-\! \xv \right> \! \leq \!  \left( i-\frac{1}{2} \right)t  \; \middle | \; \Xm=\xv+\uv_{i} \right) P_{\Xm-\uv_{i}}( {\rm{d}}\xv) \right\} \notag \\
&\stackrel{{\rm{(c)}}}{=}   \sum_{i=1}^{\mathsf{M}-1} \left \{  \int  p_{i-1} \Pr \left( \left< \av, \phi(\Ym) \!-\! \xv \right>  \!\geq \!  \left( i-\frac{1}{2} \right)t\; \middle | \; \Xm=\xv+\uv_{i-1}  \right) \mu_\Uc({\rm d}\xv)  \right. \notag\\
& \qquad \qquad +\left .    \int  p_i   \Pr \left( \left< \av, \phi(\Ym) \!-\! \xv \right> \! \leq \! \left( i-\frac{1}{2} \right)t  \; \middle | \; \Xm=\xv+\uv_{i} \right)  \mu_\Uc({\rm d}\xv) \right\} \notag \\
&\stackrel{{\rm{(d)}}}{=} \int  P_e^\psi \left(\xv,t ; \Pc_\Uc(\xv), \Uc \right) \ \mu_\Uc({\rm d}\xv),
\end{align}
\hrule
\end{figure*}
where the labeled equalities follow from:
$\rm{(a)}$ applying  the change of variable $ P_{\Xm}({\rm{d}}\xv+\uv_{i-1}) =  P_{\Xm-\uv_{i-1} }({\rm{d}}\xv)$ in the first integral and the change of variable $ P_{\Xm}({\rm{d}}\xv+\uv_{i}) =  P_{\Xm-\uv_{i} }({\rm{d}}\xv)$ in the second integral;
 $\rm{(b)}$ using the assumption that $\left<\av, \uv_i\right> = it$;
 $\rm{(c)}$  using $\mu_\Uc$ in~\eqref{eq:def_mu_U} and $p_i\in\Pc_\Uc(\xv)$ in~\eqref{eq:def_P_U};
 and $\rm{(d)}$ defining $P_e^\psi \left(\xv,t ; \Pc_\Uc(\xv), \Uc \right)$ as the error probability of an $\mathsf{M}$-ary hypothesis testing problem as in Definition~\ref{def:MHT} associated with the (possibly sub-optimal) decision rule $\psi$ such that\footnote{We note that $\psi$ in~\eqref{eq:SubDecZZB} might not be unique; if this is the case, then we randomly select one of these possible choices.}
\begin{equation}
\label{eq:SubDecZZB}
	\psi(\Ym) = \Hc_i, \text{ where } i = \argmin_{ j \in [0:\mathsf{M}-1] } | \left< \av , \phi(\Ym) -\xv \right> - jt |.
\end{equation}
Combining~\eqref{eq:gen_ZZB_proof1} and~\eqref{eq:Getting_to_the_error_prob_step}, we obtain
\begin{align}
	\Pr\left( |\left< \av, \epsilonv \right>| \geq \frac{t}{2} \right)
	&=\frac{\int   P_e^\psi \left(\xv,t ; \Pc_\Uc(\xv), \Uc \right) \ \mu_\Uc({\rm d}\xv)}{\mathsf{M}-1}  \notag\\
	& \ge \frac{ \int   P_e \left(\xv; \Pc_\Uc(\xv), \Uc \right)  \ \mu_\Uc( {\rm d}\xv) }{\mathsf{M}-1},
\label{eq:Before_valley}
\end{align}
where we have used the fact that $P_e \left(\xv ; \Pc_\Uc(\xv), \Uc \right) $ is the error probability associated with an optimal decision rule (e.g., the MAP decision rule). The lower bound in~\eqref{eq:Before_valley} can be further tightened by optimizing $\Uc = \{\uv_i\}_{i=0}^{\mathsf{M}-1}$ such that $\left<\av,\uv_i\right> = it,~\forall i\in[0:\mathsf{M}-1]$, which gives
\begin{align}\label{eq:Before_valley2}
	\Pr\left( |\left< \av, \epsilonv \right>| \geq \frac{t}{2} \right)
	& \geq\! \sup\limits_{\substack{  \mathcal{U}  \subset \mathcal{X}: \\ \mathcal{U}= \{ \uv_k \}_{k=0}^{\mathsf{M}-1},  \\  \left< \av, \uv_k \right> = kt,~\forall k }} \!\! \frac{ \int    P_e \left ({\xv}; \mathcal{P}_\mathcal{U}(\xv), \mathcal{U}  \right )   \ \mu_{\mathcal{U}} ({\rm{d}}\xv) }{\mathsf{M}-1}  \nonumber \\
	& = \frac{ \hsf_{\Msf}(t,\av,P_{\Xm,\Ym}) }{\mathsf{M}-1},
\end{align}
where the equality follows from using $\hsf_{\Msf}(t,\av,P_{\Xm,\Ym}) $ in Definition~\ref{def:h_func}.
Applying the valley-filling function to the right-hand side of~\eqref{eq:Before_valley2} is valid since $\Vc_t\{\Pr\left( {\left | \left< \av,  \epsilonv \right> \right |} \geq \frac{t}{2} \right)\} = \Pr\left( {\left | \left< \av,  \epsilonv \right> \right |} \geq \frac{t}{2} \right)$ due to the monotonicity of $\Pr\left( {\left | \left< \av,  \epsilonv \right> \right |} \geq \frac{t}{2}\right)$ with respect to $t$. Thus, we obtain 
\begin{equation}
\label{eq:BZZ_Mary_scalar_proof7}
	\Pr\left( {\left | \left< \av,  \epsilonv \right> \right |} \geq \frac{t}{2} \right)   \geq \Vc_t \left\{ \frac{\mathsf{h}_{\Msf}(t,\av,{P_{\Xm,\Ym}})}{\mathsf{M}-1}  \right\}. 
\end{equation}
Substituting~\eqref{eq:BZZ_Mary_scalar_proof7} into~\eqref{eq:BZZ_Mary_vec_proof1} concludes the proof of Theorem~\ref{thm:BZZ_Mary_vec}.
\end{IEEEproof}

\subsection{Generalized SZZB}\label{app:GSZZB}
\begin{theorem}
\label{thm:SZZB_general}
Let $\epsilonv = \phi(\Ym) - \Xm$ be the estimation error when an estimator $\phi(\Ym)\in \mathcal{X}$ is used to estimate $\Xm\in \mathcal{X}$.
Then, for every $ \av \in \mathcal{X}$, any estimator $\phi:\Yc\to\Xc$, and any integer $\mathsf{M}\geq2$, it holds that
\begin{equation}\label{eq:general_sp_zzb}
	\EE\left[ \left< \av,\epsilonv\right>^2 \right]
	\geq \sup_{\Delta>0} \frac{\Delta^2 \hsf_\Msf(\Delta,\av,P_{\Xm,\Ym})}{2(\mathsf{M}-1)}   ,
\end{equation}
where $\hsf_\Msf(\Delta,\av,P_{\Xm,\Ym})$ is defined in Definition~\ref{def:h_func}.
\end{theorem}
\begin{IEEEproof}
Similar to the proof of Theorem~\ref{thm:BZZ_Mary_vec}, we start by using the alternative expression of the 
second moment,
\begin{align}\label{eq:sp_zzb_proof1}
	& \mathbb{E} \!\left[\left< \av,\phi(\Ym)- \Xm \right>^2 \right] \nonumber \\
	&\!=\! \int_0^\infty \frac{t}{2} \Pr\!\left( | \left< \av,\phi(\Ym)- \Xm \right>| \geq \frac{t}{2} \right) {\rm d}t  \nonumber\\
    & \!\ge \!\int_0^{2 \Delta} \frac{t}{2} \Pr\!\left( | \left< \av,\phi(\Ym)- \Xm \right>| \geq \frac{t}{2} \right) {\rm d}t ,
\end{align} 
for some $\Delta >0$.
We note that the probability in~\eqref{eq:sp_zzb_proof1} can be written as
\begin{align}\label{eq:sp_zzb_proof2}
	& \Pr\left( |\left< \av, \phi(\Ym)-\Xm \right>| \geq \frac{t}{2} \right) \nonumber \\
	& =  \frac{1}{\mathsf{M}-1} \!\sum_{i=1}^{\mathsf{M}-1} \left \{ \! \int \! \Pr\! \left( \left< \av, \phi(\Ym) -\Xm \right>  \!\geq \! \frac{t}{2}\! \; \middle | \;\! \Xm=\xv  \!\right) \! P_{\Xm}({\rm{d}}\xv) \right. \nonumber \\
	& \quad \left. +\!   \int \!    \Pr \left( \left< \av, \phi(\Ym) \!-\! \Xm \right> \! \leq \! - \frac{t}{2} \; \middle | \; \Xm\!=\!\xv \right)   P_{\Xm}({\rm{d}}\xv)\right\} .
\end{align}
Now, we choose a  set of vectors $\Uc = \{ \uv_i \}_{i=0}^{\mathsf{M}-1} \subset \mathcal{X}$ such that $\left<\av,\uv_i \right> = i\Delta$. Then, we observe that the summand in~\eqref{eq:sp_zzb_proof2} can be written as follows,
\begin{align}\label{eq:sp_zzb_proof3}
    & \int  \!\Pr \left( \!\left< \av, \phi(\Ym) \!-\! \Xm \right>  \!\geq \! \frac{t}{2} \; \middle | \; \Xm=\xv  \right)  P_{\Xm}({\rm{d}}\xv)  \nonumber \\
    &   +  \int  \Pr \left( \!\left< \av, \phi(\Ym) \!-\! \Xm\right> \! \leq \! -\frac{t}{2} \; \middle | \; \Xm=\xv \right)  P_{\Xm}({\rm{d}}\xv)   \notag \\
    & \overset{\rm (a)}{=} \!\! \int \! \Pr \left( \left< \av, \phi(\Ym) \!-\! \Xm \right>  \!\geq \! \frac{t}{2} \; \middle | \; \Xm=\xv+\uv_{i-1}  \right) \!\! P_{\Xm-\uv_{i-1} }({\rm{d}}\xv)   \nonumber \\
    &  \quad  +   \int \! \Pr \left( \left< \av, \phi(\Ym) \!-\! \Xm \right> \! \leq \! - \frac{t}{2} \; \middle | \; \Xm=\xv+\uv_{i} \right)  P_{\Xm-\uv_{i}}({\rm{d}}\xv)  \notag \\
    &\overset{\rm (b)}{=}  \!\!\! \int \!\! \Pr\! \!\left( \!\! \left< \av, \phi(\Ym) \!-\! \xv \right>  \!\!\geq \!\! (i\!-\!1)\Delta \!+\! \frac{t}{2} \!\!\; \middle | \;  \! \Xm \!\!=\!\! \xv \!+\! \uv_{i-1} \!\! \right) \!\!  P_{\Xm\!-\!\uv_{i-1} }\!( {\rm{d}}\xv)   \notag\\
    & \quad  \!+\!  \int \! \Pr \! \left( \left< \av, \phi(\Ym) \!-\! \xv \right> \! \leq \! i\Delta -\frac{t}{2}  \! \; \middle | \; \! \Xm=\xv+\uv_{i}\! \right) \!\!P_{\Xm-\uv_{i}}( {\rm{d}}\xv)  \notag \\
    &\overset{\rm (c)}{=}  \!\!\!  \int \!\!  p_{i-1} \!\Pr \!\!\left( \!\! \left< \av, \phi(\Ym) \!-\! \xv \right>  \!\!\geq \!\!  (i\!-\!1)\Delta \!+\! \frac{t}{2} \!\! \; \middle | \; \! \Xm \!\!=\!\! \xv \!+\! \uv_{i-1}\!\!  \right)\! \mu_\Uc({\rm d}\xv)   \notag\\
    & \quad +   \! \int \!  p_i   \Pr \left( \left< \av, \phi(\Ym) \!-\! \xv \right> \! \leq \! i\Delta \!-\! \frac{t}{2} \! \; \middle | \; \!\Xm\!=\! \xv \!+\! \uv_{i} \right) \! \mu_\Uc({\rm d}\xv)  ,
\end{align}
where the labeled equalities follow from:
$\rm{(a)}$ applying  the change of variable $ P_{\Xm}({\rm{d}}\xv+\uv_{i-1}) =  P_{\Xm-\uv_{i-1} }({\rm{d}}\xv)$ in the first integral and the change of variable $ P_{\Xm}({\rm{d}}\xv+\uv_{i}) =  P_{\Xm-\uv_{i} }({\rm{d}}\xv)$ in the second integral;
$\rm{(b)}$ using the assumption that $\left<\av, \uv_i\right> = i\Delta$;
and $\rm{(c)}$  using $\mu_\Uc$ in~\eqref{eq:def_mu_U} and $p_i\in\Pc_\Uc(\xv)$ in~\eqref{eq:def_P_U}.

Combining~\eqref{eq:sp_zzb_proof1},~\eqref{eq:sp_zzb_proof2}, and~\eqref{eq:sp_zzb_proof3}, we arrive at~\eqref{eq:sp_zzb_proof4}, at the top of the next page,
\begin{figure*}
\begin{align}\label{eq:sp_zzb_proof4}
    (\Msf-1)\mathbb{E} \left[\left< \av,\phi(\Ym)- \Xm \right>^2 \right] 
    & \geq \sum_{i=1}^{\mathsf{M}-1}  \left\{ \int_0^{2\Delta} \frac{t}{2} \int p_{i-1} \Pr\left(  \left< \av,\phi(\Ym)- \xv \right>  \geq (i-1)\Delta + \frac{t}{2} \; \middle|\; \Xm = \xv + \uv_{i-1} \right) \mu_\Uc({\rm d} \xv)  {\rm d}t \right. \nonumber \\
    & \qquad\qquad + \left. \int_0^{2\Delta} \frac{t}{2} \int p_i \Pr\left(  \left< \av,\phi(\Ym)- \xv \right> \leq i\Delta - \frac{t}{2} \; \middle|\; \Xm = \xv +\uv_i \right) \mu_\Uc({\rm d} \xv) {\rm d}t \right\} \nonumber \\
    & \overset{\rm (a)}{=}  \sum_{i=1}^{\mathsf{M}-1}  \left\{ \int_0^{2\Delta} \frac{t}{2} \int p_{i-1} \Pr\left(  \left< \av,\phi(\Ym)- \xv \right>  \geq (i-1)\Delta + \frac{t}{2} \; \middle|\; \Xm = \xv + \uv_{i-1} \right)  \mu_\Uc({\rm d} \xv)   {\rm d}t \right. \nonumber \\
	& \qquad\qquad + \left. \int_0^{2\Delta} \frac{2\Delta - t}{2} \int p_i \Pr\left(  \left< \av,\phi(\Ym)- \xv \right> \leq (i-1)\Delta + \frac{t}{2} \; \middle|\; \Xm = \xv +\uv_i \right) \mu_\Uc({\rm d} \xv) {\rm d}t \right\} \nonumber \\
    & \overset{\rm (b)}{\geq} \int_0^{2\Delta} \frac{\min\{t, 2\Delta - t\}}{2 }   \int  \sum_{i=1}^{\mathsf{M}-1}  \left\{  p_{i-1}  \Pr\left(  \left< \av,\phi(\Ym)- \xv \right>  \geq (i-1)\Delta + \frac{t}{2} \; \middle|\; \Xm = \xv + \uv_{i-1} \right) \right. \nonumber \\
	& \qquad\qquad + \left. p_{i} \Pr\left(  \left< \av,\phi(\Ym)- \xv \right> \leq (i-1)\Delta + \frac{t}{2} \; \middle|\; \Xm = \xv +\uv_i \right)\right\}  \mu_{\Uc}({\rm d}\xv) {\rm d}t  \nonumber\\
    & \overset{\rm (c)}{=} \int_0^{2\Delta} \frac{\min\{t, 2\Delta - t\}}{2 } \int  P_e^{\psi_{\rm sp}} \left(\xv,t, \Delta ; \Pc_\Uc(\xv), \Uc \right) \mu_\Uc({\rm d}\xv) {\rm d} t,
\end{align}
\hrule
\end{figure*}
where the labeled (in)equalities follow from:
$\rm (a)$ using the change of variable  $t = 2\Delta-t^\prime$ in the second integral and then relabeling $t^\prime$ as $t$;
$\rm (b)$ the facts that $t\geq \min\{t,2\Delta-t\}$ and $2\Delta-t \geq \min\{t,2\Delta-t\}$;
and $\rm (c)$ the fact that $P_e^{\psi_{\rm sp}} \left(\xv, t,\Delta ; \Pc_\Uc(\xv), \Uc \right)$ is the error probability of an $\mathsf{M}$-ary hypothesis testing problem as in Definition~\ref{def:MHT} associated with the (possibly sub-optimal) decision rule $\psi_{\rm sp}$ such that\footnote{We note that $\psi_{\rm sp}$ in~\eqref{eq:sp_zzb_decision_rule} might not be unique; if this is the case, then we randomly select one of these possible choices.}
\begin{subequations}\label{eq:sp_zzb_decision_rule}
\begin{align}
	& \psi_{\rm sp}(\Ym) = \Hc_i,  \\
	&  i = \argmin_{ j \in [0:\mathsf{M}-1] } \left| \left< \av , \phi(\Ym) -\xv \right> - \left(j-\frac{1}{2}\right)\Delta - \frac{t}{2}  \right|.
\end{align}
\end{subequations}
With this, we further lower bound $P_e^{\psi_{\rm sp}} \left(\xv, t,\Delta ; \Pc_\Uc(\xv), \Uc \right)$ by the minimum error probability $P_e(\xv; \Pc_\Uc(\xv),\Uc)$, which is associated with an optimal decision rule (e.g., the MAP decision rule), and we obtain that
\begin{align}\label{eq:sp_zzb_proof5}
    & \mathbb{E}[\left< \av,\phi(\Ym)- \Xm \right>^2]  \nonumber \\
    & \geq \int_0^{2\Delta} \frac{\min\{t, 2\Delta - t\}}{2(\Msf-1) } \int  P_e(\xv; \Pc_\Uc(\xv),\Uc) \mu_\Uc({\rm d}\xv) {\rm d} t \nonumber \\
    & = \frac{\Delta^2}{2(\Msf-1)} \int  P_e(\xv; \Pc_\Uc(\xv),\Uc) \mu_\Uc({\rm d}\xv) ,
\end{align}
where the equality follows by the fact that $P_e(\xv; \Pc_\Uc(\xv),\Uc)$ is independent of $t$.
The lower bound above can be further tightened by optimizing $\Uc = \{\uv_i\}_{i=0}^{\mathsf{M}-1}$ such that $\left<\av,\uv_i\right> = i \Delta,~\forall i\in[0:\mathsf{M}-1]$, which gives
\begin{align}\label{eq:sp_zzb_proof6}
	& \EE\left[ \left< \av,\epsilonv\right>^2 \right] \nonumber \\
	&\! \geq \sup_{\Delta>0} \frac{\Delta^2}{2(\mathsf{M}-1)} \!\!  \sup\limits_{\substack{  \mathcal{U}  \subset \mathcal{X}: \\ \mathcal{U}= \{ \uv_k \}_{k=0}^{\mathsf{M}-1}  \\  \left< \av, \uv_k \right> = k\Delta, \\ k \in [0:\mathsf{M}-1] }} \!\! \int \!\! P_e \left(\xv ; \Pc_\Uc(\xv), \Uc \right)  \mu_\Uc({\rm d}\xv).
\end{align}
This concludes the proof of Theorem~\ref{thm:SZZB_general}. 
\end{IEEEproof}

\section{Proof of Proposition~\ref{prop:tensorization}}
\label{app:sec:tensorization}

To prove the property of tensorization, it suffices to show that $\hsf_{\Msf}(t,\ev_i,P_{\Xm,\Ym}) $ in~\eqref{eq:BZZ_Mary_vec_mmse_h} depends only on $P_{X_i}$ and $P_{\Ym_i|X_i}$ instead of $P_{\Xm}$ and $P_{\Ym|\Xm}$.
We start by observing that
\begin{align}
\label{eq:tensorization_proof0}
	&  \hsf_{\Msf}(t,\ev_i,P_{\Xm,\Ym}) \nonumber\\
	& = \sup\limits_{\substack{  \mathcal{U}  \subset \mathcal{X}: \\ \mathcal{U}= \{ \uv_k \}_{k=0}^{\mathsf{M}-1} , \\  \left<\uv_k,\ev_i \right>= kt,~\forall k}}  \, \int_\Xc    P_e \left ({\xv}; \mathcal{P}_\mathcal{U}(\xv), \mathcal{U}  \right )  \,    \mu_{\mathcal{U}} ( {\rm{d}} \xv) \nonumber \\
	& =  \mathsf{M} - \!\!\!\!\!\!\! \inf\limits_{\substack{  \mathcal{U}  \subset \mathcal{X}: \\ \mathcal{U}= \{ \uv_k \}_{k=0}^{\mathsf{M}-1}  \\  (\uv_k )_i = kt,~\forall k }} \underset{\Xc\times \Yc}{\int}  \!\!\max_{\ell\in[0:\mathsf{M}-1]} \!\{ P_{\Ym|\Xm-\uv_\ell}( {\rm d} \yv | \xv ) P_{\Xm-\uv_\ell}({\rm d}\xv) \},
\end{align}
where we have used Lemma~\ref{lem:Pe_different_form} in Appendix~\ref{app:lem:Pe_different_form}, and $p_{\ell} = \frac{P_{\Xm-\uv_\ell}({\rm d} \xv)}{ \mu_\Uc({\rm d}\xv)}$. 
Note that the assumptions that $P_\Xm = \prod_{i=1}^d P_{X_i}$ and $P_{\Ym|\Xm} = \prod_{i=1}^d P_{\Ym_i|X_i}$ imply that  $P_{\Xm-\uv} = \prod_{i=1}^d P_{X_i-u_i}$ and $P_{\Ym|\Xm-\uv} = \prod_{i=1}^d P_{\Ym_i|X_i-u_i}$.
Let $X_i\in\Xc_i$ and $\Ym_i\in\Yc_i$ for all $i\in[1:d]$, which implies that $\prod_{i=1}^d \Xc_i = \Xc$ and $\prod_{i=1}^d \Yc_i = \Yc$.
Then, we can write the integral in~\eqref{eq:tensorization_proof0} as
\begin{align}\label{eq:tensorization_proof1}
	& \int_{\Xc\times \Yc}  \max_{\ell\in[0:\mathsf{M}-1]} \left \{ P_{\Ym|\Xm-\uv_\ell}( {\rm d} \yv | \xv ) P_{\Xm-\uv_\ell}({\rm d}\xv) \right \}\nonumber \\
	&\! = \! \underset{\Xc\times \Yc}{\int}  \!\!\! \!\!\max_{\ell\in[0:\mathsf{M}-1]} \! \Biggl\{ \!\prod_{j=1}^d \left( P_{\Ym_j|X_j-(\uv_\ell)_j}( {\rm d} \yv_j | x_j ) P_{X_j-(\uv_\ell)_j}({\rm d}x_j) \right ) \!\Biggr\} \nonumber \\
	&\! \!\geq \!\!\!\! \underset{ \underset{j\neq s}{\prod} \Xc_j \times \Yc_j}{\int} \!\!\!\! \!\!\max_{\ell\in[0:\mathsf{M}-1]} \! \Biggl\{ \prod_{j\neq s} \! P_{\Ym_j|X_j-(\uv_\ell)_j}( {\rm d} \yv_j | x_j ) P_{X_j-(\uv_\ell)_j}({\rm d}x_j)\! \Biggr\} \! ,
\end{align}
where the inequality follows by exchanging the $\max$ with the integral for some $s\in[1:d]$, and using the fact that the probability measure over the sample space is equal to one.
By iteratively doing this for all $j$'s except for $j=i$ (recall that $i$ is the parameter for $\hsf_{\Msf}(t,\ev_i,P_{\Xm,\Ym})$), we obtain
\begin{align}\label{eq:tensorization_proof2}
	& \int_{\Xc\times \Yc}  \max_{\ell\in[0:\mathsf{M}-1]} \left \{ P_{\Ym|\Xm-\uv_\ell}( {\rm d} \yv | \xv ) P_{\Xm-\uv_\ell} ({\rm d}\xv) \right \} \nonumber \\
	&  \geq   \underset{\Xc_i\times \Yc_i}{\int}  \max_{\ell\in[0:\mathsf{M}-1]} \left \{ P_{\Ym_i|X_i-\ell t}( {\rm d} \yv_i| x_i ) P_{X_i-\ell t}({\rm d}x_i) \right \}.
\end{align}
Since the lower bound in~\eqref{eq:tensorization_proof2} can be achieved by setting $\Uc = \{\uv_k\}_{k=0}^{\mathsf{M}-1}$ such that $\uv_k = kt \ev_i,~k\in[0:\mathsf{M}-1]$ in~\eqref{eq:tensorization_proof0}, a solution for the minimization problem in~\eqref{eq:tensorization_proof0} is given by $\Uc = \{\uv_k: \uv_k = kt \ev_i,~k\in[0:\mathsf{M}-1] \}$. Hence, we have that
\begin{align}\label{eq:tensorization_proof3}
	& \!\hsf_{\Msf}(t,\ev_i,P_{\Xm,\Ym}) \nonumber\\
	&\! \! = \mathsf{M} \! - \! \!\!\! \underset{\Xc_i\times \Yc_i}{\int} \!\! \max_{\ell\in[0:\mathsf{M}-1]} \! \left \{ P_{\Ym_i|X_i-\ell t}( {\rm d} \yv_i| x_i ) P_{X_i-\ell t}({\rm d}x_i) \right \},
\end{align}
which concludes the proof of Proposition~\ref{prop:tensorization} by noting that~\eqref{eq:tensorization_proof3} depends only on $P_{X_i}$ and $P_{\Ym_i|X_i}$.

\section{Non-Tensorization of the Bayesian Cramér-Rao Bound}
\label{app:CRBNoTens}
Let $\Ym = \Xm +\Nm, \text{ where } \Nm\sim\Nc(\zerov_d, {\eta I_d}),$ with $d$ being the dimension of $\Xm$ and $\eta$ representing the noise level. For this channel model, the Bayesian Cram\'er-Rao bound is given by~\cite{GuoInfoEst2013}
\begin{equation}
\label{eq:CRBGaussian}
{\mathsf{CRB}}(P_{\Xm,\Ym}) = \frac{d^2}{J(\Xm)+d/\eta},
\end{equation}
where $J(\Xm)$ is the Fisher information of $\Xm$, i.e., $J(\Xm) = \EE \left [\| \nabla_{\mathbf{X}} \log f_{\Xm}(\Xm) \|^2\right ]$ with $f_{\mathbf{X}}(\cdot)$ being the PDF of $\mathbf{X}$.
Now, if $f_\Xm(\mathbf{x}) = \prod_{i=1}^d f_{X_i}(x_i)$, it is a simple exercise (see~\cite[Chapter 11.10]{Cover:InfoTheory}) to show that the Fisher information tensorizes, that is,
\begin{equation}
\label{eq:FisherTens}
J(\Xm) = \sum_{i=1}^d \EE \left [ \left(\frac{\partial \log f_{X_i}(X_i)}{\partial X_i} \right )^2 \right ].
\end{equation}
Thus, substituting~\eqref{eq:FisherTens} inside~\eqref{eq:CRBGaussian}, we arrive at
\begin{equation}
{\mathsf{CRB}}(P_{\Xm,\Ym}) = \frac{d^2}{\sum_{i=1}^d \EE \left [ \left(\frac{\partial \log f_{X_i}(X_i)}{\partial X_i} \right )^2 \right ] + d/\eta},
\end{equation}
which clearly does not tensorize since the sum appears in the denominator instead of the numerator.

\section{Proof of Theorem~\ref{thm:zzb_sig_inf_M}}\label{app:high_noise}

At first, we observe that if $f(t)\leq g(t)$ for all $t\in\RR$, then 
\begin{equation}\label{eq:valley_upper}
	\Vc_t\{f(t)\} \leq \Vc_t\{g(t)\},~\forall t\in\RR.
\end{equation}
Second, the valley-filling function is lower semicontinuous, i.e., we have that
\begin{equation}\label{eq:valley_liminf}
	\liminf_{n\to\infty} \Vc_t\{ f_n(t)\} \geq \Vc_t\{\liminf_{n\to\infty} f_n(t)\}.
\end{equation}
To highlight the dependency of $\mathsf{h}_\Msf(t,\ev_i,P_{\Xm,\Ym})$ in~\eqref{eq:BZZ_Mary_vec_mmse_h} on $\eta$, we use $\mathsf{h}_\Msf(t,\ev_i,P_{\Xm,\Ym};\eta)$.
We also note that $\mathsf{h}_\Msf(t,\ev_i,P_{\Xm,\Ym};\eta)$ is non-decreasing with respect to the noise level $\eta$ since $ P_e \left (\eta, {\xv}; \mathcal{P}_\mathcal{U}(\xv), \mathcal{U}  \right )  $ is non-decreasing in $\eta$ as assumed in {\bf A2}.
We then have that the ZZB in~\eqref{eq:BZZ_Mary_vec_mmse_sub1} can be written as
\begin{align}\label{eq:BZZ_Mary_vec_mmse_sub1_proof}
	\overline{\mathsf{ZZ}}(P_{\Xm,\Ym},\mathsf{M})
	& = \sum_{i=1}^{d} \! \int_0^\infty \frac{t}{2} \Vc_t\!\left\{ \frac{ \mathsf{h}_\Msf(t,\ev_i,P_{\Xm,\Ym};\eta)}{\mathsf{M}-1}\! \right\} \! {\rm d} t \nonumber \\
	& \leq \!\sum_{i=1}^{d} \!\int_0^\infty \!\frac{t}{2} \Vc_t\!\left\{ \frac{ \mathsf{h}_\Msf(t,\ev_i,P_{\Xm,\Ym};\infty)}{\mathsf{M}-1}\! \right\} \! {\rm d} t.
\end{align}
For the sake of space, we abbreviate the constraint  $\Cc=  \{ \Uc \subset \Xc:  \mathcal{U}= \{ \uv_k \}_{k=0}^{\mathsf{M}-1} , (\uv_k )_i = kt,~\forall k \}$.
Due to the assumption {\bf A3}, we can write $\mathsf{h}_\Msf(t,\ev_i,P_{\Xm,\Ym};\infty)$ as
\begin{align}\label{eq:BZZ_Mary_vec_mmse_h_infty}
	\mathsf{h}_\Msf(t,\ev_i,P_{\Xm,\Ym};\infty)
	& = \sup\limits_{\substack{  \mathcal{U} \in \Cc }}  \int    P_e \left (\infty, {\xv}; \mathcal{P}_\mathcal{U}(\xv), \mathcal{U}  \right )  \mu_\Uc({\rm d}\xv) \nonumber\\
	& = \sup\limits_{\substack{  \mathcal{U} \in \Cc }} \left\{  \mathsf{M} - \int  \max_{j\in[0:\mathsf{M}-1]} P_{\Xm-\uv_{j} }({\rm d}\xv) \right\} \nonumber \\
	& =   \mathsf{M} \! -\! \inf\limits_{\substack{  \mathcal{U} \in \Cc}}  \int  \! \! \max_{j\in[0:\mathsf{M}-1]} P_{\Xm-\uv_{j} }({\rm{d}}\xv),
\end{align}
which,  substituted inside~\eqref{eq:BZZ_Mary_vec_mmse_sub1_proof}, leads to
\begin{equation}
\label{eq:zzb_sig_inf_upper}
	\overline{\mathsf{ZZ}}(P_{\Xm,\Ym},\mathsf{M}) 
	\leq \sum_{i=1}^{d}\!  \int_0^\infty \! \frac{t}{2} \Vc_t  \left\{\! \frac{\mathsf{M}  - \mathsf{H}_\Msf(t,\ev_i,P_\Xm)}{\mathsf{M}-1} \!\right\}\! {\rm d} t.
\end{equation}
The above provides an upper bound on $\overline{\mathsf{ZZ}}(P_{\Xm,\Ym},\mathsf{M})$. We next derive a lower bound on $\overline{\mathsf{ZZ}}(P_{\Xm,\Ym},\mathsf{M})$.
We have that
\begin{align}\label{eq:BZZ_Mary_vec_mmse_sub1_lower}
	& \liminf_{\eta\to\infty}   \overline{\mathsf{ZZ}}(P_{\Xm,\Ym},\mathsf{M}) \nonumber
	\\& \overset{\rm (a)}{\geq} \!\sum_{i=1}^{d}  \int_0^\infty \!\! \liminf_{\eta\to\infty}  \frac{t}{2} \Vc_t \left\{ \! \frac{ \mathsf{h}_\Msf(t,\ev_i,P_{\Xm,\Ym};\eta)}{\mathsf{M}-1} \! \right\}  {\rm d} t \nonumber \\
	& \overset{\rm (b)}{\geq} \sum_{i=1}^{d}  \int_0^\infty \frac{t}{2} \Vc_t \left\{ \! \frac{ \liminf_{\eta\to\infty} \mathsf{h}_\Msf(t,\ev_i,P_{\Xm,\Ym};\eta)}{\mathsf{M}-1} \! \right\}  {\rm d} t,
\end{align}
where $\rm (a)$ is due to Fatou's lemma and $\rm (b)$ follows from~\eqref{eq:valley_liminf}.
Moreover,  we have that
\begin{align}\label{eq:BZZ_Mary_vec_mmse_h_liminf}
	& \liminf_{\eta\to\infty} \mathsf{h}_\Msf(t,\ev_i,P_{\Xm,\Ym};\eta)  \nonumber \\
	& =  \liminf_{\eta\to\infty}  {\sup \limits_{\substack{  \mathcal{U} \in \Cc }}} \!  \int  \!\!  P_e \left (\eta, {\xv}; \mathcal{P}_\mathcal{U}(\xv), \mathcal{U}  \right ) \mu_\Uc({\rm d}\xv) \nonumber \\
	&  \overset{\rm (a)}{\geq} \! {\sup\limits_{\substack{  \mathcal{U} \in \Cc }}} \!\liminf_{\eta\to\infty} \! \! \int  \!\! P_e \left (\eta, {\xv}; \mathcal{P}_\mathcal{U}(\xv), \mathcal{U}  \right ) \mu_\Uc({\rm d}\xv)  \nonumber \\
	&  \overset{\rm (b)}{\geq} { \sup\limits_{\substack{  \mathcal{U} \in \Cc }}}  \int   P_e \left (\infty, {\xv}; \mathcal{P}_\mathcal{U}(\xv), \mathcal{U}  \right ) \mu_\Uc({\rm d}\xv) \nonumber \\
	&  \overset{\rm (c)}{=} \mathsf{M} \!-\!\! {\inf\limits_{\substack{  \mathcal{U} \in \Cc }}} \int  \!\!\!\!\max_{j\in[0:\mathsf{M}-1]} P_{\Xm-\uv_{j} }({\rm d}\xv),
\end{align}
where 
$\rm (a)$ follows by exchanging the $\sup$ and the $\lim\inf$, $\rm (b)$ is due to Fatou's lemma, and $\rm (c)$ is from the assumption {\bf A3}.
Substituting~\eqref{eq:BZZ_Mary_vec_mmse_h_liminf} into~\eqref{eq:BZZ_Mary_vec_mmse_sub1_lower}, we obtain that, for $\eta\to\infty$, it holds that
\begin{equation}
	\!\! \overline{\mathsf{ZZ}}(P_{\Xm,\Ym},\mathsf{M})
	 \geq \sum_{i=1}^{d}  \!\int_0^\infty \frac{t}{2} \Vc_t\left\{ \frac{ \mathsf{M} \! -\! \mathsf{H}_\Msf(t,\ev_i,P_\Xm)  }{\mathsf{M}-1} \right\} {\rm d} t,
\end{equation}
which agrees with \eqref{eq:zzb_sig_inf_upper}; this proves~\eqref{eq:high_noise_valley_ZZB}.
With similar steps as in the proof of~\eqref{eq:high_noise_valley_ZZB}, one can easily show~\eqref{eq:high_noise_ZZB} and~\eqref{eq:high_noise_SZZB}.
This completes the proof of Theorem~\ref{thm:zzb_sig_inf_M}.

\section{Proof of Theorem~\ref{thm:low_noise}}\label{app:low_noise_proof}
Using the change of variable $t = \sqrt{\eta} \tau$ in~\eqref{eq:BZZ_Mary_vec_mmse_sub2},  we obtain that
\begin{equation}\label{eq:low_noise_proof0}
	\frac{ {\mathsf{ZZ}}(P_{\Xm,\Ym},2 )}{\eta} 
	 = \sum_{i=1}^{d}  \int_0^\infty \frac{\tau}{2} \mathsf{h}_{2}(\sqrt{\eta}\tau,\ev_i,P_{\Xm,\Ym}) \, {\rm d} \tau.
\end{equation}
Moreover, since we can write 
\begin{align}
	\mu_\Uc 
	& = \alpha\sum_{k=0}^{1} P_{\Xm_C - \uv_k} + (1-\alpha)\sum_{k=0}^{1} P_{\Xm_D - \uv_k}  \nonumber\\
	& = \alpha \mu_{\Uc_C} + (1-\alpha)\mu_{\Uc_D},
\end{align}
we can lower bound $\mathsf{h}_{2}(\sqrt{\eta}\tau,\ev_i,P_{\Xm,\Ym})$ in~\eqref{eq:low_noise_proof0} as follows,
 \begin{align}\label{eq:low_noise_proof_cont}
	& \mathsf{h}_{2}(\sqrt{\eta}\tau,\ev_i,P_{\Xm,\Ym}) \nonumber \\
	& \overset{\rm (a)}{\geq} \!\!  \sup\limits_{\substack{  \mathcal{U}  \subset \mathcal{X}: \\ \mathcal{U}= \{ \uv_k \}_{k=0}^{1}  \\  ( \uv_k )_i = \sqrt{\eta} k\tau, ~\forall k}}  \!\!\! \alpha \int    P_e \left (\eta,\xv; \mathcal{P}_{\mathcal{U}}(\xv), \mathcal{U}  \right ) \mu_{\mathcal{U}_C} ( {\rm{d}} \xv) \nonumber \\
	& \overset{\rm (b)}{=} \!\!\!\!  \sup\limits_{\substack{  \mathcal{U}  \subset \mathcal{X}: \\ \mathcal{U}= \{ \uv_k \}_{k=0}^{1}  \\  ( \uv_k )_i = \sqrt{\eta} k\tau, ~\forall k }} \!\!\!\!\! \alpha \int    P_e \left (\eta,\xv; \mathcal{P}_{\mathcal{U}_C}(\xv), \mathcal{U}  \right )  \mu_{\mathcal{U}_C} ({\rm{d}}\xv) \nonumber \\
	& = \alpha \mathsf{h}_2 (\sqrt{\eta}\tau, \ev_i,P_{\Xm_C,\Ym}),
\end{align}
where
$\rm (a)$ follows by dropping the integral with respect to $\mu_{\Uc_D}$
and $\rm (b)$ is due to the fact that, almost surely, with respect to $\mu_{\mathcal{U}_C} $,
\begin{equation}
	\mathcal{P}_{\mathcal{U}}(\xv) 
	= \mathcal{P}_{\mathcal{U}_C}(\xv),
\end{equation} 
where 
\begin{equation}
\mathcal{P}_{\mathcal{U}_C}(\xv) \!=\! \left \{p_i : p_i \!=\!   \frac{P_{\Xm_C-\uv_{i}}({\rm d}\xv) }{ \mu_{\mathcal{U}_C} ({\rm d} \xv)} ,  \uv_i \in \mathcal{U}\!=\! \{ \uv_k \}_{k=0}^{\mathsf{M}-1} \right \};
\end{equation} 
to see this note that $P_{\Xm_D}$ and $\mu_{\Uc_D}$ are singular with respect to $\mu_{\mathcal{U}_C}$ and hence, almost surely, with respect to $\mu_{\mathcal{U}_C} $ we have that
\begin{equation}
	\frac{ P_{\Xm-\uv_{i}}({\rm d}\xv) }{\mu_{\mathcal{U}}  ({\rm d}\xv) } 
	= \frac{ P_{\Xm_C-\uv_{i} } ({\rm d}\xv)}{    \mu_{\mathcal{U}_C}({\rm d}\xv)   } .
\end{equation} 
Now, we can leverage Lemma~\ref{lem:continuous_gaussian_h} in Appendix~\ref{app:continuous_gaussian_h}  to further lower bound $ \mathsf{h}_2 (\sqrt{\eta}\tau, \ev_i,P_{\Xm_C,\Ym})$ in~\eqref{eq:low_noise_proof_cont} as
\begin{align}\label{eq:low_noise_proof2}
	 & \mathsf{h}_2 (\sqrt{\eta}\tau, \ev_i,P_{\Xm_C,\Ym}) \nonumber \\
	& \!\geq \! \int \! \biggl(  \int_{\yv  \in \Rc_0(\eta)}   \! \! f_\Zm(\yv \!-\! \tau \ev_i ) f_{\Xm_C}(\xv\!+ \! \sqrt{\eta} \tau \ev_i) \ {\rm  d} \yv   \nonumber \\
	&~~~~  +   \int_{\yv  \in \Rc_1(\eta)} \! f_\Zm(\yv  ) f_{\Xm_C}(\xv) \ {\rm  d} \yv  \,   \biggr) {\rm d}\xv,
\end{align}
where 
\begin{align*}
	& \Rc_0(\eta)  = \left\{ \yv :   \ln \frac{f_{\Xm_C}(\xv )}{f_{\Xm_C}(\xv \!+\! \sqrt{\eta}\tau \ev_i)} +  \frac{\tau^2 }{2} >  \tau y_i \right\}, \text{ and } \nonumber \\
	& \Rc_1(\eta)  = \left\{ \yv :   \ln \frac{f_{\Xm_C}(\xv)}{f_{\Xm_C}(\xv \!+\! \sqrt{\eta}\tau \ev_i)} +  \frac{\tau^2 }{2} <  \tau y_i \right\}.
\end{align*}
In particular,  to obtain~\eqref{eq:low_noise_proof2}, we have let $\vv_0 = \zerov_d$ and $\vv_1 = \tau \ev_i$ in Lemma~\ref{lem:continuous_gaussian_h} in Appendix~\ref{app:continuous_gaussian_h}, and we have used $f_\Zm$ to denote the PDF of $\Zm\sim\Nc(\zerov_d,I_d)$. 

Then,  by using Fatou's lemma with~\eqref{eq:low_noise_proof2}, we arrive at
\begin{align}\label{eq:low_noise_proof4}
	& \liminf_{\eta\to0} \mathsf{h}_2 (\sqrt{\eta}\tau, \ev_i,P_{\Xm_C,\Ym})  \nonumber \\
	& \geq  \int  \biggl(  \int_{\yv  \in \Rc_0(0)}    f_\Zm(\yv - \tau \ev_i ) f_{\Xm_C}(\xv) \ {\rm  d} \yv   \nonumber \\
	&~~~~~~  +   \int_{\yv  \in \Rc_1(0)} \! f_\Zm(\yv  ) f_{\Xm_C}(\xv) \  {\rm  d} \yv  \,   \biggr) {\rm d}\xv \nonumber \\
	& =  \Pr \left( |Z_i| > \frac{\tau}{2} \right).
\end{align}
Moreover, always using Fatou's lemma, we have that
\begin{align}
	& \liminf_{\eta\to0} \frac{  {\mathsf{ZZ}}(P_{\Xm,\Ym},2)}{\eta} \nonumber \\
	& \geq  \sum_{i=1}^{d} \!\int_0^\infty  \liminf_{\eta\to0} \frac{\tau}{2} \mathsf{h}_{2}(\sqrt{\eta}\tau,\ev_i,P_{\Xm,\Ym}) \ {\rm d} \tau.
\end{align}
Using~\eqref{eq:low_noise_proof_cont} and~\eqref{eq:low_noise_proof4} inside the above expression, we obtain
\begin{align}\label{eq:low_noise_proof5}
	\liminf_{\eta\to0} \frac{ {\mathsf{ZZ}}(P_{\Xm,\Ym},2)}{\eta} 
	&  \geq \alpha  \sum_{i=1}^{d} \! \int_0^\infty \frac{\tau}{2}  \Pr\left(|Z_i| > \frac{\tau}{2} \right) \, {\rm d} \tau \nonumber \\
	& \! = \alpha \!\! \sum_{i=1}^{d} \EE[ Z_i^2] 
	= \alpha d.
\end{align}
To conclude the proof of Theorem~\ref{thm:low_noise}, we note~\cite{David2016,MMSEdim} that
\begin{equation}\label{eq:low_noise_proof6App}
	\lim_{\eta\to0} \frac{{\rm mmse}(\Xm|\Ym)}{\eta}
	 = \alpha d.
\end{equation}
Since ${\rm mmse}(\Xm|\Ym) \geq  \! {\mathsf{ZZ}}(P_{\Xm,\Ym},\mathsf{M}) \geq  {\mathsf{ZZ}}(P_{\Xm,\Ym},2)$,~\eqref{eq:low_noise_proof5} and~\eqref{eq:low_noise_proof6App} imply that
\begin{equation}
	\lim_{\eta\to0} \frac{ {\mathsf{ZZ}}(P_{\Xm,\Ym},2)}{\eta}
	 = \alpha d,
\end{equation}
which completes the proof of Theorem~\ref{thm:low_noise}.

\section{Proof of Proposition~\ref{prop:assum_low_SZZB}}\label{app:low_noise_szzb}
In this proof, we will use the same notation used in the proof of Theorem~\ref{thm:low_noise} in Appendix~\ref{app:low_noise_proof}.
Recall that Theorem~\ref{thm:low_noise} shows that
\begin{equation}
    \lim_{\eta\to0} \frac{ {\mathsf{ZZ}}(P_{\Xm,\Ym},2)}{\eta}
	 = \alpha d.
\end{equation}
This demonstrates that in the proof of Theorem~\ref{thm:low_noise} the inequalities~\eqref{eq:low_noise_proof_cont},~\eqref{eq:low_noise_proof2}, and~\eqref{eq:low_noise_proof4} hold with equality with $\lim_{\eta\to0}$ instead of $\liminf_{\eta\to0}$.
 Thus, for $ \tau >0$, it holds that
\begin{equation}\label{eq:low_noise_szzb_proof1}
    \lim_{\eta\to0} \hsf_2(\sqrt{\eta}\tau,\ev_i,P_{\Xm,\Ym})
     = \alpha \Pr\left(|Z|>\frac{\tau}{2}\right),
\end{equation}
where $Z\sim\Nc(0,1)$.
Since for any $\eta>0$ it holds that $\sup_{\Delta>0} f(\Delta) = \sup_{\tau>0} f(\eta \tau)$, we write $\Zsf\Zsf_{\rm sp}(P_{\Xm,\Ym},2)$ in Theorem~\ref{thm:SZZB} by setting $\Delta = \sqrt{\eta}\tau$ as follows,
\begin{equation}
    \Zsf\Zsf_{\rm sp}(P_{\Xm,\Ym},2)
     = \sum_{i=1}^d \sup_{\tau>0} \frac{\eta \tau^2}{2} \hsf_2(\sqrt{\eta}\tau,\ev_i,P_{\Xm,\Ym}).
\end{equation}
With this, we arrive at
\begin{align}
    \lim_{\eta\to0} \frac{\Zsf\Zsf_{\rm sp}(P_{\Xm,\Ym},2)}{\eta}
    & = \sum_{i=1}^d \lim_{\eta\to0} \sup_{\tau>0} \frac{\tau^2 \hsf_2(\sqrt{\eta}\tau,\ev_i,P_{\Xm,\Ym})}{2} \nonumber \\
    & \overset{\rm (a)}{\geq}  \sup_{\tau>0} \frac{\alpha d \tau^2 }{2} \Pr\left(|Z|>\frac{\tau}{2}\right) \nonumber \\
    & \overset{\rm (b)}{=} 4 \, \alpha \, d    \sup_{\tau>0} \tau^2 Q(\tau) \nonumber \\
    & \overset{\rm (c)}{=} \alpha d \gamma ,
\end{align}
where the labeled (in)equalities follow from:
$\rm (a)$ exchanging the $\lim$ and the $\sup$, and using~\eqref{eq:low_noise_szzb_proof1}; 
$\rm (b)$ using $Q(x) = \int_{x}^{\infty} f_Z(u)\ {\rm d}u$;
and $\rm (c)$ letting $\gamma =4 \sup_{\tau>0} \tau^2 Q(\tau)$. The above shows the lower bound in Proposition~\ref{prop:assum_low_SZZB}.

We now focus on the upper bound in Proposition~\ref{prop:assum_low_SZZB}.
Similar to the proof of Theorem~\ref{thm:low_noise},
it is known~\cite{David2016,MMSEdim} that
\begin{equation}
	\lim_{\eta\to0} \frac{{\rm mmse}(\Xm|\Ym)}{\eta}
	 = \alpha d.
\end{equation}
Hence, the facts that ${\rm mmse}(\Xm|\Ym) \geq  {\mathsf{ZZ}}_{\rm sp}(P_{\Xm,\Ym},\mathsf{M}) \geq  {\mathsf{ZZ}}_{\rm sp}(P_{\Xm,\Ym},2)$ implies that
\begin{equation}
    \alpha d \gamma 
     \leq \lim_{\eta\to0} \frac{\Zsf\Zsf_{\rm sp}(P_{\Xm,\Ym},2)}{\eta} 
    \leq \alpha d.
\end{equation}
This concludes the proof of Proposition~\ref{prop:assum_low_SZZB}.

\section{Proof of Proposition~\ref{prop:discrete_zero_zzb}}
\label{app:discrete_zero_zzb}
Let $p_{\Xm}$ denote the PMF of $\Xm$.
For a discrete $\Xm\in\Sc_\Xm$, where $\Sc_\Xm$ is the support of $\Xm$, from~\eqref{eq:BZZ_Mary_vec_mmse_h} we have that
\begin{align}\label{eq:discrete_zero_proof3}
	& \mathsf{h}_\Msf(t,\ev_i,P_{\Xm,\Ym}) \nonumber \\
	& = \sup\limits_{\substack{  \mathcal{U}  \subset \Xc: \\ \mathcal{U}= \{ \uv_k \}_{k=0}^{\mathsf{M}-1} , \\  \left<\uv_k,\ev_i\right>= kt,~\forall k} }  \sum_{\xv\in \overline{\Sc}_{\Uc} }     \sum_{{k=0}}^{\mathsf{M}-1}  p_{\Xm-\uv_{k} } \!(\xv)  P_e \left(\xv;\Pc_\Uc(\xv),\Uc \right),
\end{align}
where $\overline{\Sc}_{\Uc} = \bigcup_{i=1}^d\{\Sc_{\Xm}-\uv_i\}$.\footnote{Note that $\Uc$ depends on $t$, we omit this dependency in order not to overload the notation.} We will demonstrate that $   P_e \left(\xv;\Pc_\Uc(\xv),\Uc \right)$ in~\eqref{eq:discrete_zero_proof3} is equal to zero almost surely for $t \geq 0$, which, via \eqref{eq:BZZ_Mary_vec_mmse_sub2} and~\eqref{eq:discrete_zero_proof3},  implies that ${\mathsf{ZZ}}(P_{\Xm,\Ym},\mathsf{M})= 0$.

We note that, given  $t \geq 0$ and $\xv  \in \overline{\Sc}_{\Uc}$,  a sufficient condition for $   P_e \left(\xv;\Pc_\Uc(\xv),\Uc \right)$ to be zero is that there exists a $j \!\in\! [0:\mathsf{M}-1]$ for which $p_j\!=\! 1$, where $p_j\in\Pc_\Uc(\xv)$. Under this condition,  in fact, the optimal decision rule would always declare the correct hypothesis leading to a zero error probability. 
From the definition of $p_i\in\Pc_\Uc(\xv)$ in~\eqref{eq:def_P_U}, we have that
\begin{equation}
	p_i =\frac{  p_{\Xm-\uv_{i} }(\xv) }{   \sum_{{k=0}}^{\mathsf{M}-1}  p_{\Xm-\uv_{k} }(\xv) }, 
\end{equation}
and hence, there exists a $j \in [0:\mathsf{M}-1]$ for which $p_j  = 1$ if and only if 
\begin{equation}
\bigcup_{\Ic\subseteq[0:\mathsf{M}-1]:|\Ic|\geq2} \bigcap_{k\in\Ic} \{\Sc_\Xm - \uv_k\} = \varnothing. 
\end{equation}
Thus, we have the following inclusion: for a given $\xv  \in \overline{\Sc}_{\Uc}$
\begin{equation}
\Ac \subseteq  \left\{ t \ge 0:   P_e \left(\xv;\Pc_\Uc(\xv),\Uc \right) = 0  \right\} ,
\end{equation} 
where 
\begin{align}\label{eq:discrete_zero_proof4}
	\Ac = \biggl\{t\geq0 : & 	\bigcup_{\substack{\Ic\subseteq[0:\mathsf{M}-1] : \\ |\Ic| \geq 2}} \bigcap_{k\in\Ic} \{\Sc_\Xm - \uv_k\} = \varnothing, \nonumber \\
	& \uv_k\in \Xc,   (\uv_k)_i = kt, ~k\in[0:\mathsf{M}-1] \biggr \}.
\end{align}
We now want to show that $\mathcal{A}$ above has a full measure, i.e., the complement of $\Ac$ on $t \in [0,\infty)$ has a zero Lebesgue measure. 
To this end, in Lemma~\ref{lem:discrete_zero_lemma}  in Appendix~\ref{app:discrete_zero_lemma}, we show that  $\mathcal{A}^c$ (i.e., the complement of $\mathcal{A})$ is countable and hence, $\mathcal{A}^c$ is of measure zero as desired.  
This implies that $P_e \left(\xv;\Pc_\Uc(\xv),\Uc \right)$ in~\eqref{eq:discrete_zero_proof3} is equal to zero almost surely for $t \geq 0$ and hence,  the integral in~\eqref{eq:BZZ_Mary_vec_mmse_sub2} is equal to zero, leading to ${\mathsf{ZZ}}(P_{\Xm,\Ym},\mathsf{M}) = 0$. This concludes the proof of Proposition~\ref{prop:discrete_zero_zzb}.

\section{Proof of Theorem~\ref{thm:Discrete_not_tight}}
\label{app:Discrete_not_tight}

When $\EE[X] = 0$, an alternative form of $\var(X)$ (to which the MMSE converges in the high-noise regime -- see~\eqref{eq:HighNoiseMMSEVar}) is given by
\begin{equation}\label{eq:Discrete_not_tight_proof2}
	\var(X)
	= \EE[X^2] 
	 = \int_0^\infty \frac{t}{2} \Pr\left(|X| \geq \frac{t}{2} \right) \ {\rm d} t.
\end{equation}
Now, recall from the proof of Theorem~\ref{thm:BZZ_Mary_vec_mmse} (see~\eqref{eq:BZZ_Mary_scalar_proof7} in Appendix~\ref{app:gen_zzb}) that the ZZB was derived by establishing the following lower bound: for $t >0$, 
\begin{equation}
\label{eq:AppHInt}
	\Pr\left( {\left | \left< \av,  \epsilonv \right> \right |} \geq \frac{t}{2} \right)   \geq \Vc_t \left\{ \frac{\mathsf{h}_{\Msf}(t,\av,{P_{\Xm,\Ym}})}{\mathsf{M}-1}  \right\}. 
\end{equation}
By letting $\av=\ev_1=1$ (since we are considering the univariate case) and $\phi(Y) = \EE[X | Y ] = \EE[X] = 0$ (since we are considering the high-noise regime), we have that
the left-hand side of the above inequality is equal to the probability term inside the integral in~\eqref{eq:Discrete_not_tight_proof2}.
Moreover, in the high-noise regime, we have that $\mathsf{h}_{\Msf}(\cdot,\cdot,\cdot)$ can be written as in~\eqref{eq:BZZ_Mary_vec_mmse_h_infty}. With this, we can rewrite~\eqref{eq:AppHInt} as follows,
\begin{equation}
\Pr\left(|X| \!\geq \!\frac{t}{2} \right) \! \!\ge \! \!\Vc_t\left\{  \frac{ \mathsf{M} \!- \!\sum_{x\in \overline{\Sc}_t }\max_{k\in[0:\mathsf{M}-1]} p_X(x\!+\!kt) }{\mathsf{M}-1} \right\} \!,  \label{eq:Discrete_not_tight_proof3}
\end{equation} 
where $\overline{\Sc}_t = \bigcup_{j=0}^{\mathsf{M}-1} \left\{ \Sc_X - jt \right \}$ is the union of the supports $\Sc_X - jt$ of $X-jt,~j\in[0:\Msf-1]$. 
We next show that~\eqref{eq:Discrete_not_tight_proof3} does not hold with equality and thus,  $ \var(X) > \overline{\mathsf V}(P_X, \mathsf{M})$.
First, let $x_0 = \inf_{x\in \Sc_X} |x|$ and note that the left-hand side of~\eqref{eq:Discrete_not_tight_proof3} is
\begin{equation}\label{eq:Discrete_not_tight_proof33}
	\Pr\left(|X| \geq \frac{t}{2} \right) 
	= 1, \text{ if } t \leq x_0.
\end{equation}
Second,  assume that $t>0$ and consider the sum inside the right-hand side of~\eqref{eq:Discrete_not_tight_proof3}. 
Since $p_X(x+kt) = p_{X-kt}(x)$, we have
\begin{equation}\label{eq:Discrete_not_tight_proof5}
	\sum_{x\in \overline{\Sc}_t}  \max_{k\in[0:\mathsf{M}-1]} p_X(x+kt) 
	\overset{\rm (a)}{>} \sum_{x\in \Sc_X} \max_{k\in[0:\mathsf{M}-1]} p_{X-kt}(x)
	\overset{\rm (b)}{\geq} 1,
\end{equation}
where $\rm (a)$ follows by the fact that $\overline{\Sc}_t\setminus \Sc_X \neq \varnothing$ if $t>0$, and $\rm (b)$ follows by dropping the $\max$ and choosing $k=0$.
Note that~\eqref{eq:Discrete_not_tight_proof5} is true for any $t > 0$ and hence,  from the definition of valley-filling function in~\eqref{eq:VF}, it follows that the right-hand side of~\eqref{eq:Discrete_not_tight_proof3} is strictly smaller than one for all $t>0$.  This, together with~\eqref{eq:Discrete_not_tight_proof33}, shows that there exists a range of $t$, namely $ 0< t \leq x_0$, such that the condition in~\eqref{eq:Discrete_not_tight_proof3} does not hold with equality.
This concludes the proof of Theorem~\ref{thm:Discrete_not_tight}.

\section{Proof of Proposition~\ref{prop:mmse=ZZB}}
\label{app:mmse=ZZB}

We start by noting that the ZZB in Theorem~\ref{thm:BZZ_Mary_vec_mmse}  was derived by individually lower bounding each of the $d$ elements that contribute to ${\rm mmse}(\Xm|\Ym)$ (see~\eqref{eq:def_mmse}).  Thus,~\eqref{eq:mmse=ZZB} holds if and only if the $i$th element that contributes to ${\rm mmse}(\Xm|\Ym)$ is equal to the $i$th element that contributes to ${\mathsf{ZZ}}(P_{\Xm,\Ym},\mathsf{M})$, where $i\in[1: d]$.

Now, consider the $i$th element that contributes to ${\mathsf{ZZ}}(P_{\Xm,\Ym},\mathsf{M})$ in~\eqref{eq:BZZ_Mary_vec_mmse_sub2}.
This term was derived by bounding only the probability of error in $\hsf_\Msf(t,\ev_i,P_{\Xm,\Ym})$ (see~\eqref{eq:Before_valley} in Appendix~\ref{app:gen_zzb}) as
\begin{equation}
\label{eq:ProbIneq}
   P_e^{\phi} \left ({\xv} ,t; \mathcal{P}_\mathcal{U}(\xv), \mathcal{U} \right )  
 \ge  P_e \left ({\xv}; \mathcal{P}_\mathcal{U}(\xv), \mathcal{U} \right ),
\end{equation}
where $P_e^{\phi} \left ({\xv},t; \mathcal{P}_\mathcal{U}(\xv), \mathcal{U} \right )$ is the error probability incurred by a possibly sub-optimal decision rule $\phi(\Ym)$ such that
\begin{equation}\label{eq:phi_sub_opt}
	\phi(\Ym) = \Hc_j, \text{ where } j \in \!\! \argmin_{k\in[0:\mathsf{M}-1]} \! |\EE[X_i |\Ym] - x_i - kt|,
\end{equation}
and $P_e\left ({\xv}; \mathcal{P}_\mathcal{U}(\xv), \mathcal{U} \right )$ is the minimum error probability, i.e.,  incurred by the optimal MAP decision rule $\phi^\star(\Ym)$, that is,
\begin{equation}\label{eq:phi_opt}
	\phi^\star(\Ym)
	 = 
	\Hc_j,  \text{ where } j \in \argmax_{k\in[0:\Msf-1]} f_{\Xm|\Ym}(\xv+\uv_k | \Ym ).
\end{equation}
In other words,  $P_e^{\phi} \left ({\xv},t; \mathcal{P}_\mathcal{U}(\xv), \mathcal{U} \right )$ leads to ${\mathsf{ZZ}}(P_{\Xm,\Ym},\mathsf{M})$ and $P_e \left ({\xv}; \mathcal{P}_\mathcal{U}(\xv), \mathcal{U} \right )$ leads to ${\rm mmse}(\Xm|\Ym)$.
Thus, the equality in~\eqref{eq:mmse=ZZB} holds if and only if~\eqref{eq:ProbIneq} holds with equality for all $i \in [1:d]$ and hence, if and only if $\phi=\phi^\star$ for all $i \in [1:d]$.

Now, since $\hsf_\Msf(t,\ev_i,P_{\Xm,\Ym})$ is maximized over $\Uc = \{\uv_k\}_{k=0}^{\mathsf{M}-1}\subset \Xc$, for all $t>0$, we need at least one set $\Uc$ (recall that $t$ is a constraint for $\Uc$) such that $\phi(\yv)=\phi^\star(\yv)$ for all $\yv\in\Yc$. 
It therefore follows that a sufficient and necessary condition for the tightness of the $i$th element that contributes to ${\mathsf{ZZ}}(P_{\Xm,\Ym},\mathsf{M})$ is given by that for all $t>0$, there exist $\Uc_i = \{\uv_{i,k}\}_{k=0}^{\mathsf{M}-1}$ such that $(\uv_{i,k} )_i = kt,~k\in[0:\mathsf{M}-1]$ and for all $ (\xv,\yv)\in\Xc\times \Yc$,
\begin{align}
	  & \argmax_{k\in[0:\mathsf{M}-1]} f_{\Xm|\Ym}(\xv+\uv_{i,k} | \Ym = \yv)  \nonumber \\
	  & \bigcap  \argmin_{k\in[0:\mathsf{M}-1]} |\EE[X_i|\Ym = \yv] - x_i - kt | \neq \varnothing.
\end{align}
Hence,~\eqref{eq:mmse=ZZB} holds if and only if there exist such $\Uc_i$ for all $i \in[1:d]$. This concludes the proof of Proposition~\ref{prop:mmse=ZZB}.

\section{Proof of Corollary~\ref{cor:unimodal_symmetric}}
\label{app:CorIff}
Proposition~\ref{prop:mmse=ZZB} gives the sufficient and necessary condition that, for all $x,\yv$ and $t\!>\!0$, there exist $\hat{k}_1 \!=\! \hat{k}_2$ such that
\begin{subequations}
\label{eq:IffUni}
\begin{align}
	& \hat{k}_1 \in \argmax_{k\in[0:\mathsf{M}-1]}  f_{X|\Ym}(x + kt |  \yv), \text{ and} \label{eq:snc_p2_1}  \\
	& \hat{k}_2 \in \argmin_{k\in[0:\mathsf{M}-1]}  |\EE[X | \Ym = \yv] - x - kt |.  \label{eq:snc_p2_2} 
\end{align}
\end{subequations}
We first assume that $f_{X|\Ym}$ has at least two modes denoted as $m_1$ and $m_2$.
Then,  there exists at least one mode, say $m_1$, that is different from $\EE[X | \Ym = \yv]$. 
Now,  let $x= m_1 - \varepsilon$ and $t = \frac{2\varepsilon}{\Msf-1}$ where $\varepsilon$ is such that $f_{X|\Ym}$
is non-decreasing in $[m_1-\varepsilon,m_1)$ and non-increasing in  $(m_1,m_1-\varepsilon]$, and
$0 < \varepsilon < |\EE[X|\Ym=\yv] - m_1|$.  This choice of $x$ and $t$ implies that $\hat{k}_1 \notin \{0, \Msf-1\}$ and $\hat{k}_2 \in \{0, \mathsf{M}-1\}$. Thus, $\hat{k}_1$ and $\hat{k}_2$ are always different, which implies that the multimodal assumption does not satisfy the condition in Proposition~\ref{prop:mmse=ZZB}. Hence, the PDF $f_{X|\Ym}$ has to be unimodal. Moreover, by using a similar argument as above, it is not difficult to show that for a unimodal PDF $f_{X|\Ym}$ with mode $m$, we need $m = \EE[X | \Ym = \yv]$ to satisfy the condition in Proposition~\ref{prop:mmse=ZZB}. 

Now, let $f_{X|\Ym}$ be unimodal and asymmetric with respect to its mode $m$.
For such a PDF, we can find $\varepsilon_0>0$ and $\varepsilon_1>0$ such that $\varepsilon_0 \neq \varepsilon_1$ and $f_{X|\Ym}(m-\varepsilon_0|\yv) = f_{X|\Ym}(m+\varepsilon_1|\yv)$. Consider the following choice for $x$ and $t>0$,
\begin{equation}
x  = m-\varepsilon_0+(-1)^{\ell}\delta~ \text{  and  }~
t = \varepsilon_0 + \varepsilon_1,
\end{equation}
where $\ell = \argmax_{i \in \{0,1\}} \varepsilon_i$ and $\delta>0$ is a small enough number. With this choice, we obtain that $\hat{k}_1 = \ell$ and $\hat{k}_2 = 1-\ell$ and hence, $\hat{k}_1\neq \hat{k}_2$, which implies that the asymmetric assumption does not satisfy the condition in Proposition~\ref{prop:mmse=ZZB}.  Hence, the PDF $f_{X|\Ym}$ has to be symmetric.

In summary, the above analysis shows that unimodality and symmetry are necessary conditions. The proof of Corollary~\ref{cor:unimodal_symmetric} is concluded by noting that~\cite{bell1995performance} showed that these are sufficient.

\section{Proof of Proposition~\ref{prop:SZZB_condtion}}\label{app:SZZB_condtion}
With reference to Appendix~\ref{app:GSZZB}, the SZZB is obtained by multiple steps of bounding several terms. Specifically, the steps are: 
\begin{enumerate}
    \item Restricting the integration area from $t\in[0,\infty)$ to $t\in[0,2\Delta]$ in~\eqref{eq:sp_zzb_proof1};
    \item $t\geq \min\{t,2\Delta-t\}$ and $2\Delta-t\geq \min\{t,2\Delta-t\}$ in $\rm(b)$ in~\eqref{eq:sp_zzb_proof4};
    \item $ P_e^{\phi_{\rm sp}}(\xv,t, \Delta; \Pc_\Uc(\xv),\Uc) \geq P_e(\xv; \Pc_\Uc(\xv),\Uc)$ in~\eqref{eq:sp_zzb_proof5}.
\end{enumerate}
The above steps of bounding lead to the following bound,
\begin{align}\label{eq:szzb_tightness_proof2}
    & \sum_{i=1}^d \int_0^\infty \frac{t}{2} \Pr\left(|\EE[X_i|\Ym] - X_i| \geq \frac{t}{2}\right) {\rm d} t \nonumber \\
    & = {\rm mmse}(\Xm|\Ym) \nonumber \\
    & \geq \Zsf\Zsf_{\rm sp}(P_{\Xm,\Ym},\Msf) \nonumber \\
    & = \sum_{i=1}^d \int_0^{\Delta_i^\star} \frac{t}{2} \frac{2\hsf_\Msf(\Delta_i^\star,\ev_i,P_{\Xm,\Ym})}{\Msf-1} {\rm d} t,
\end{align} 
where in the last equality we denote by $\Delta_i^\star$ the solution for $\sup_{\Delta>0} \frac{\Delta^2\hsf_\Msf(\Delta,\ev_i,P_{\Xm,\Ym})}{2(\Msf-1)}$ for each $i\in[1:d]$. We note that such a $\Delta_i^\star$ always exists in $(0,\infty)$ due to the following facts:
\begin{itemize}
    \item From Lemma~\ref{lem:proof_h_rate} in Appendix~\ref{app:lem:proof_h_rate}, $\lim_{\Delta\to\infty} \frac{\Delta^2}{2(\Msf-1)} \hsf_\Msf(\Delta,\ev_i,P_{\Xm,\Ym}) = 0$, which shows that $\Delta\to\infty$ gives only the trivial bound;
    \item Due to the fact that $\hsf_\Msf(\Delta,\av,P_{\Xm,\Ym}) \leq \Msf$ (see~\eqref{eq:alt_h}), $\lim_{\Delta\to0} \frac{\Delta^2}{2(\Msf-1)} \hsf_\Msf(\Delta,\ev_i,P_{\Xm,\Ym}) = 0$, which shows that $\Delta\to0$ gives only the trivial bound.
\end{itemize}
  We now establish necessary conditions for the equality in~\eqref{eq:szzb_tightness_proof2} to hold. Let $\Delta_i^\star,~i\in[1:d],$ yield a SZZB tight to the MMSE. Then, from step 1), it must follow that for all $i\in[1:d]$,
\begin{align}
    & \int_0^\infty \frac{t}{2}\Pr\left(|\EE[X_i|\Ym] - X_i| \geq \frac{t}{2}\right) {\rm d}t \nonumber \\
    &  = \int_0^{2\Delta_i^\star} \frac{t}{2}\Pr\left(|\EE[X_i|\Ym] - X_i| \geq \frac{t}{2}\right) {\rm d}t,
\end{align}
which implies that
\begin{equation}\label{eq:szzb_tightness_proof2.5}
    \Pr\left(|\EE[X_i|\Ym] - X_i| \geq \Delta_i^\star\right)
     = 0,
\end{equation}
is a necessary condition. 

Step 2) indicates that a SZZB tight to the MMSE must have equality for the inequality $\rm (b)$ in~\eqref{eq:sp_zzb_proof4}. Note that, for all $i \in [1:d]$, we can write the equality in $\rm (a)$ in~\eqref{eq:sp_zzb_proof4} by~\eqref{eq:szzb_tightness_proof2_a}, at the top of the next page,
\begin{figure*}
\begin{align}\label{eq:szzb_tightness_proof2_a}
    & \sum_{j=1}^{\mathsf{M}-1}  \left\{ \int_0^{2\Delta_i^\star} \frac{t}{2} \int p_{j-1}  \Pr\left(  \left< \av,\phi(\Ym)- \xv \right>  \geq (j-1)\Delta_i^\star + \frac{t}{2} \; \middle|\; \Xm = \xv + \uv_{j-1} \right)  \mu_\Uc({\rm d} \xv)   \ {\rm d}t \right. \nonumber \\
	& \quad + \left. \int_0^{2\Delta_i^\star} \frac{2\Delta_i^\star-t}{2} \int p_j \Pr\left(  \left< \av,\phi(\Ym)- \xv \right> \leq (j-1)\Delta_i^\star + \frac{t}{2} \; \middle|\; \Xm = \xv +\uv_j \right) \mu_\Uc({\rm d} \xv) \ {\rm d}t \right\} \nonumber \\
    & \overset{\rm (\alpha)}{=} \int_0^{2\Delta_i^\star} \frac{t}{2} \sum_{j=1}^{\mathsf{M}-1}\left\{ \Pr\left(  \left< \av,\phi(\Ym)- \Xm \right>  \geq \frac{t}{2}\right)   \right\} {\rm d}t + \int_0^{2\Delta_i^\star} \frac{2\Delta_i^\star-t}{2} \sum_{j=1}^{\mathsf{M}-1}\left\{ \Pr\left(  \left< \av,\phi(\Ym)- \Xm \right> \leq -\Delta_i^\star + \frac{t}{2} \right) \right\}{\rm d}t  \nonumber \\
    & \overset{\rm (\beta)}{=} \int_0^{2\Delta_i^\star} \frac{t}{2} (\Msf-1) \Pr\left( \EE[X_i|\Ym] - X_i \geq \frac{t}{2}\right) {\rm d}t + \int_0^{2\Delta_i^\star} \frac{2\Delta_i^\star-t}{2} (\Msf-1) \Pr\left( \EE[X_i|\Ym] - X_i \leq -\Delta_i^\star + \frac{t}{2} \right) {\rm d}t ,
\end{align}
\hrule
\end{figure*}
where the equality in $\rm (\alpha)$ follows from the facts that for $j\in[0:\Msf-1]$, $p_j = \frac{ P_{\Xm - \uv_j}({\rm d} \xv)}{ \mu_\Uc ({\rm d} \xv)}$,
and $\left<\av,\uv_j\right> = j\Delta_i^\star$, and the equality in $\rm (\beta)$ follows by using $\av=\ev_i$ and $\phi(\Ym) = \EE[\Xm|\Ym]$ for the MMSE.
Similarly, the expression in $\rm (b)$ in~\eqref{eq:sp_zzb_proof4} is given by~\eqref{eq:szzb_tightness_proof2_b}, at the top of the next page.
\begin{figure*}
\begin{align}\label{eq:szzb_tightness_proof2_b}
    & \int_0^{2\Delta_i^\star} \frac{\min\{t,2\Delta_i^\star-t\}}{2} (\Msf-1) \Pr\left( \EE[X_i|\Ym] - X_i  \geq \frac{t}{2}\right) {\rm d}t \nonumber \\
	& \quad + \int_0^{2\Delta_i^\star} \frac{\min\{t,2\Delta_i^\star-t\}}{2} (\Msf-1) \Pr\left( \EE[X_i|\Ym] - X_i \leq -\Delta_i^\star + \frac{t}{2} \right) {\rm d}t .
\end{align}
\hrule
\end{figure*}
Step 2) holds if and only if~\eqref{eq:szzb_tightness_proof2_a} is equal to~\eqref{eq:szzb_tightness_proof2_b}. Equating~\eqref{eq:szzb_tightness_proof2_a} to~\eqref{eq:szzb_tightness_proof2_b} and evaluating $\min\{t,2\Delta_i^\star -t\}$ in~\eqref{eq:szzb_tightness_proof2_b}, we obtain the following equivalent condition,
\begin{align}\label{eq:szzb_tightness_proof2_c}
   &\! \int_{\Delta_i^\star}^{2\Delta_i^\star} (t-\Delta_i^\star) \Pr \!\left( \EE[X_i|\Ym] - X_i  \geq \frac{t}{2}\right) \!{\rm d}t  \nonumber \\
    &\! = \int_0^{\Delta_i^\star} \! (t-\Delta_i^\star) \Pr \!\left( \EE[X_i|\Ym] - X_i \leq -\Delta_i^\star + \frac{t}{2} \right)\! {\rm d} t.
\end{align}
Since~\eqref{eq:szzb_tightness_proof2_c} has a non-negative term in the left-hand side and a non-positive term in the right-hand side, both these terms have to be equal to zero for~\eqref{eq:szzb_tightness_proof2_c} to hold. 
Equating the left-hand side of~\eqref{eq:szzb_tightness_proof2_c} to zero, we obtain
\begin{align}\label{eq:szzb_tightness_proof5}
    & \int_{\Delta_i^\star}^{2\Delta_i^\star} t \Pr\left(  \EE[X_i|\Ym] - X_i \geq \frac{t}{2} \right) {\rm d}t \nonumber \\
    & = \Delta_i^\star \int_{\Delta_i^\star}^{2\Delta_i^\star} \Pr\left(  \EE[X_i|\Ym] - X_i \geq  \frac{t}{2} \right)  {\rm d} t,
\end{align}
which implies that $\Pr\left(  \EE[X_i|\Ym] - X_i \geq  \frac{t}{2} \right ) = 0$, for all $t\in(\Delta_i^\star, 2\Delta_i^\star]$.
Similarly, equating the right-hand side of~\eqref{eq:szzb_tightness_proof2_c} to zero and using the change of variable $t=2\Delta_i^\star-v$, we arrive~at 
\begin{align}\label{eq:szzb_tightness_proof6}
     & \int_{\Delta_i^\star}^{2\Delta_i^\star} v \Pr\left(  \EE[X_i|\Ym] - X_i \leq- \frac{v}{2} \right)  {\rm d} v \nonumber \\
     & = \Delta_i^\star \int_{\Delta_i^\star}^{2\Delta_i^\star} \Pr\left(  \EE[X_i|\Ym] - X_i \leq - \frac{v}{2} \right)  {\rm d} v ,
\end{align}
which implies that $\Pr\left(  \EE[X_i|\Ym] - X_i \leq - \frac{t}{2} \right)=0$, for all $t\in(\Delta_i^\star, 2\Delta_i^\star]$.
Hence,~\eqref{eq:szzb_tightness_proof5} and~\eqref{eq:szzb_tightness_proof6} are the necessary conditions for the SZZB to be tight to the MMSE. Equivalently, we have that
\begin{equation}\label{eq:szzb_tightness_proof7}
    \Pr\left( |\EE[X_i|\Ym] - X_i | \geq  \frac{t}{2} \right) = 0,~\forall t\in(\Delta_i^\star, 2\Delta_i^\star].
\end{equation}
Combining the two necessary conditions in~\eqref{eq:szzb_tightness_proof2.5} and in~\eqref{eq:szzb_tightness_proof7}, we can conclude that to have a SZZB tight to the MMSE the following must hold,
\begin{equation}\label{eq:szzb_tightness_proof8}
    \Pr\left( |\EE[X_i|\Ym] - X_i | \geq  \frac{t}{2} \right) = 0,~\forall t > \Delta_i^\star.
\end{equation}
From step 3), together with~\eqref{eq:sp_zzb_proof4},~\eqref{eq:sp_zzb_proof5}, and~\eqref{eq:sp_zzb_proof6}, it must hold~that
\begin{align}
    & \int_0^{2\Delta_i^\star} \frac{\min\{t, 2\Delta_i^\star - t\}}{2(\Msf-1) } \int  P_e^{\phi_{\rm sp}} (\xv, t, \Delta_i^\star ; \Pc_\Uc(\xv), \Uc ) \mu_\Uc({\rm d}\xv)  {\rm d} t \nonumber\\
    & = \int_0^{2\Delta_i^\star} \frac{\min\{t, 2\Delta_i^\star - t\}}{2(\Msf-1) } \hsf_{\Msf}(\Delta_i^\star,\ev_i,P_{\Xm,\Ym}) \ {\rm d} t.
\end{align}
Note that $\int  P_e^{\phi_{\rm sp}} \left(\xv, t, \Delta_i^\star ; \Pc_\Uc(\xv), \Uc \right) \mu_\Uc({\rm d}\xv)$ is a function of $t$ (see~\eqref{eq:sp_zzb_decision_rule}), whereas $\hsf_{\Msf}(\Delta_i^\star,\ev_i,P_{\Xm,\Ym})$ is independent of $t$. 
Thus, for some constant~$c>0$, we must have that
\begin{equation}
\label{eq:CondStep3Cost}
    \int  P_e^{\phi_{\rm sp}} \left(\xv,t, \Delta_i^\star ; \Pc_\Uc(\xv), \Uc \right) \mu_\Uc({\rm d}\xv)
     = c,~\forall t\in(0,2\Delta_i^\star).
\end{equation}
Now, recall from~\eqref{eq:sp_zzb_proof4} and~\eqref{eq:szzb_tightness_proof2_b} that
\begin{align}
    & \int_0^{2\Delta_i^\star} \frac{\min\{t, 2\Delta_i^\star - t\}}{2(\Msf-1) }\! \int \!  P_e^{\phi_{\rm sp}} (\xv, t, \Delta_i^\star ; \Pc_\Uc(\xv), \Uc ) \mu_\Uc({\rm d}\xv) {\rm d} t \nonumber \\
    & = \int_0^{2\Delta_i^\star} \frac{\min\{t, 2\Delta_i^\star - t\}}{2 } \left(\Pr \!\left( \EE[X_i|\Ym] - X_i \geq \frac{t}{2} \right) \right. \nonumber \\ 
    & \qquad \quad \left. + \Pr \!\left(\EE[X_i|\Ym] - X_i \leq -\Delta_i^\star + \frac{t}{2} \right) \right) {\rm d}t,
\end{align}
which, together with~\eqref{eq:CondStep3Cost}, implies that for all $ t\in(0,2\Delta_i^\star)$,
\begin{align}\label{eq:szzb_tightness_proof9}
    & \Pr \!\left( \EE[X_i|\Ym] \!-\! X_i \geq \frac{t}{2} \right) \!+ \Pr \!\left(\EE[X_i|\Ym] \! - \! X_i \leq -\Delta_i^\star + \frac{t}{2} \right) \nonumber \\
    & = c.
\end{align}
Together with the necessary condition in~\eqref{eq:szzb_tightness_proof8}, partitioning $(0,2\Delta_i^\star)$ into $(0,\Delta_i^\star)$, $\{\Delta_i^\star\}$ and $(\Delta_i^\star,2\Delta_i^\star)$
indicates that~\eqref{eq:szzb_tightness_proof9} holds if and only if
\begin{align}
    & \Pr\left( \EE[X_i|\Ym] - X_i \geq \frac{t}{2} \right) = c,~\forall t\in(0,\Delta_i^\star), \text{ and} \\
    &\Pr\left( |\EE[X_i|\Ym] - X_i| \geq \frac{t}{2} \right) = c,~\text{ for }t=\Delta_i^\star,\text{ and}
    \\
    &\Pr\left(\EE[X_i|\Ym] - X_i \leq -\Delta_i^\star + \frac{t}{2} \right) = c,~\forall t\in(\Delta_i^\star,2\Delta_i^\star).
\end{align}
Thus, the above implies that the following is also a necessary condition,
\begin{equation}\label{eq:szzb_tightness_proof10}
    \Pr\left( |\EE[X_i|\Ym] - X_i| \geq \frac{t}{2} \right) = c,~\forall t\in(0,\Delta_i^\star].
\end{equation}
Note that the necessary condition in~\eqref{eq:szzb_tightness_proof8} and the condition in~\eqref{eq:szzb_tightness_proof10} imply that the distribution of $|\EE[X_i|\Ym] - X_i|$ must be discrete with at most two mass points. 
This concludes the proof of Proposition~\ref{prop:SZZB_condtion}.

\section{Ancillary Lemmas}
\label{app:Lemmas}

\subsection{Lemma~\ref{lem:Pe_different_form}}
\label{app:lem:Pe_different_form}

\begin{lemma}\label{lem:Pe_different_form}
The minimum probability of error for the $\mathsf{M}$-ary hypothesis testing problem in Definition~\ref{def:MHT} can be written as
\begin{equation}
	P_e \left ({\xv}; \mathcal{P}_\mathcal{U}(\xv), \mathcal{U}  \right )
	 = 1 - \int  \max_{i\in[0:\mathsf{M}-1]} \left \{ p_i  P_{\Ym|\Xm}( {\rm d} \yv | \xv+\uv_i ) \right \}.
\end{equation}

\end{lemma}

\begin{IEEEproof}
The minimum error probability is attained when the MAP decision rule is used~\cite{Kay1998}. Then, for an $\mathsf{M}$-ary hypothesis testing problem as in Definition~\ref{def:MHT}, the minimum error probability is given by~\cite{sason2017arimoto},
\begin{equation}
	P_e \left ({\xv}; \mathcal{P}_\mathcal{U}(\xv), \mathcal{U}  \right )
	= 1 - \int   \max_{i\in[0:\mathsf{M}-1]} \Pr(\Hc_i|\Ym=\yv)  P_\Ym({\rm d}\yv) .
\end{equation}
Using the Bayes' rule, we obtain
\begin{align}
	& P_e \left ({\xv}; \mathcal{P}_\mathcal{U}(\xv), \mathcal{U}  \right ) \nonumber \\
	& = 1 - \int   \max_{i\in[0:\mathsf{M}-1]} \Pr(\Hc_i|\Ym=\yv) P_\Ym({\rm d}\yv)  \nonumber \\
	& = 1 - \int  \max_{i\in[0:\mathsf{M}-1]} \left \{\Pr(\Hc_i) P_{\Ym|\Hc_i}({\rm d}\yv | \Hc_i )  \right \} \nonumber \\
	& = 1 - \int  \max_{i\in[0:\mathsf{M}-1]} \left \{p_i  P_{\Ym|\Xm}( {\rm d} \yv | \xv+\uv_i ) \right \},
\end{align}
which concludes the proof of Lemma~\ref{lem:Pe_different_form}.
\end{IEEEproof}

\subsection{Lemma~\ref{lem:continuous_gaussian_h}}\label{app:continuous_gaussian_h}

\begin{lemma}\label{lem:continuous_gaussian_h}
Let $\Ym = \Xm + \Nm$ with $\Nm\sim\Nc(\zerov_d,\eta I_d)$ and let $\Xm$ be continuous. Let $f_\Zm$ be the PDF of $\Zm\sim\Nc(\zerov_d,I_d)$. Then, for any $\eta>0$, it holds that
\begin{align}
	& \mathsf{h}_2(t,\ev_i,P_{\Xm,\Ym})   \nonumber \\
	& =\!\!\!\!\!\!  \sup\limits_{\substack{  \mathcal{U}  \subset \mathcal{X}: \\ \mathcal{U}= \{ \vv_k \}_{k=0}^{1}  \\  ( \vv_k)_i= \frac{kt}{\sqrt{\eta}},~\forall k }}  \, \!\!\! \!\int \!  \left(  \int_{\yv  \in \Rc_0}   \! \!f_\Zm(\yv \!-\! \vv_1 ) f_\Xm(\xv\!+\!\sqrt{\eta} \vv_1)  {\rm  d} \yv  \right. \nonumber \\
	&\qquad\  \left. +   \int_{\yv  \in \Rc_1}  f_\Zm(\yv - \vv_0 ) f_\Xm(\xv+\sqrt{\eta} \vv_0)  {\rm  d} \yv  \,   \right) {\rm d}\xv, 
\end{align}
where 
\begin{align*}
	\Rc_0  \!=\! \left\{ \yv \!:  \! \ln \frac{f_\Xm(\xv \!+\! \sqrt{\eta}\vv_0)}{f_\Xm(\xv \!+\! \sqrt{\eta}\vv_1)} \!+\!  \frac{\|\vv_1\|_2^2 \!-\! \| \vv_0 \|_2^2}{2} \!>\!  (\vv_1 \!-\! \vv_0)^\mathsf{T}\yv \right\}, \nonumber \\
	\Rc_1  \!=\! \left\{ \yv \!: \!  \ln \frac{f_\Xm(\xv \!+\! \sqrt{\eta}\vv_0)}{f_\Xm(\xv \!+\! \sqrt{\eta}\vv_1)} \!+\!  \frac{\|\vv_1\|_2^2 \!-\! \| \vv_0 \|_2^2}{2} \!<\!  (\vv_1 \!-\! \vv_0)^\mathsf{T}\yv \right\}.
\end{align*}
\end{lemma}

\begin{IEEEproof}
The error probability for the binary hypothesis testing problem is given by
\begin{align}\label{eq:Pe_Gauss_BHT_expression}
	\!\! P_e \left ( \eta, {\xv}; \mathcal{P}_\mathcal{U}(\xv), \mathcal{U}  \right ) 
	& = \EE\left[ \min_{i\in\{0,1\}} \Pr(\Hc_i|\Ym) \right] \nonumber \\
	& \overset{\rm (a)}{=} \EE\left[ \min_{i\in\{0,1\}} \frac{f_{\Ym|\Xm}(\Ym|\xv+\uv_i) }{f_\Ym(\Ym)} \frac{q_i}{q_0+q_1 } \right] \nonumber \\
	& = \int  \min_{i\in\{0,1\}} \! f_\Nm(\yv - \xv - \uv_i)   \frac{q_i }{q_0 + q_1} {\rm  d} \yv \nonumber \\
	& \overset{\rm (b)}{=} \! \int \! \min_{i\in\{0,1\}} \!f_\Nm(\zv - \uv_i)   \frac{q_i }{q_0 \!+\! q_1} {\rm  d} \zv ,
\end{align}
where $\rm (a)$ follows from the Bayes' rule and letting $f_\Xm(\xv+\uv_i) = q_i,~i\in\{0,1\}$, which leads to
\begin{equation}
\Pr(\Hc_i) = \frac{f_\Xm(\xv+\uv_i)}{f_\Xm(\xv+\uv_0)+f_\Xm(\xv+\uv_1)} = \frac{q_i}{q_0+q_1};
\end{equation}
and $\rm (b)$ follows from the change of variable $\zv = \yv - \xv$.
It is not difficult to see that~\eqref{eq:Pe_Gauss_BHT_expression} can be written as
\begin{align}\label{eq:P_e_expression_lemma_proof}
	& P_e \left (\eta,{\xv}; \mathcal{P}_\mathcal{U}(\xv), \mathcal{U}  \right ) \nonumber \\
	& = \frac{1}{q_0\!+\!q_1}  \int_{\zv  \in \Rc_0^\prime}   f_\Zm\left(\zv - \frac{\uv_1}{\sqrt{\eta}} \!\right) q_1  \ {\rm  d} \zv  \nonumber \\
	& \quad +  \frac{1}{q_0\!+\!q_1}  \int_{\zv  \in \Rc_1^\prime}  f_\Zm\left(\zv - \frac{\uv_0}{\sqrt{\eta}} \!\right) q_0 \ {\rm  d} \zv,
\end{align}
where $\Zm\sim\Nc(\zerov_d, I_d)$,
\begin{align*}
	& \Rc_0^\prime  = \left\{ \zv \in {\RR^d} :  \ln\frac{q_0}{q_1}  +  \frac{\|\uv_1\|_2^2 - \| \uv_0 \|_2^2}{2\eta} > \frac{(\uv_1-\uv_0)^\mathsf{T}\zv}{\sqrt{\eta}} \right\}, \\ & \text{and} \nonumber \\
	& \Rc_1^\prime = \left\{ \zv \in {\RR^d}  :  \ln\frac{q_0}{q_1}  +  \frac{\|\uv_1\|_2^2 - \| \uv_0 \|_2^2}{2\eta} < \frac{(\uv_1-\uv_0)^\mathsf{T}\zv}{\sqrt{\eta}} \right\}.
\end{align*}
The proof of Lemma~\ref{lem:continuous_gaussian_h} is concluded by substituting~\eqref{eq:P_e_expression_lemma_proof} inside $ \mathsf{h}_2(t,\ev_i,P_{\Xm,\Ym})$ in~\eqref{eq:BZZ_Mary_vec_mmse_h} with the change of variable $\uv_i = \sqrt{\eta} \vv_i$ and noticing that $\mu_\Uc(\xv) = q_0+q_1$.
\end{IEEEproof}

\subsection{Lemma~\ref{lem:discrete_zero_lemma}}\label{app:discrete_zero_lemma}
\begin{lemma}\label{lem:discrete_zero_lemma}
Assume that $\Sc_\Xm$ is countable. Then, it holds that
\begin{align}
	\Bc = \Biggl\{t\geq0 : &
	\bigcup_{\substack{\Ic\subseteq[0:\mathsf{M}-1]: \\ |\Ic| \geq 2}}  \bigcap_{k\in\Ic} \{ \Sc_\Xm - \uv_k\} \neq \varnothing, \uv_k\in\Xc,  \nonumber \\
	& \, (\uv_k)_i = kt, ~k\in[0:\mathsf{M}-1] \Biggr \}
\end{align}
is countable.
\end{lemma}

\begin{IEEEproof}
Let $(a,b) \in [0:\mathsf{M}-1]^2$, $a\neq b$ and define
\begin{align}\label{eq:discrete_zero_lemma_proof1}\Bc_{a,b} 
	 = \{  t\geq 0 : &  \{\Sc_\Xm - \uv_a\} \cap  \{\Sc_\Xm - \uv_b\} \neq \varnothing, \uv_a\in\Xc,  \nonumber \\
	 & \, \uv_b\in\Xc, ( \uv_a)_i  = at, ( \uv_b )_i = bt  \}.
\end{align}
Now, observe that
\begin{equation}
	\Bc  
	 = \bigcup_{\substack{(a,b)\in[0:\mathsf{M}-1]^2 \\ a\neq b}} \Bc_{a,b}.
	\label{eq:discrete_zero_lemma_proof2}
\end{equation}
Since $\Sc_\Xm$ is countable by assumption, then without loss of generality we can assume that $\Sc_\Xm = \{\xv_w : w\in\NN\}$.\footnote{When $\Sc_\Xm$ is finite, the proof follows by replacing $\NN$ with $[1:|\Sc_\Xm|]$.} 
Now note that, if $t\in\Bc_{a,b}$, then we have that $\xv_w - \uv_a = \xv_z - \uv_b$ for some $(w,z)\in \NN^2$. 
Thus, $\Bc_{a,b}$ in~\eqref{eq:discrete_zero_lemma_proof1} can be written as
\begin{align}\label{eq:discrete_zero_lemma_proof3}
	\Bc_{a,b} 
	& = \{  t\geq 0 : \exists (w,z)\in\NN^2, \xv_w - \xv_z = \uv_a - \uv_b,\uv_a\in\Xc,  \nonumber \\
	& \qquad\qquad\quad \uv_b\in\Xc, (\uv_a )_i= at, (\uv_b )_i = bt  \},
\end{align}
and
\begin{align}
	\Bc_{a,b} 
	& \subseteq \{  t\geq 0 : \exists (w,z)\in\NN^2, \xv_w - \xv_z = \uv_a - \uv_b, \uv_a\in\Xc, \notag \\
	& \qquad\qquad\quad \uv_b\in\Xc, ( \uv_a)_i - ( \uv_b )_i = (a-b)t \} \notag \\
	& \subseteq \left\{  t\geq 0 : \exists (w,z)\in\NN^2, (\xv_w - \xv_z)_i = (a-b)t  \right\} \nonumber \\
	& := \overline{\Bc_{a,b} }.
	\label{eq:discrete_zero_lemma_proof4}
\end{align}
Now, note that
\begin{equation}
\overline{\Bc_{a,b} } \subseteq \left\{ \frac{ (\xv_w - \xv_z )_i}{a-b} : (w,z)\in\NN^2 \right\},
\end{equation}
which implies that 
$\overline{\Bc_{a,b} }$ is countable and hence, from~\eqref{eq:discrete_zero_lemma_proof4} we have that ${\Bc_{a,b} }$ is countable.
Since the union of countably many countable sets is still countable~\cite{folland1999real}, $\mathcal{B}$ in~\eqref{eq:discrete_zero_lemma_proof2} is countable, which concludes the proof of Lemma~\ref{lem:discrete_zero_lemma}.
\end{IEEEproof}

\subsection{Lemma~\ref{lem:proof_h_rate}}
\label{app:lem:proof_h_rate}
\begin{lemma}\label{lem:proof_h_rate}
Assume that ${\rm mmse}(\Xm|\Ym) <\infty$. Then,
\begin{equation}
    \lim_{t\to\infty}t^2\hsf_\Msf(t,\av,P_{\Xm,\Ym})
     = 0.
\end{equation}
\end{lemma}
\begin{IEEEproof}
Consider the MMSE and the ZZB without the valley-filling function,
\begin{align}\label{eq:proof_h_rate}
    \infty 
    & > {\rm mmse}(\Xm|\Ym) \nonumber \\
    & = \sum_{i=1}^d \int_0^\infty \frac{t}{2} \Pr\left(|\EE[X_i|\Ym] - X_i| \geq \frac{t}{2}\right) {\rm d} t \nonumber\\
    & \geq \Zsf\Zsf(P_{\Xm,\Ym},\Msf) \nonumber\\
    & = \sum_{i=1}^{d}  \int_0^\infty \frac{t}{2}  \frac{{\mathsf{h}_\Msf }(t,\ev_i,P_{\Xm,\Ym})}{\mathsf{M}-1}  {\rm d} t .
\end{align}
Now, assume that $\lim_{t\to\infty}t^2 \hsf_\Msf(t,\av,P_{\Xm,\Ym}) = c > 0$. Then, for some large enough $t$, the integrand has a tail $ \frac{c}{2(\mathsf{M}-1)t}$, which is not integrable in $t$ over $(a,\infty)$ for any $a>0$. This contradicts the assumption, and since $\hsf_\Msf(t,\av,P_{\Xm,\Ym})\geq0$, we obtain
\begin{align}
    \lim_{t\to\infty}t^2 \hsf_\Msf(t,\av,P_{\Xm,\Ym})
    & = 0.
\end{align}
This concludes the proof of Lemma~\ref{lem:proof_h_rate}.
\end{IEEEproof}

\section{Proof of Examples}\label{app:proof_examples}

\subsection{Proof of Example~\ref{ex:szzb_low_noise_gaussian}}\label{app:szzb_low_noise_gaussian}
We start by noting that the PDFs of $X$ and $Y|X=x$ are given by
\begin{equation}
    f_X(x) 
    = \frac{1}{\sqrt{2\pi} }{\rm{e}}^{-\frac{x^2}{2}} \ \text{ and }  \
    f_{Y|X}(y|x)
    = \frac{1}{\sqrt{2\pi\eta}} {\rm{e}}^{-\frac{(y-x)^2}{2\eta}}.
\end{equation}
For the $1$-dimensional case, we have $\Uc = \{0,t,2t,\cdots,(\Msf-1)t\}$ for $\hsf_\Msf (t, \ev_1, P_{X,Y})$ in~\eqref{eq:BZZ_Mary_vec_mmse_h}. Then, we can write $\hsf_\Msf (t, \ev_1, P_{X,Y})$ in~\eqref{eq:BZZ_Mary_vec_mmse_h} as~\eqref{eq:low_noise_szzb_Gauss_ex_proof1}, at the top of the next page,
\begin{figure*}
\begin{align}\label{eq:low_noise_szzb_Gauss_ex_proof1}
    \hsf_\Msf (t, \ev_1, P_{X,Y}) 
    & = \int_{-\infty}^\infty P_e(x;\Pc_\Uc(x),\Uc) \mu_\Uc({\rm d}x) \nonumber \\
    & \overset{\rm (a)}{=} \int_{-\infty}^\infty \EE\left[1 - \max_{k\in[0:\Msf-1]} \Pr(\Hc_k|Y)\right] \left(\sum_{j=0}^{\Msf-1}f_X(x+jt)\right){\rm d}x  \nonumber \\
    & \overset{\rm (b)}{=} \int_{-\infty}^\infty \EE\left[1 - \max_{k\in[0:\Msf-1]} \frac{f_{Y|X}(Y|x+kt)\Pr(\Hc_k)}{f_Y(Y)} \right] \left(\sum_{j=0}^{\Msf-1}f_X(x+jt)\right){\rm d}x  \nonumber \\
    & \overset{\rm (c)}{=} \Msf -  \int_{-\infty}^\infty \EE\left[\max_{k\in[0:\Msf-1]} \frac{f_{Y|X}(Y|x+kt) f_X(x+kt)}{f_Y(Y)} \right] {\rm d}x  \nonumber \\
    & = \Msf - \int_{-\infty}^\infty \int_{-\infty}^\infty \frac{1}{2\pi \sqrt{\eta}} {\rm{e}}^{-\frac{1}{2\eta}\left (\min_{k\in[0:\Msf-1]} \left \{ (y-x-kt)^2+\eta(x+kt)^2 \right \} \right )} {\rm d}y \ {\rm d}x ,
\end{align}
\hrule
\end{figure*}
where the labeled equalities in~\eqref{eq:low_noise_szzb_Gauss_ex_proof1} follow from:
$\rm (a)$ using the hypothesis testing problem defined in Definition~\ref{def:MHT} with $\mu_\Uc=\sum_{j=0}^{\Msf-1}P_{X-jt}$ in Definition~\ref{def:h_func};
$\rm (b)$ applying the Bayes' rule;
and
$\rm (c)$ the fact that $\Pr(\Hc_k) = \frac{f_X(x+kt)}{\sum_{j=0}^{\Msf-1}f_X(x+jt)}$,  for all $k \in [0:\Msf-1]$.

Solving $\min_{k\in[0:\Msf-1]} \{(y-x-kt)^2+\eta(x+kt)^2 \}$ for $k$ (i.e., taking the first derivative with respect to $k$, setting it equal to zero, and solving for $k$ keeping in mind that $k$ has to be integer), we obtain
\begin{equation}\label{eq:low_noise_szzb_Gauss_ex_proof2}
    k^\star
    = \begin{cases}
        0 & \text{ if } \frac{y-(1+\eta)x}{(1+\eta)t} \leq \frac{1}{2}, \\
        j & \text{ if } j-\frac{1}{2} \!<\! \frac{y\!-\!(1\!+\!\eta)x}{(1+\eta)t} \leq j \!+\! \frac{1}{2},j\in[1:\Msf\!-\!2], \\
        \Msf-1 & \text{ if } \Msf-\frac{3}{2} < \frac{y-(1+\eta)x}{(1+\eta)t}.
    \end{cases}
\end{equation}
Using~\eqref{eq:low_noise_szzb_Gauss_ex_proof2}, it is a simple exercise to show that~\eqref{eq:low_noise_szzb_Gauss_ex_proof1} becomes
\begin{align}
    & \hsf_\Msf (t, \ev_1, P_{X,Y}) \nonumber \\
    & =  (\Msf-1) \left(1- F_Z\left(\frac{\sqrt{\eta+1}}{2\sqrt{\eta}}t\right) + F_Z\left(-\frac{\sqrt{\eta+1}}{2\sqrt{\eta}}t\right) \right), 
    \label{eq:NewExample1}
\end{align}
where we have let $f_Z(\cdot)$ and $F_Z(\cdot)$ be the PDF and the cumulative distribution function (CDF) of $Z\sim\Nc(0,1)$, respectively, and we have used the following identity,
\begin{equation}
    \int_{-\infty}^\infty f_Z(x) F_Z(a+bx) \ {\rm d}x
    = F_Z\left(\frac{a}{\sqrt{1+b^2}}\right).
\end{equation} 
By substituting~\eqref{eq:NewExample1} inside the SZZB in Theorem~\ref{thm:SZZB} yields
\begin{equation}
    \Zsf\Zsf_{\rm sp}(P_{X,Y},\Msf)
     = \sup_{t>0} t^2 Q\left(\frac{\sqrt{\eta+1}}{2\sqrt{\eta}} t\right)  = \gamma \frac{\eta}{\eta+1},
\end{equation} 
where $Q(x) = \int_{x}^{\infty} f_Z(u){\rm d}u$ and $\gamma = 4 \sup_{t>0}t^2 Q(t)$. Therefore, we obtain
\begin{equation}
    \lim_{\eta\to0}\frac{\Zsf\Zsf_{\rm sp}(P_{X,Y},\Msf)}{\eta}
     = \lim_{\eta\to0} \frac{\gamma}{\eta+1}  = \gamma,
\end{equation}
which concludes the proof of Example~\ref{ex:szzb_low_noise_gaussian}.

\subsection{Proof of Example~\ref{ex:bern}}\label{app:ex_discrete_ber}
From Theorem~\ref{thm:zzb_sig_inf_M}, it follows that
\begin{align}\label{eq:ex_ber_proof1}
	& \mathsf{H}_\Msf(t,1,P_X) \nonumber \\
	& =  \sum_{x\in\overline{\Sc}_t}  \max_{j\in[0:\Msf-1]} p_{X }(x+jt) \nonumber \\
    & =  \sum_{x\in\overline{\Sc}_t}  \max_{j\in[0:\Msf-1]} \left\{(1-p)\mathbbm{1}\{x=-jt\} + p\mathbbm{1}\{x = 1-jt\} \right\},
\end{align}
where $\overline{\Sc}_t = \cup_{i\in[0:\Msf-1]} \{-it, 1 - it\}$,
with $|\overline{\Sc}_t| \leq 2\Msf$. We now consider two cases separately.

\noindent {\bf{Case~1:}} $|\overline{\Sc}_t| = 2\Msf$. In this case, the supports of the random variables $X-it, i \in [0:\Msf-1]$ are all disjoint. Thus, from~\eqref{eq:ex_ber_proof1}, we obtain $\mathsf{H}_\Msf(t,1,P_X) = \Msf$ since the right-hand side of~\eqref{eq:ex_ber_proof1} is the sum of $\Msf$ PDFs of shifted Bernoulli random variables. 

\noindent {\bf{Case~2:}} $|\overline{\Sc}_t| < 2\Msf$. In this case, there exists a non-empty intersection between $\{-kt,1-kt\}$ and $\{\ell t,1-\ell t\}$ for some $k\neq\ell,$ $(k,\ell)^2\in[0:\Msf-1]^2$. 
Without loss of generality, we assume that $k<\ell$.
Then, since $t \geq 0$, there exists a non-empty intersection only when $1-\ell t = - kt$.
For example, if $t=1$, we have that $1-\ell t = - kt$ for all $k=\ell-1$, which results in $\overline{\Sc}_1 = \{-(\Msf-1), - (\Msf-2), \cdots, 0,1\}$. Thus, the condition $1-\ell t = - kt$, where $k < \ell$, implies that
\begin{equation}
   t = \frac{1}{\ell-k}.
\end{equation}
The two cases above show that $|\overline{\Sc}_t|<2\Msf$ if $t \in \left \{ \frac{1}{\Msf-\kappa}: \kappa\in[1:\Msf-1] \right \}$, and $|\overline{\Sc}_t|=2\Msf$ otherwise.
Thus, we can rewrite~\eqref{eq:ex_ber_proof1} as follows,
\begin{equation}
    \mathsf{H}_\Msf(t,1,P_X)
    = \Msf,\text{ if }t \notin \left\{ \frac{1}{\Msf-\kappa}: \kappa\in[1:\Msf-1] \right\}.
\end{equation}
Thus, it remains to understand the value of $\mathsf{H}_\Msf(t,1,P_X)$ when $t=\frac{1}{\Msf-\kappa}$ for some $\kappa\in[1:\Msf-1]$. In this case, there are $\kappa$ pairs of $(k,\ell)\in[0:\Msf-1]^2$ satisfying $k<\ell$ and $t= \frac{1}{\Msf-\kappa} = \frac{1}{\ell-k}$. Specifically, for each $k\in[0:\kappa-1]$, we have that $\ell=\Msf-\kappa+k\in[\Msf-\kappa:\Msf-1]$.
With such a pair $(k,\ell)$, we obtain
\begin{align}\label{eq:ex_ber_proof1_a}
    \mathbbm{1}\{x = -kt\} 
    & = \mathbbm{1}\left\{x  = -\frac{k}{\ell-k}\right\} \nonumber \\
    & = \mathbbm{1}\left\{x  = 1 - \frac{\ell}{\ell-k}\right\} \nonumber \\
    & = \mathbbm{1}\{x  = 1 - \ell t\} .
\end{align}
Thus, we observe that for $t=\frac{1}{\Msf-\kappa}$, a total number $\kappa$ of $x\in\overline{\Sc}_t$ are such that
\begin{equation}
    \mathbbm{1}\{x = -kt\} 
     = \mathbbm{1}\{x  = 1 - \ell t\}.
\end{equation}
We group these $x$'s inside ${\Sc}^{(\kappa)}_t \subseteq \overline{\Sc}_t$ with $ |{\Sc}^{(\kappa)}_t|= \kappa$.
From~\eqref{eq:ex_ber_proof1}, we obtain
\begin{align}
	& \mathsf{H}_\Msf(t,1,P_X) \nonumber\\
	& = \sum_{x\in{\Sc}^{(\kappa)}_t}  \max_{j\in[0:\Msf-1]} \left\{(1-p)\mathbbm{1}\{x=-jt\} + p\mathbbm{1}\{x = 1-jt\} \right\} \notag
 \\& \quad + \!\!\!\!\sum_{x\in\overline{\Sc}_t \setminus {\Sc}^{(\kappa)}_t} \! \max_{j\in[0:\Msf-1]} \left\{(1-p)\mathbbm{1}\{x=-jt\} \!+\! p\mathbbm{1}\{x = 1-jt\} \right\} \notag 
 \\& = \sum_{x\in{\Sc}^{(\kappa)}_t} \max \{1-p,p\} + (\Msf - \kappa) \notag.
 \\& = \Msf - \kappa \min\{p,1-p\}.
\end{align}
In summary, by putting together the two cases $|\overline{\Sc}_t| = 2\Msf$ and $|\overline{\Sc}_t| < 2\Msf$, we obtain
\begin{align}\label{eq:ex_ber_proof2}
    & \mathsf{H}_\Msf(t,1,P_X) \nonumber \\
    & = \begin{cases}
        \Msf - j\min\{p,1-p\} & \text{ if } t = \frac{1}{\Msf-j},~j\in[1:\Msf-1], \\
        \Msf & \text{ otherwise}.
    \end{cases}
\end{align}
From the above, we have that
\begin{align}
    & \Vc_t\{\Msf - \mathsf{H}_\Msf(t,1,P_X)\} \nonumber \\
    & = \sup_{u:u\geq t} \{\Msf - \mathsf{H}_\Msf(t,1,P_X) \} \nonumber \\
    & = \begin{cases}
        (\Msf-1)\min\{p,1-p\} & \text{ if } t \leq 1, \\
        0 & \text{ otherwise}.
    \end{cases}
\end{align}
Thus, from Theorem~\ref{thm:zzb_sig_inf_M}, we arrive at
\begin{align}
    \overline{\mathsf V}(P_X,\Msf)
    & = \int_0^\infty \frac{t}{2} \frac{\Vc_t\{\Msf - \mathsf{H}_\Msf(t,1,P_X)\}}{\Msf-1} \ {\rm d} t \nonumber \\
    & = \frac{1}{4}\min\{p,1-p\},
\end{align}
and
\begin{align}
    \Vsf_{\rm sp}(P_X,\Msf)
    & = \sup_{t>0} \frac{t^2}{2(\Msf-1)} (\Msf - \mathsf{H}_\Msf(t,1,P_X)) \nonumber \\
    & = \frac{1}{2} \min\{p,1-p\},
\end{align}
where the supremum is attained at $t=1$.
This concludes the proof of Example~\ref{ex:bern}.

\subsection{Proof of Example~\ref{ex:unif_mixed_unif}}\label{app:no_ineq}

\noindent {\bf Case 1} ($X\sim f_{X_1}$):
Consider the uniform random variable $X\in[0,1]$ with its PDF $f_X(x) = \mathbbm{1}\{x\in[0,1]\}$.
We start by noting that $\EE[X] = 1/2$ and $\EE[X^2] = 1/3$. Thus, $\var(X) = 1/12$.
To compute $\overline{\Vsf}(P_X, \Msf)$, ${\Vsf}(P_X, \Msf)$, and ${\Vsf}_{\rm sp}(P_X, \Msf)$, we only need to know $\mathsf{H}_\Msf(t,1,P_X)$ in~\eqref{eq:h_high_noise}. For $t\geq0$, we have that
\begin{align}
	\mathsf{H}_\Msf(t,1,P_X)
	& = \int_{-\infty}^\infty  \max_{k\in[0:\Msf-1]} f_{X-kt}(x) \ {\rm d} x \nonumber \\
	& = \int_{-\infty}^\infty  \max_{k\in[0:\Msf-1]} \mathbbm{1}\{x\in[-kt,1-kt] \} \ {\rm d} x \nonumber \\
	& = \begin{cases}
		1 + (\Msf-1)t & \text{ if } 0 \leq t  \leq 1 ,\\
		\Msf & \text{ if } 1 < t.
	\end{cases}
\end{align}
from which we obtain
$\overline{\Vsf}(P_X, \Msf) = {\Vsf}(P_X, \Msf)$ (since the valley-filling function does not affect the bound) and 
\begin{equation}
	\overline{\Vsf}(P_X, \Msf)
	 =  \int_0^1 \frac{t}{2} (1-t) \ {\rm d} t  = \frac{1}{12}.
\end{equation}
Moreover, 
\begin{equation}
	\Vsf_{\rm sp}(P_X, \Msf)
	 = \sup_{0<\Delta \leq1} \frac{\Delta^2}{2} (1-\Delta) = \frac{2}{27}.
\end{equation}
Therefore, $\overline{\Vsf}(P_X, \Msf) = {\Vsf}(P_X, \Msf) > \Vsf_{\rm sp}(P_X, \Msf)$ for any integer $\Msf\geq 2$.

Furthermore, an MMSE lower bound with an infinite number of hypotheses is given by~\cite[eq.~(28) and eq.~(32)]{GenOutageBoundsApproach}
\begin{align}
    & {\rm mmse}(X|\Ym) \nonumber \\
    & \geq \frac{1}{2} \int_{0}^\infty t\left(1 - \EE\left[\int_0^t \max_{l\in\ZZ}\{f_{X|\Ym}(x+lt|\Ym)\} {\rm d}x \right] \right) {\rm d}t.
\end{align}
In the high-noise regime, we regard $f_{X|\Ym}$ to be independent of $\Ym$, i.e., $f_{X|\Ym} = f_{X}$, leading to 
\begin{align}
    \var(X)
    \geq \frac{1}{2} \int_{0}^\infty \!\! t \! \left(1 - \int_0^t \max_{l\in\ZZ}\{f_{X}(x+lt) \} {\rm d}x  \right) {\rm d}t.
    \label{eq:NewBoundInfHypCase1}
\end{align}
For $t\geq 0$, the considered uniform random variable $X$ yields
\begin{align}
    \int_0^t \max_{l\in\ZZ}\{f_{X}(x+lt) \} {\rm d}x
    & = \begin{cases}
        t & \text{ if } 0\leq t \leq 1, \\
        1 & \text{ if } 1\leq t . 
    \end{cases}
\end{align}
Then, the ZZB with infinite number of hypotheses~\cite{GenOutageBoundsApproach}, i.e., the right-hand side of~\eqref{eq:NewBoundInfHypCase1}, is
\begin{align}
    \frac{1}{2} \int_{0}^\infty t\left(1 - \int_0^t \max_{l\in\ZZ}\{f_{X}(x+lt) \} {\rm d}x  \right) {\rm d}t
    & = \frac{1}{12},
\end{align}
which is equal to $\var(X)$. Thus, also the valley-filling version of such a bound is equal to $\frac{1}{12}$.

\noindent {\bf Case 2} ($X\sim f_{X_2}$):
Consider $f_X(x) = \mathbbm{1}\{x\in[0,1/2]\} + \mathbbm{1}\{x\in[1,3/2]\} $. We start by noting that $\EE[X] = 3/4$ and $\EE[X^2] = 5/6$. Thus, $\var(X) = 13/48$.
Then, by partitioning $[0,\infty)$ into several regions (since $t \geq 0$), we obtain
\begin{align}
\label{eq:HmExample3Case2}
	&\! \mathsf{H}_\Msf(t,1,P_X) \nonumber \\
	& \!= \int_{-\infty}^\infty \max_{k\in[0:\Msf-1]} f_X(x+kt) \ {\rm d}x \nonumber \\
	& \!= \begin{cases}
		2(\Msf-1) t + 1 & \text{ if } 0 \leq t < \frac{1}{2(\Msf-1)} , \\
		(\Msf-1)t + \frac{3}{2} & \text{ if } \frac{1}{2(\Msf-1)} \leq t < \frac{1}{2} , \\
		(\Msf-3)t + \frac{5}{2} & \text{ if } \frac{1}{2} \leq t < \frac{3}{4} , \\
		- (\Msf-1)t + \frac{3(\Msf-1)}{2} \!+\! 1 & \text{ if } \frac{3}{4} \leq t < 1 , \\
		(\Msf-1)t -\frac{1}{2}(\Msf-1) \!+\! 1 & \text{ if } 1 \leq t < \frac{3}{2} , \\
		\Msf & \text{ if } \frac{3}{2} \leq t.
	\end{cases}
\end{align}
It is a simple exercise to show that, for $\Msf = 2 $, we have that
\begin{equation}
\Vc_t\left \{ \frac{2 - \mathsf{H}_2(t,1,P_X)}{1}\right \}
= \begin{cases}
1-2t & \text{ if } 0 \leq t < \frac{1}{4}, \\
\frac{1}{2} & \text{ if } \frac{1}{4} \leq t < 1, \\
\frac{3}{2}-t & \text{ if } 1 \leq t < \frac{3}{2}, \\
0 & \text{ if } \frac{3}{2} \leq t,
\end{cases}
\end{equation}
and for $\Msf \geq 3$, we have that
\begin{align}
	& \Vc_t\left\{\frac{\Msf - \mathsf{H}_\Msf(t,1,P_X)}{\Msf-1}\right\} \nonumber \\
	&  = \begin{cases}
		1 - 2t & \text{ if } 0 \leq t < \frac{1}{2(\Msf-1)} , \\
		-t + \frac{\Msf-3/2}{\Msf-1} & \text{ if } \frac{1}{2(\Msf-1)} \leq t < \frac{\Msf-2}{2\Msf-2} , \\
		\frac{1}{2} & \text{ if } \frac{\Msf-2}{2\Msf-2} \leq t < 1 , \\
		-t + \frac{3}{2}& \text{ if } 1 \leq t < \frac{3}{2} , \\
		0 & \text{ if } \frac{3}{2} \leq t  .
	\end{cases}
\end{align}
With the above, the ZZB in Theorem~\ref{thm:zzb_sig_inf_M}  with the valley-filling function is given by
\begin{align}
	\overline{\Vsf}(P_X, \Msf) 
	& =  \int_0^\infty \frac{t}{2} \Vc_t\left\{\frac{\Msf - \mathsf{H}_\Msf(t,1,P_X)}{\Msf-1}\right\} \ {\rm d}t \nonumber \\
	& = \begin{cases}
 \frac{77}{384} & \text{ if } \Msf=2,\\
 \frac{20\Msf^2 - 43\Msf + 26}{96(\Msf-1)^2}  & \text{ if } \Msf \geq 3,
 \end{cases}
\end{align}
where note that $\overline{\Vsf}(P_X, \Msf) \leq \frac{5}{24}$ for $\Msf \geq 3$, where the inequality holds with equality if $\Msf\to\infty$.

By using~\eqref{eq:HmExample3Case2}, the ZZB in Theorem~\ref{thm:zzb_sig_inf_M} without the valley-filling function is given by
\begin{align}
    \Vsf(P_X, \Msf)
    & = \int_0^\infty \frac{t}{2} \left(\frac{\Msf - \mathsf{H}_\Msf(t,1,P_X)}{\Msf-1}\right)  {\rm d}t \nonumber \\
    & = \frac{71\Msf^3-232\Msf^2 + 251\Msf - 86}{384(\Msf-1)^3}  \leq \frac{71}{384},
\end{align}
where the inequality holds with equality if $\Msf\to\infty$.

Next, using~\eqref{eq:HmExample3Case2}, we have that the SZZB in Theorem~\ref{thm:zzb_sig_inf_M} is given by
\begin{align}
	{\Vsf}_{\rm sp}(P_X, \Msf) 
	& = \sup_{t>0} \frac{t^2}{2} \frac{\Msf - \mathsf{H}_\Msf(t,1,P_X)}{\Msf-1} \nonumber \\
	& = \frac{1}{4},
\end{align}
where the supremum is attained at $t=1$. 

Hence, ${\Vsf}_{\rm sp}(P_X, \Msf)  > \overline{\Vsf}(P_X, \Msf) > {\Vsf}(P_X, \Msf)$ for $\Msf\geq 2$.

Similar to Case~1, an MMSE lower bound with an infinite number of hypotheses~\cite[eq.~(28) and (32)]{GenOutageBoundsApproach} in the high-noise regime is given by~\eqref{eq:NewBoundInfHypCase1}.
For $t\geq0$, the inner integral of the right-hand side of~\eqref{eq:NewBoundInfHypCase1} with respect to $x$ is equal to
\begin{align}
    \int_0^t \max_{l\in\ZZ}\{f_{X}(x+lt) \} {\rm d}x
    & = \begin{cases}
        t & \text{ if } 0\leq t \leq \frac{3}{4}, \\
        \frac{3}{2} - t & \text{ if } \frac{3}{4} < t \leq 1, \\
        t - \frac{1}{2} & \text{ if } 1< t \leq \frac{3}{2}, \\
        1 & \text{ if } \frac{3}{2} < t.
    \end{cases}
\end{align}
With this, the right-hand side of~\eqref{eq:NewBoundInfHypCase1} is given by
\begin{align}
    \frac{1}{2} \int_{0}^\infty t\left(1 - \int_0^t \max_{l\in\ZZ}\{f_{X}(x+lt) \} {\rm d}x  \right) {\rm d}t
    & = \frac{71}{384}.
\end{align}
Furthermore, the valley-filling version of this bound is given by
\begin{align}
    \frac{1}{2} \int_{0}^\infty t\Vc_t\left\{1 - \int_0^t \max_{l\in\ZZ}\{f_{X}(x+lt) \} {\rm d}x  \right\} {\rm d}t
    & = \frac{5}{24}.
\end{align}

\section*{Acknowledgment}
The authors would like to thank the Associate Editor and the Reviewers for their suggestions.

\bibliographystyle{IEEEtran}
\bibliography{MMSE_LB}

\begin{IEEEbiographynophoto}{Minoh Jeong} is currently a Postdoctoral Research Fellow in the Department of Electrical Engineering and Computer Science at the University of Michigan. In 2024, he received a Ph.D. degree from the Department of Electrical and Computer Engineering at the University of Minnesota. He received his B.S. degree in 2017 from Inha University, and M.S. degree in 2019 from Ajou University. His current research interest are in the areas of statistical signal processing, machine learning, and information theory.
\end{IEEEbiographynophoto}


\begin{IEEEbiographynophoto}{Alex Dytso}
(Senior Member, IEEE) received the Ph.D. degree from the Department of Electrical and Computer Engineering, University of Illinois, Chicago, in 2016. From September 2016 to August 2020, he was a PostDoctoral Associate with the Department of Electrical Engineering, Princeton University. From 2020 to 2022, he was an Assistant Professor with the Department of Electrical and Computer Engineering, New Jersey Institute of Technology (NJIT). Currently, he is a Staff Engineer with Qualcomm Flarion Technologies Inc. His current research interests include the areas of multi-user information theory and estimation theory, and their applications in wireless networks.
\end{IEEEbiographynophoto}

\begin{IEEEbiographynophoto}{Martina Cardone} (Senior Member, IEEE) received the Ph.D. degree in electronics and communications from T\'el\'ecom ParisTech (with work done at Eurecom in Sophia Antipolis, France) in 2015. She is currently an Assistant Professor with the Electrical and Computer Engineering Department, University of Minnesota. From July 2015 to August 2017, she was a Postdoctoral Research Fellow with the Electrical and Computer Engineering Department, UCLA Henry Samueli School. Her main research interests are in estimation theory, network information theory, network coding, and wireless networks with a special focus on their capacity, security, and privacy aspects. She is a recipient of the 2022 McKnight Land-Grant Professorship, the NSF CAREER Award in 2021, the NSF CRII Award in 2019, the Outstanding Ph.D. Award at T\'el\'ecom ParisTech (Paris, France), and the Qualcomm Innovation Fellowship in 2014.
\end{IEEEbiographynophoto}

\end{document}